\documentclass[aps,prb,twocolumn,superscriptaddress,showpacs,english]{revtex4-1}
\usepackage[T1]{fontenc}
\usepackage[latin9]{inputenc}
\usepackage{babel}
\usepackage{amsmath}
\usepackage{amssymb}
\usepackage{graphicx}
\usepackage{xcolor}
\usepackage{esint}
\definecolor{mydarkgreen}{rgb}{0.0,0.5,0.0}
\usepackage[linktocpage=true,
  colorlinks=true, 
  pdfborder={0 0 0},
  linkcolor=blue,
  citecolor=red,
  filecolor=yellow,
  urlcolor=mydarkgreen,
  bookmarks,
  pdftitle={paper},
  pdfauthor={},
]{hyperref} 

\makeatletter

\usepackage{babel}
\usepackage{upgreek}

\makeatother

\begin{document}

\def\EliashbergGF{$\bf{I}$.113}
\def\EliashbergOne{$\bf{I}$.130}
\def\EliashbergTwo{$\bf{I}$.132}
\def\NonLinearGapEq{$\bf{I}$.95}
\def\SubDerivationXCPotential{$\bf{I}$.C.}
\def\LinearSSOperator{$\bf{I}$.80}
\def\LinearSSOperatorM{$\bf{I}$.91}
\def\LinearSSOperatorD{$\bf{I}$.92}
\def\LinearSSOperatorCph{$\bf{I}$.93}
\def\LinearSSOperatorCc{$\bf{I}$.94}
\def\ExcitationDiscussion{$\bf{I}$.III.A.2.d}
\def\SubDerivationXCPotentialNonLinear{$\bf{I}$.C.2}
\def\aSqareFDiagonal{$\bf{I}$.122}
\def\aSqareFOffDiagonal{$\bf{I}$.123}
\def\CoulombCoupl{$\bf{I}$.124}
\def\OrderParameterDefinition{$\bf{I}$.4}
\def\AvaragingProcedure{$\bf{I}$.120}
\def\KSSEPhOneOne{$\bf{I}$.66}
\def\KSSEPhOneMinusOne{$\bf{I}$.67}
\def\KSSEPhMinusOneOne{$\bf{I}$.68}
\def\KSSEPhMinusOneMinusOne{$\bf{I}$.69}
\def\KSSECoulOneMinusOne{$\bf{I}$.75}
\def\KSSECoulMinusOneOne{$\bf{I}$.76}
\def\MatsubaraIntegralMph{$\bf{I}$.73}
\def\SpinSCDFTSDA{$\bf{I}$.III.A.2.c}
\def\EliashbergCalc{$\bf{I}$.IV.A.1}

\title{Ab-Initio Theory of Superconductivity in a Magnetic Field II. : Numerical
solution.}

\author{A. Linscheid}
\affiliation{Max Planck Institute of Microstructure Physics, Weinberg 2, D-06120 Halle, Germany.}
\author{A. Sanna}
\affiliation{Max Planck Institute of Microstructure Physics, Weinberg 2, D-06120 Halle, Germany.}
\author{E.K.U. Gross}
\affiliation{Max Planck Institute of Microstructure Physics, Weinberg 2, D-06120 Halle, Germany.}

\begin{abstract}
We numerically investigate the Spin Density Functional theory for superconductors (SpinSCDFT) and
the approximated exchange-correlation functional, derived and presented
in the preceding paper $\bf I$.
As a test system we employ a free electron gas featuring an
exchange-splitting, a phononic pairing field and a Coulomb
repulsion. SpinSCDFT results are compared with Sarma, the Bardeen
Cooper and Schrieffer theory and with an Eliashberg type of approach.
We find that the spectrum of the superconducting Kohn-Sham SpinSCDFT
system is not in agreement with the true quasi particle structure.
Therefore, starting from the Dyson equation, we derive a scheme
that allows to compute the many body excitations of the superconductor
and represents the extension to superconductivity of the $\rm{G}_{0}\rm{W}_{0}$
method in band structure theory. This superconducting $\rm{G}_{0}\rm{W}_{0}$ method
vastly improves the predicted spectra.
\end{abstract}

\maketitle

\section{Introduction}

Interaction between the magnetic (M) and superconducting (SC) order
leads to complex and fascinating phenomena. Apart from the
Meissner effect as the most apparent aspect of this interaction
on macroscopic length scales, for singlet
superconductors, the ferromagnetic parallel spin alignment competes with
spin anti-parallel Cooper pair formation. While for triplet superconductors
such as $\rm{UGe}_{2}$ a ferromagnetic (F) order is possible even in a bulk geometry
\cite{SaxenaSCOnTheBorderOfItinerantElectronFerromagnetismInUGe2_2000},
F/SC interfaces or SC surfaces in an external magnetic field
allow to study the microscopic competition of a large spin-splitting also for singlet superconductors.
This may lead to spatial inhomogeneities of the SC order parameter,
such as the phase predicted by Fulde and Ferrell and Larkin and Ovchinnikov\cite{LarkinOvchinnikovInhomogeneousStateOfSuperconductors1965,FuldeFerrellSuperconductivityInAStrongSpinExchangeField1964}.
Furthermore, the spin valve behavior of complex F/SC structures\cite{Tagirov1999,Buzdin1999,Gu2002} may
provide opportunities for novel devices making use of the unique electronic
configuration that appears due to the vicinity of these two competing phases
(see Ref.~\onlinecite{BuzdinProximityEffectsInSCFerromagnetHeterostructures2005} for a review).

These effects are addressed in the theoretical literature so far mostly within
model or semi-empirical calculations due to the lack of a complete
and efficient ab-initio theory. This leaves the prediction of essential
material dependent properties as critical temperature and excitation gap
in the presence of a magnetic field out of reach.
The Spin Density Functional theory for superconductors (SpinSCDFT) approach presented
by Ref.~\onlinecite{LinscheidSpinSCDFTI} (hereafter
referred to as $\bf I$) may fill this gap, as the theory has the computational
convenience of a Kohn-Sham density functional framework and allows to calculation
of material dependent SC parameters from the crystal structure.
The SpinSCDFT is in principle exact, but relies on the approximation of
the exchange-correlation ($xc$) potential. A first approach to 
derive such an $xc$ potential relies, in turn, on the Sham-Schl\"uter equation
\cite{ShamSchlueterDFTOfTheEnergyGap1983} for a SC \cite{MarquesPhDThesis2000}
and is presented in $\bf I$.

In this work, we present numerical results for SpinSCDFT, aiming
to achieve a deeper understanding of this theoretical framework and
to characterize and validate the $xc$ potential,
as derived in $\bf I$. In particular we will investigate the properties
of the Kohn-Sham pairing function, that is a key object in SCDFT \cite{LuedersSCDFTI2005,MarquesSCDFTIIMetals2005}.
The test system we adopt for this analysis is a spin-splitted free
electron gas with a phononic and Coulomb coupling. Details of the
model will be presented in Sec.~\ref{sec:The-model-system}. One
advantage of this simplified model with a homogeneous exchange splitting
is its similarity to the starting point of Ref.~\onlinecite{VonsovskySuperconductivityTransitionMetals}
and \onlinecite{SarmaOnTheInfluenceOfAUniformExchangeFieldActingOnSC1963}
for their discussion of the Eliashberg equations and BCS theory, respectively.
As compared to Ref.~\onlinecite{VonsovskySuperconductivityTransitionMetals},
we use a different notation (compare $\bf I$) and
take a more general route which reduces to the earlier results in
the case that the magnetic field homogeneously splits the electronic
states. We will compute the temperature vs exchange splitting diagram of the model using,
apart from SpinSCDFT, the BCS theory and the Eliashberg equations.
Then, in Sec.~\ref{sec:Results-of-SpinSCDFT} we will compare our
SpinSCDFT results with the BCS approach (reviewed in Sec.~\ref{sec:BCSwithJ})
and with the reference Eliashberg method (Sec.~\ref{sec:Eliashberg}).

The SpinSCDFT Kohn-Sham system proves to give qualitatively correct
results for the $J-T$ diagram. However, we find in Sec.~\ref{sec:Results-of-SpinSCDFT}
that it does not show a physical excitation spectrum. A similar problem
is very well known in conventional DFT, and is usually called the
band gap problem. Since the excitation gap is a very important property
of superconductors, it is important to devolve methods to compute
it. Therefore, the last part of this work will be devoted to describe
an extension of the $\rm{G}_{0}\rm{W}_{0}$ method to our superconducting system and
show that it entirely solves the problem, similar to its normal state
counterpart\cite{HybertsenTheoryQuasiparticlesG0W01985}.

\section{A test system\label{sec:The-model-system}}

The model system which we will use to investigate the SpinSCDFT formalism
is based on a non interacting electron gas under the influence of
an homogeneous magnetic field $B_{{\rm {\scriptscriptstyle 0}}}$.
The energy of its electronic states $\varepsilon_{k\sigma}$, relative
to the Fermi energy $E_{f}$ ($k=\boldsymbol{k},n$ where here $n$
is a band index and we use the notation $-k=-\boldsymbol{k},n$),
reads 
\begin{equation}
\varepsilon_{k\sigma}=\frac{1}{2}\boldsymbol{k}^{2}-E_{f}-{\rm sign}(\sigma)\mu_{{\rm {\scriptscriptstyle B}}}B_{{\rm {\scriptscriptstyle 0}}}\,.
\end{equation}
The Fermi energy is defined by integrating the density of states (DOS)
up to $E_{f}$ to have $N_{{\scriptscriptstyle {\rm e}}}$ electrons
in the system. We set the density to $N_{{\scriptscriptstyle {\rm e}}}/\Omega_{{\rm {\scriptscriptstyle uc}}}=1\, a_{0}^{-3}$
($a_{0}$ is the Bohr radius and $\Omega_{{\rm {\scriptscriptstyle uc}}}$
the unit cell volume) which leads to a relatively large $E_{f}$ of
$4.78\,\mathrm{Ha}$ ignoring the small imbalance in up and down spin
occupations. We also define a center of energy between spin splitted
states $\varepsilon(k)=\frac{1}{2}(\varepsilon_{k\uparrow}+\varepsilon_{-k\downarrow})=\frac{1}{2}\boldsymbol{k}^{2}-E_{f}$
and the splitting $J(k)=\frac{1}{2}(\varepsilon_{k\uparrow}-\varepsilon_{-k\downarrow})=-\mu_{{\rm {\scriptscriptstyle B}}}B_{{\rm {\scriptscriptstyle 0}}}$.
This will prove useful since, as seen in I, many SpinSCDFT entities
depend on $k$ only via these two parameters $\varepsilon$ and $J$.

\begin{figure}
\begin{centering}
\includegraphics[width=0.8\columnwidth]{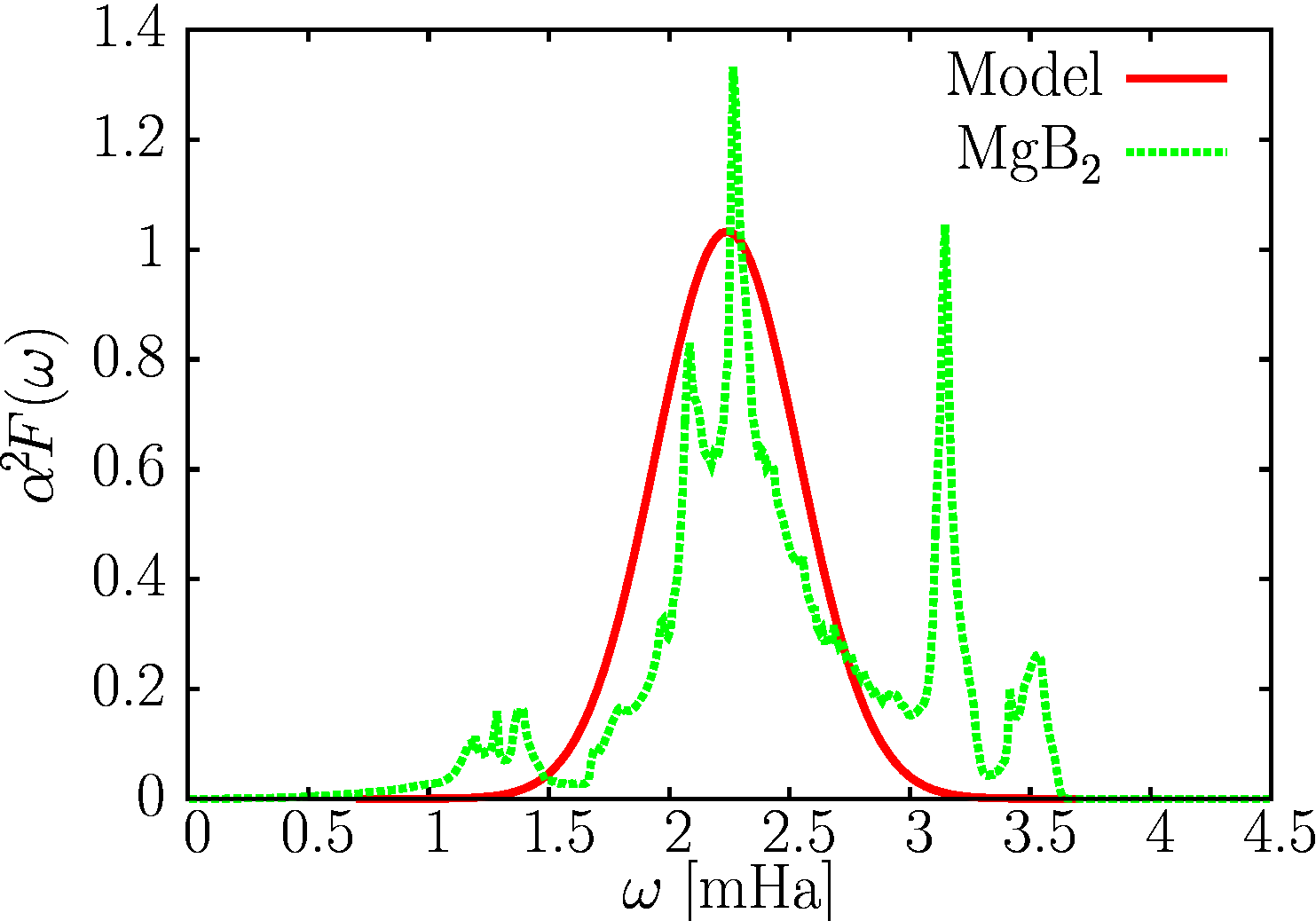}
\par\end{centering}

\caption{(color online) Model $\alpha^{\!2}\! F(\omega)$ function used in
this work (full red line) as compared to that of $\text{MgB}_{2}$
(green dotted line). \label{fig:Model_a2F}}
\end{figure}

Superconductivity is induced in this model by an electron-phonon like
attractive interaction, expressed by the Gaussian Eliashberg function\cite{AllenTheoryOfSCTc1983}:
\begin{equation}
\alpha^{\!2}\! F(\omega)=\lambda\frac{\omega}{2}\frac{1}{\omega_{w}\sqrt{\pi}}\text{e}^{-\frac{1}{2}(\frac{\omega-\omega_{0}}{\omega_{w}})^{2}}\,.\label{eq:a2f}
\end{equation}
This model depends on three parameters: $\lambda$ the electron-phonon
coupling constant \cite{AllenTheoryOfSCTc1983}; $\omega_{0}$ the
center of mass of the phonon spectrum, and $\omega_{w}$ the width
of the optical branch. In the calculations we fix these numbers to
$\omega_{0}=2.2\mbox{mHa}$ $\omega_{W}=0.5\mbox{mHa}$ and $\lambda=0.7$
which lead to coupling properties that are loosely similar to those
of ${\rm MgB}_{2}$\cite{Bohnen2001}. The resulting spectrum is plotted in Fig.~\ref{fig:Model_a2F}
\begin{figure}
\begin{centering}
\begin{minipage}[t]{0.5\columnwidth}%
\begin{center}
\includegraphics[width=0.95\columnwidth]{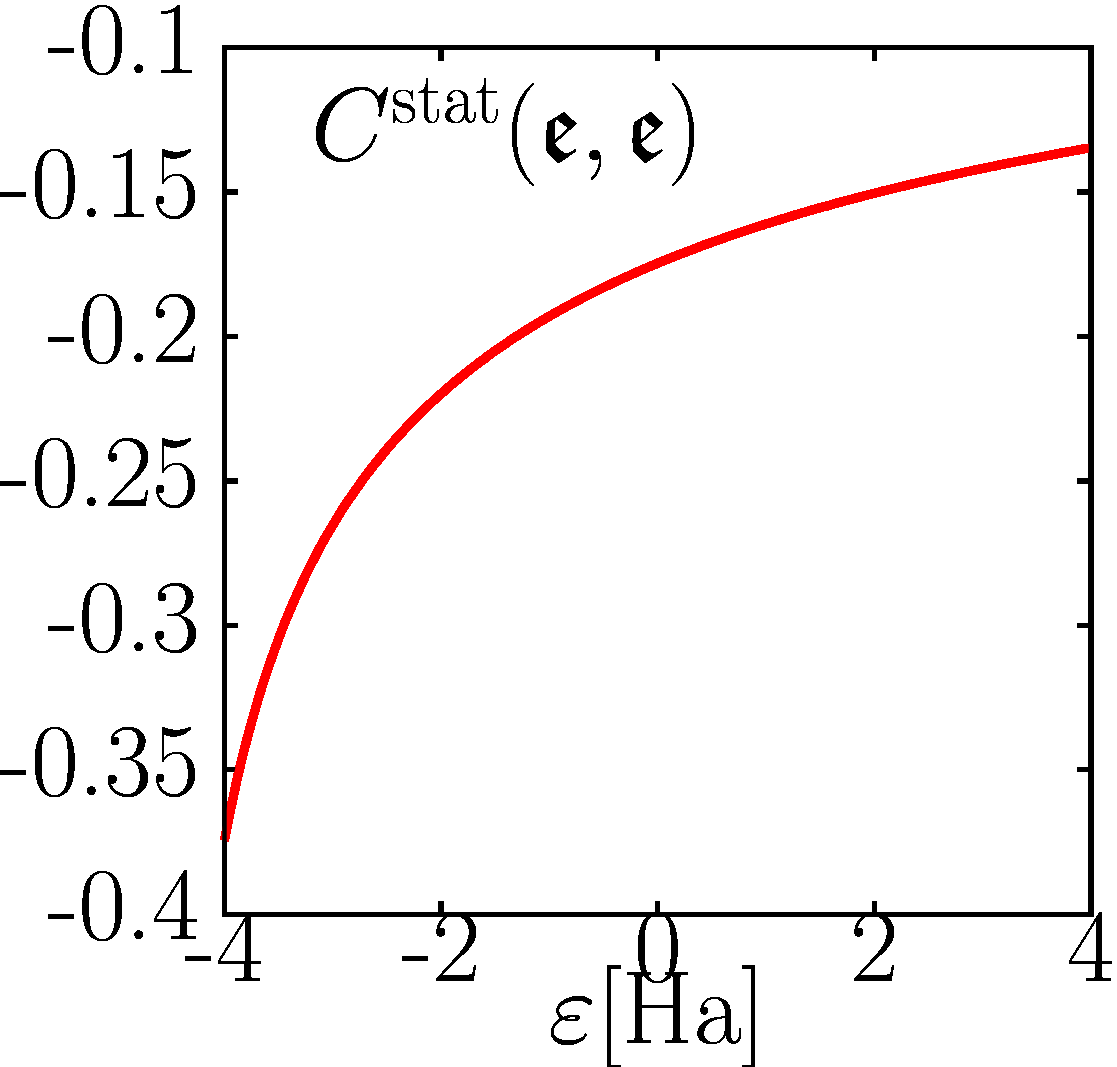}
\par\end{center}%
\end{minipage}\nolinebreak%
\begin{minipage}[t]{0.5\columnwidth}%
\begin{center}
\includegraphics[width=0.87\columnwidth]{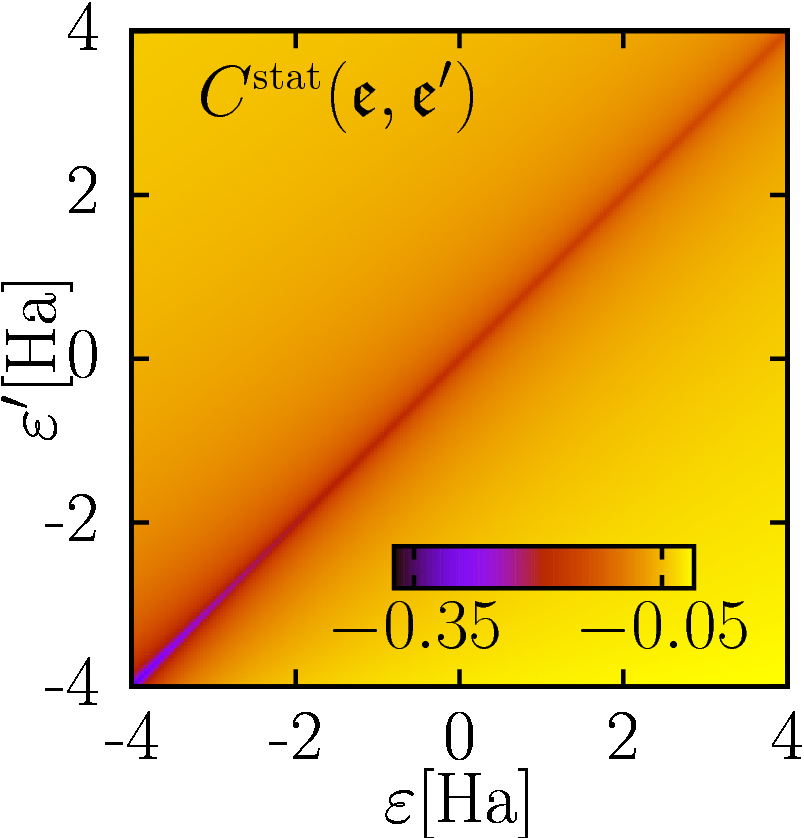}
\par\end{center}%
\end{minipage}
\par\end{centering}

\caption{(color online) Screened Coulomb interaction function $C^{{\scriptscriptstyle \text{stat}}}(\mathfrak{e},\mathfrak{e}^{\prime})$
as given by the model expression in Eq.~(\ref{eq:CoulombTFModel}).
The chosen model parameters are $E_{f}=4.78\,\mathrm{Ha}$ and $k_{\mathrm{{\scriptscriptstyle TF}}}^{2}=(0.005)^{2}\,\mathrm{Ha}$.\label{fig:CoulombTFModel}}
\end{figure}
and compared with a recomputed one of ${\rm MgB}_{2}$.  In SpinSCDFT one can
consider a general Coulomb coupling on the same footing as the phonon
interaction\cite{LuedersSCDFTI2005}.  Here, we use a simple Thomas-Fermi
based model that was used before in SCDFT\cite{LuedersSCDFTI2005,MassiddaRoleOfCoulombInSCPropOfCaC62009}.
In this model the screened Coulomb matrix elements between a state
of energy $\varepsilon$ and one of energy $\varepsilon^{\prime}$
is given by 
\begin{eqnarray}
 &  & \hspace{-0.7cm}C^{{\scriptscriptstyle \text{stat}}}(\varepsilon,\varepsilon^{\prime})\approx-\frac{\pi\rho^{{\scriptscriptstyle \text{EG}}}(\varepsilon^{\prime})}{2\sqrt{(\varepsilon+E_{f})(\varepsilon^{\prime}+E_{f})}}\times\nonumber \\
 &  & \hspace{-0.5cm}\times\ln\!\biggl(\frac{\varepsilon\!+\!\varepsilon^{\prime}\!+\!2E_{f}+2\!\sqrt{(\varepsilon\!+\! E_{f})(\varepsilon^{\prime}\!+\! E_{f})}+\frac{1}{2}k_{\mathrm{{\scriptscriptstyle TF}}}^{2}}{\varepsilon\!+\!\varepsilon^{\prime}\!+\!2E_{f}-2\!\sqrt{(\varepsilon\!+\! E_{f})(\varepsilon^{\prime}\!+\! E_{f})}+\frac{1}{2}k_{\mathrm{{\scriptscriptstyle TF}}}^{2}}\biggr)\,.\label{eq:CoulombTFModel}
\end{eqnarray}
The screening parameter is chosen to be $k_{\mathrm{{\scriptscriptstyle TF}}}^{2}=(0.005)^{2}{\rm Ha}$.
With this parameter, the shape of the model $C^{{\scriptscriptstyle \text{stat}}}(\varepsilon,\varepsilon^{\prime})$
is shown in Fig.~\ref{fig:CoulombTFModel} . All properties of the
test system depend on the Bloch vector $\boldsymbol{k}$ and the band
index $n$ only via the single particle energy $\varepsilon_{k}$.
For brevity we use the notation $\mathfrak{e}=(\varepsilon,J)\quad\int{\rm d}\mathfrak{e}=\int{\rm d}\varepsilon\int{\rm d}J$.
Further let $\updelta(\mathfrak{e}-\mathfrak{e}^{\prime})=\updelta(\varepsilon-\varepsilon^{\prime})\updelta(J-J^{\prime})$,
then we may cast a Brillouin zone integral into the isotropic formulation
with the double DOS
\begin{equation}
\varrho(\mathfrak{e})=\sum_{k}\updelta(\varepsilon-\frac{\varepsilon_{k\uparrow}+\varepsilon_{-k\downarrow}}{2})\updelta(J-\frac{\varepsilon_{k\uparrow}-\varepsilon_{-k\downarrow}}{2})\,.
\end{equation}
This quantities describes the number of states on equal center of
energy $\varepsilon$ and splitting $J$ surfaces. In our model $\varrho(\mathfrak{e})$
the external field is homogeneous. This means the number of states on equal splitting
surfaces has a delta distribution character that peaks at
$J_{{\scriptscriptstyle 0}}=-\mu_{{\rm {\scriptscriptstyle B}}}B_{{\rm {\scriptscriptstyle 0}}}$.
In the remainder of the paper, $J_{{\scriptscriptstyle {\rm 0}}}$
replaces the $J$ integrals almost everywhere so we simplify the notation
using $J_{{\scriptscriptstyle {\rm 0}}}\rightarrow J$.

\section{The BCS Theory with an Exchange Splitting\label{sec:BCSwithJ}}

\begin{figure*}
\begin{centering}
\begin{minipage}[t]{0.5\textwidth}%
\begin{center}
\includegraphics[width=0.8\textwidth]{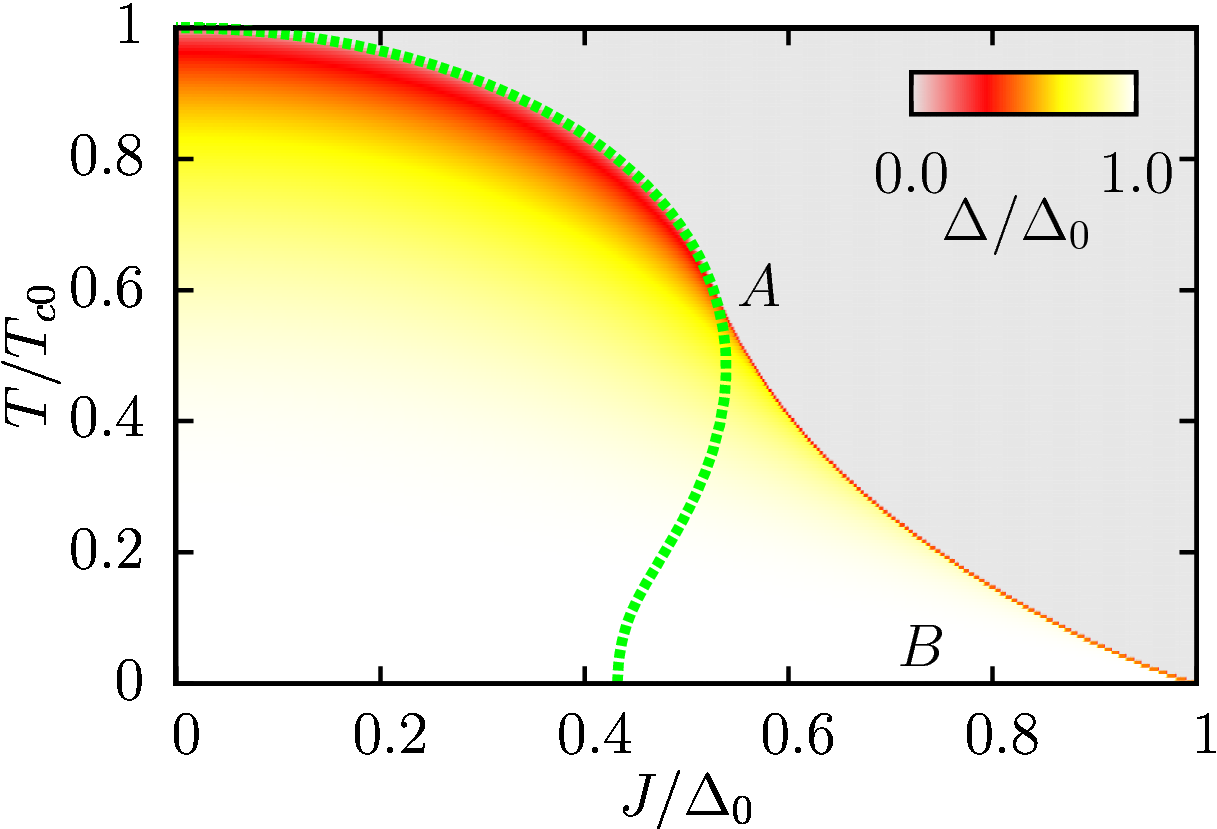}\\
a)
\par\end{center}%
\end{minipage}\nolinebreak%
\begin{minipage}[t]{0.5\textwidth}%
\begin{center}
\includegraphics[width=0.8\textwidth]{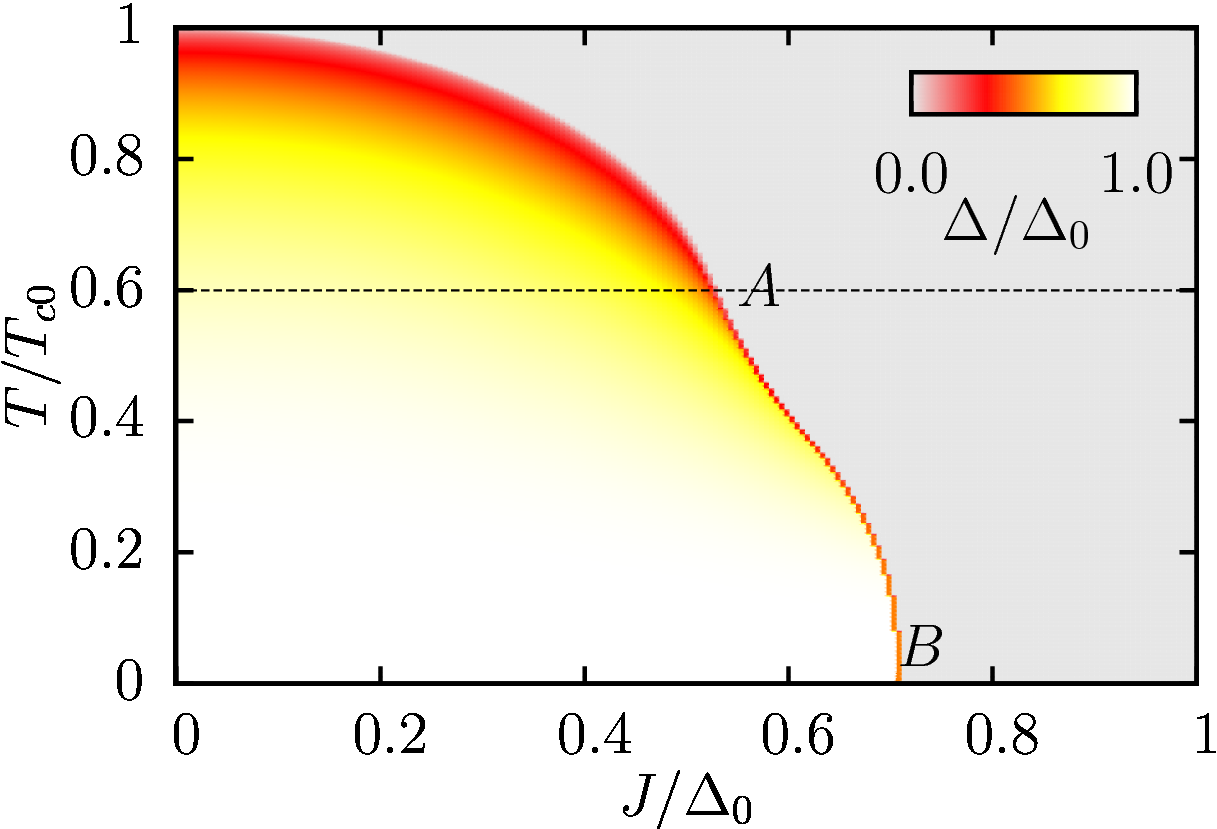}\\
b)
\par\end{center}%
\end{minipage}
\par\end{centering}

\caption{(color online) BCS solutions for a spin splitted band structure \cite{SarmaOnTheInfluenceOfAUniformExchangeFieldActingOnSC1963}.
In the panel a) we plot the solution $\varDelta$ if we can find one,
while in b) $\varDelta$ is set to zero if the free energy favors
the magnetic state. The green curve in a) shows the $T_{c}(J)$ behavior
from the linearized equation which has a curious shape that bends
inwards. Below the thin dashed line in b) at the label $A$ at $T/T_{c0}\approx0.6$
no solution with small $\varDelta$ exists and the transition is of
first order. Label $B$ at $1/\sqrt{2}$ represents the Chandrasekhar-
Clongston\cite{ChandrasekharTheMaximumCriticalFieldHighFieldSuperconductors1962,ClogstonUpperLimitCriticalFieldHardSuperconductors1962}
limit.\label{fig:BCSsolutions}}
\end{figure*}

The $J-T$ diagram of a BCS model with a homogeneous exchange splitting
parameter $J$ has been presented by Ref.~\onlinecite{SarmaOnTheInfluenceOfAUniformExchangeFieldActingOnSC1963}.
This approach, that we are going to review here, can only be used
to obtain qualitative results. Still, it will be an important guideline
in understanding the more involved Eliashberg and SpinSCDFT results
of the next sections. In a BCS model\cite{BCS1957} one replaces
the interactions among single electrons with an effective one, keeping
only the matrix elements that couple the states $k,\uparrow$ and
$-k,\downarrow$. The effective interaction is approximated with ``a
box'' centered at the Fermi level (from $-\varOmega_{{\scriptscriptstyle {\rm d}}}$
to $\varOmega_{{\scriptscriptstyle {\rm d}}}$ which is of the order
of the Debye phonon frequency to mimic phononic type of pairing and
with height $-V$). This leads to a fixed point equation for the mean
field gap $\varDelta$ \cite{SarmaOnTheInfluenceOfAUniformExchangeFieldActingOnSC1963}
\begin{eqnarray}
\frac{1}{\rho(0)V} & = & \int_{0}^{\varOmega_{{\scriptscriptstyle {\rm d}}}}\hspace{-0.2cm}\hspace{-0.2cm}\frac{\mbox{d}\varepsilon}{\sqrt{\varepsilon^{2}+\varDelta^{2}}}\bigl(f_{\beta}(J-\sqrt{\varepsilon^{2}+\varDelta^{2}})\nonumber \\
 & - & f_{\beta}(J+\sqrt{\varepsilon^{2}+\varDelta^{2}})\bigr)\,.\label{eq:BCSEquationSplitted}
\end{eqnarray}
$\rho(0)$ is the DOS at the Fermi level and $J$ is the splitting
energy between up and down states. Apart from the solutions $\varDelta$ of Eq.~\eqref{eq:BCSEquationSplitted}
there is also the trivial solution $\varDelta=0$.
We solve Eq.~(\eqref{eq:BCSEquationSplitted})
numerically as a function of $T$ and $J$%
\footnote{We use $\varOmega_{{\rm {\scriptscriptstyle d}}}=0.2$ and $\rho(0)V=1.0$
in the numerical calculation.%
}. The solutions $\varDelta(T,J)$ are presented
in Fig.~\ref{fig:BCSsolutions} a). There, we normalize $\varDelta$ to
$\varDelta_{{\scriptscriptstyle 0}}$, the solution for $T\rightarrow0$
and $J=0$. Similarly, we normalize the $J$ to $\varDelta_{{\scriptscriptstyle 0}}$
and $T$ to $T_{{\rm {\scriptscriptstyle c0}}}$, the
critical temperature for $J=0$. In this way we remove the explicit
dependence on the parameters $\rho(0)V$ and $\varOmega_{{\scriptscriptstyle {\rm d}}}$.

When one attempts to linearize Eq.~\eqref{eq:BCSEquationSplitted},
a peculiar behavior is found in that the $T_{{\rm {\scriptscriptstyle c}}}(J)$
curve bends inwards \cite{SarmaOnTheInfluenceOfAUniformExchangeFieldActingOnSC1963}.
We solve the linearized Eq.~\eqref{eq:BCSEquationSplitted} and show
the resulting $T_{{\rm {\scriptscriptstyle c}}}(J)$ as a green line
in Fig.~\ref{fig:BCSsolutions} a). As pointed out by the 
Refs.~\onlinecite{SarmaOnTheInfluenceOfAUniformExchangeFieldActingOnSC1963,PowellTheGapEquationsForSpinSingletAndTripletFerromagneticSuperconductors2003},
unlike the original BCS model at $J=0$, this equation leads to a
$J-T$ diagram in which the SC transition can be discontinuous in $\varDelta$,
i.e.~of first order. Below the temperature $T/T_{{\rm {\scriptscriptstyle c0}}}\approx0.6$
at point $A$ and the dashed line in Fig.~\ref{fig:BCSsolutions}
b) no small $\varDelta$ solution to the non-linear equation is can be found
and the initial assumption of the linearization that an arbitrary
small solutions exists is not valid.

While we can find a non-vanishing solution $\varDelta$ it may not
correspond to the stable thermodynamic phase. In Fig.~\ref{fig:BCSsolutions}
b) we remove the non-vanishing solutions $\varDelta$, if the free
energy favors the magnetic state. The resulting $T-J$ diagram shows
that for $J$ larger than to the Chandrasekhar-Clogston limit at $T=0$
\cite{ChandrasekharTheMaximumCriticalFieldHighFieldSuperconductors1962,ClogstonUpperLimitCriticalFieldHardSuperconductors1962}
of $J_{{\rm {\scriptscriptstyle c}}}=\varDelta_{{\scriptscriptstyle 0}}/\sqrt{2}$
no SC solution is stable.

Another interesting approach to describe SC in the presence of a magnetic
field is presented by Powell \textit{et al.}~\cite{PowellTheGapEquationsForSpinSingletAndTripletFerromagneticSuperconductors2003}
who use a Hubbard model in connection with a homogeneous exchange splitting.
They treat the pairing part of the interactions among electrons in
the system in the Hartree-Fock approximation, similar to BCS as described
above and consequently arrive at a similar gap equation as compared
to Eq.~(\eqref{eq:BCSEquationSplitted}). The matrix elements of the
KS system of SpinSCDFT within the spin decoupling approximation will
turn out to have a similar analytic structure.

Also, Ref.~\onlinecite{PowellTheGapEquationsForSpinSingletAndTripletFerromagneticSuperconductors2003}
discusses why the transition is of first order. They observe that
for $J<\varDelta$ and $T=0$ the gap equation (\ref{eq:BCSEquationSplitted}), and consequently
$\varDelta$, is independent on $J$. Thus $\varDelta(J)=\varDelta_{{\scriptscriptstyle 0}}$ for
$J<\varDelta_{{\scriptscriptstyle 0}}$. At $\varepsilon<\sqrt{{J}^{2}-{\varDelta}^{2}}$ on
the other hand the Fermi functions at $T=0$ in Eq.~(\ref{eq:BCSEquationSplitted})
are equal and thus cancel. Also for this type of solution $\varDelta$ must be larger than
$J$ and for $J>\varDelta$ only the trivial solution $\Delta=0$ can be found.
At this point follows that $\varDelta(T=0\rm{K},J)=\varDelta_{0}\uptheta(\varDelta_{0}-J)$ and
the transition is discontinuous $T=0$.

The above analysis will be crucial later, in Sec.~\ref{sec:Eliashberg}
and \ref{sec:Results-of-SpinSCDFT}, to guide the discussion of the
more sophisticate approaches, that feature a qualitatively similar
behavior. In the next section we will discuss results of the Eliashberg
method (as derived in $\bf I$, Sec.~IV) when applied to our test system
of Sec.~\ref{sec:The-model-system}.

\section{Solutions to the Phonon Only Eliashberg Equations\label{sec:Eliashberg}}

We solve the Eliashberg Eqs.~(\EliashbergOne) to (\EliashbergTwo).
The approximations used here, for the special case of homogeneous exchange field,
lead to equations similar to those derived by Vonsovsky \textit{et al.}\cite{VonsovskySuperconductivityTransitionMetals}.

Similar to every equation that describes a spontaneously broken symmetry,
in addition to a possible finite solution, the Eliashberg Eqs.~(\EliashbergOne)
to (\EliashbergTwo) always have the solution $\varDelta_{n}^{{\scriptscriptstyle {\rm E}}}(J)=0$.
Usually, this non-SC solution is not stable below $T_{{\rm {\scriptscriptstyle c}}}$
in the sense that small symmetry breaking fields (that in the self
consistent iteration scheme is equivalent to a small but non-zero
starting guess) lead to the finite $\varDelta_{n}^{{\scriptscriptstyle {\rm E}}}(J)$
solution via iteration of the Eliashberg Eqs.~(\EliashbergOne) to
(\EliashbergTwo). Thus, we say that in this case the $\varDelta_{n}^{{\scriptscriptstyle {\rm E}}}(J)=0$
solution has a zero basin of attraction; Only the starting value ${\varDelta}_{n}^{{\scriptscriptstyle {\rm{E\ init}}}}(J)=0$
leads to the final solution $\varDelta_{n}^{{\scriptscriptstyle {\rm E}}}(J)=0$.
Whenever $J=0$, the $\varDelta_{n}^{{\scriptscriptstyle {\rm E}}}(J)=0$
solution has a zero basin of attraction below $T_{{\rm {\scriptscriptstyle c}}}$.

\begin{figure*}
\centering{}%
\begin{minipage}[t]{1\textwidth}%
\begin{center}
\begin{minipage}[t]{0.5\textwidth}%
\begin{center}
\includegraphics[width=0.8\textwidth]{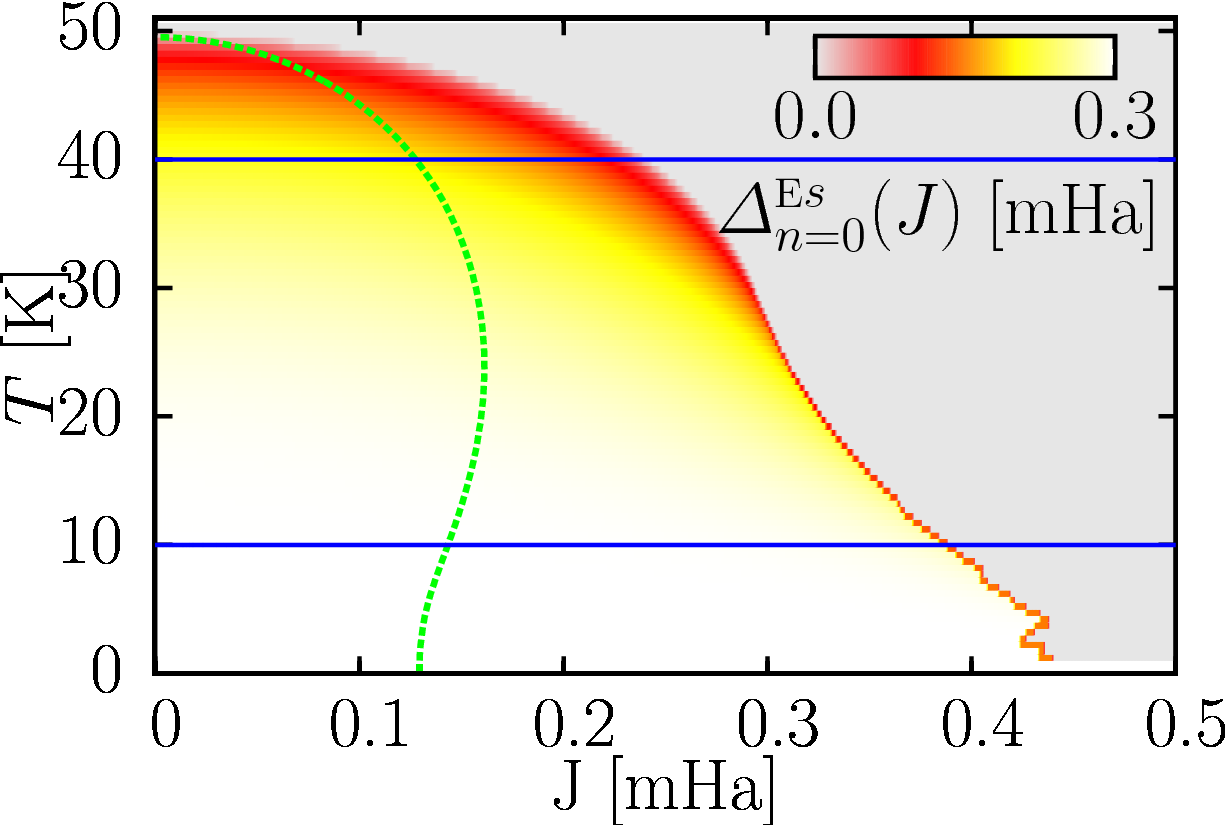}\\
a)
\par\end{center}%
\end{minipage}\nolinebreak%
\begin{minipage}[t]{0.5\textwidth}%
\begin{center}
\includegraphics[width=0.8\textwidth]{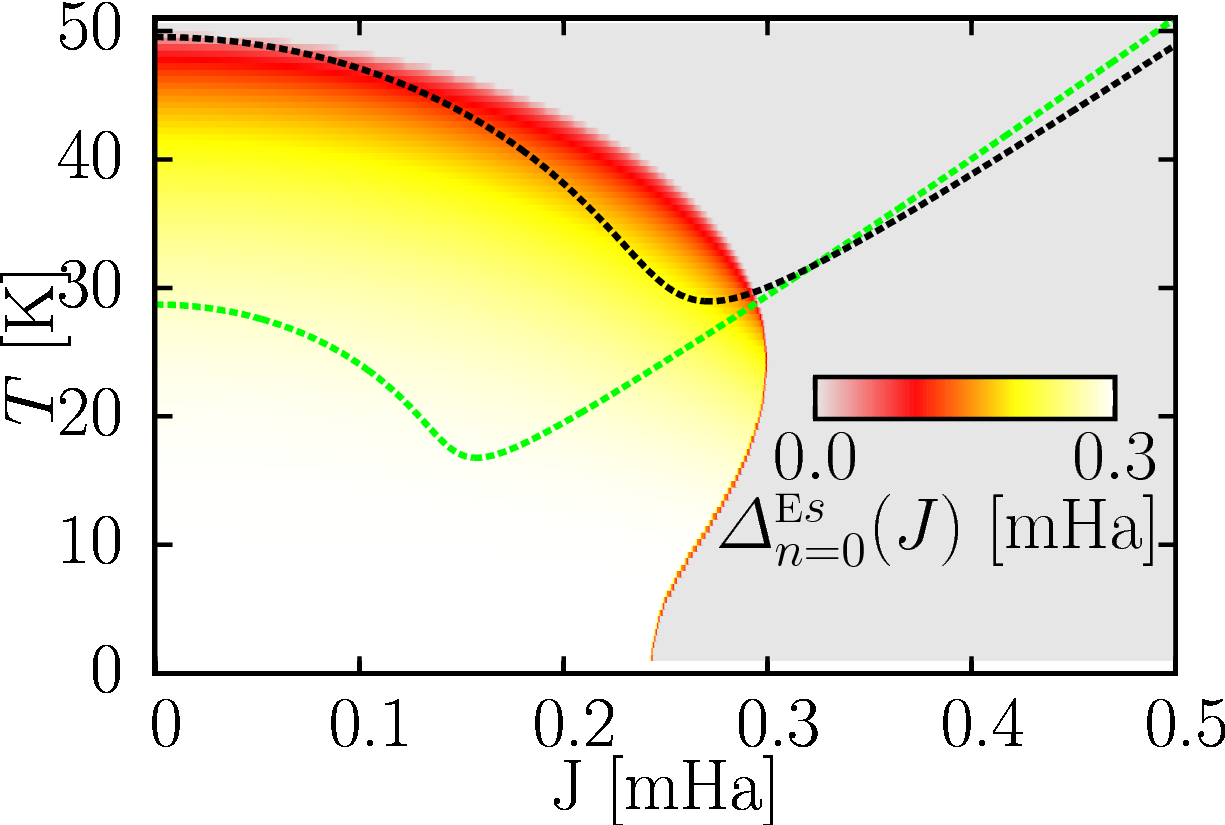}\\
b)
\par\end{center}%
\end{minipage}
\par\end{center}

\caption{(color online) $J-T$ diagram of the $n=0$ component $\varDelta_{n=0}^{{\scriptscriptstyle {\rm E}}}$
from the solution to the Eliashberg equations.
We follow the SC solution $\varDelta_{n}^{{\scriptscriptstyle {\rm E}}}(J)\neq0$
in a) or the non-SC solution $\varDelta_{n}^{{\scriptscriptstyle {\rm E}}}(J)=0$
in b) and observe that we can find a region where both are (meta)
stable. We show the full solution $\varDelta_{n}^{{\scriptscriptstyle {\rm E}}}(J)$
along the blue lines in Fig.~\ref{fig:DeltaEliashberg40K10K}. For
comparison we show the linear BCS curve as a green dotted line in
a). In b) we compare with Spin SCDFT results of Sec.~\ref{sec:Results-of-SpinSCDFT}
(green curve); the black curve is scaled on both axis by $T_{{\rm {\scriptscriptstyle c}}}^{{\rm {\scriptscriptstyle Elishberg}}}/T_{{\rm {\scriptscriptstyle c}}}^{{\rm {\scriptscriptstyle SpinSCDFT}}}$.\label{fig:PhaseDiagramEliashbergEquations}}
\end{minipage}
\end{figure*}

From Eq.~(\EliashbergGF) we know that the complex $\varDelta_{n}^{{\scriptscriptstyle {\rm E}}}$
changes the poles of the Green function. We assume the term 
$\tilde{A}_{k}^{\omega}(\omega_{n})$ to be zero, for simplicity.
Then from the analytic continuation to the real axis of Eq.~(\EliashbergGF),
we see that the energy $\omega$ of such a pole satisfies the condition
$\omega=\mbox{sign}(\sigma){J}_{k}\pm\sqrt{{\varepsilon_{k}}^{2}+{\varDelta(\omega)}^{2}}$
which is analogous to the usual Eliashberg equations
(compare also Ref.~\onlinecite{CarbottePropertiesOfBosonExchangeSC1990}).
At $T=0$, the analytic continuation of $\varDelta_{n}^{{\scriptscriptstyle {\rm E}}}$ the real axis
is purely real in the range of the Fermi energy and its value there defines the SC excitation gap \cite{CarbottePropertiesOfBosonExchangeSC1990}.
Thus, the Matsubara component $n=0$ of $\varDelta_{n}^{{\scriptscriptstyle {\rm E}}}$ is related to the
SC excitation gap of the quasi particle system. We choose this as a characteristic
property that we investigate as a function of $J$ and $T$. In the
following we generate two $J-T$ diagrams shown in Fig.~\ref{fig:PhaseDiagramEliashbergEquations}.
In a) we follow the SC solution, i.e.~we take the converged $\varDelta_{n}^{{\scriptscriptstyle {\rm E}}}(J)$
as input for the calculation at $\varDelta_{n}^{{\scriptscriptstyle {\rm E}}}(J+{\rm d}J)$,
starting at $J=0$ with ${\rm d}J$ positive. This way we compute
the diagram "from left to right" and test the stability of the $\varDelta_{n}^{{\scriptscriptstyle {\rm E}}}(J)\neq0$
solution. In b), we take the converged $\varDelta_{n}^{{\scriptscriptstyle {\rm E}}}(J)$
as input for the calculation at $\varDelta_{n}^{{\scriptscriptstyle {\rm E}}}(J-{\rm d}J)$,
starting at $J=0.5{\rm mHa}$. Thus, we generate the diagram "from right
to left". Because for large $J$ $\varDelta_{n}^{{\scriptscriptstyle {\rm E}}}(J)$
is zero, we start from a small, symmetry breaking value at $\varDelta_{n}^{{\scriptscriptstyle {\rm E}}}(J-{\rm d}J)$
instead of zero. This way we test the stability of the trivial $\varDelta_{n}^{{\scriptscriptstyle {\rm E}}}(J)=0$
solution.

Comparing a) and b) we see that the borders of stability between the
stability of $\varDelta_{n}^{{\scriptscriptstyle {\rm E}}}(J)=0$
and $\varDelta_{n}^{{\scriptscriptstyle {\rm E}}}(J)\neq0$ do not
agree. In fact, we find a region where both, the $\varDelta_{n}^{{\scriptscriptstyle {\rm E}}}(J)=0$
and the $\varDelta_{n}^{{\scriptscriptstyle {\rm E}}}(J)\neq0$ solution
have a finite basin of attraction; here the normal and the SC state
are (meta) stable. The shape of the border of the region where $\varDelta_{n}^{{\scriptscriptstyle {\rm E}}}(J)=0$
is unstable resembles closely to the linear BCS solution which we
show in Fig.~\ref{fig:PhaseDiagramEliashbergEquations} a) as a green
dashed line.
\begin{figure}
\begin{centering}
\begin{minipage}[t]{0.5\columnwidth}%
\begin{center}
\includegraphics[width=1\textwidth]{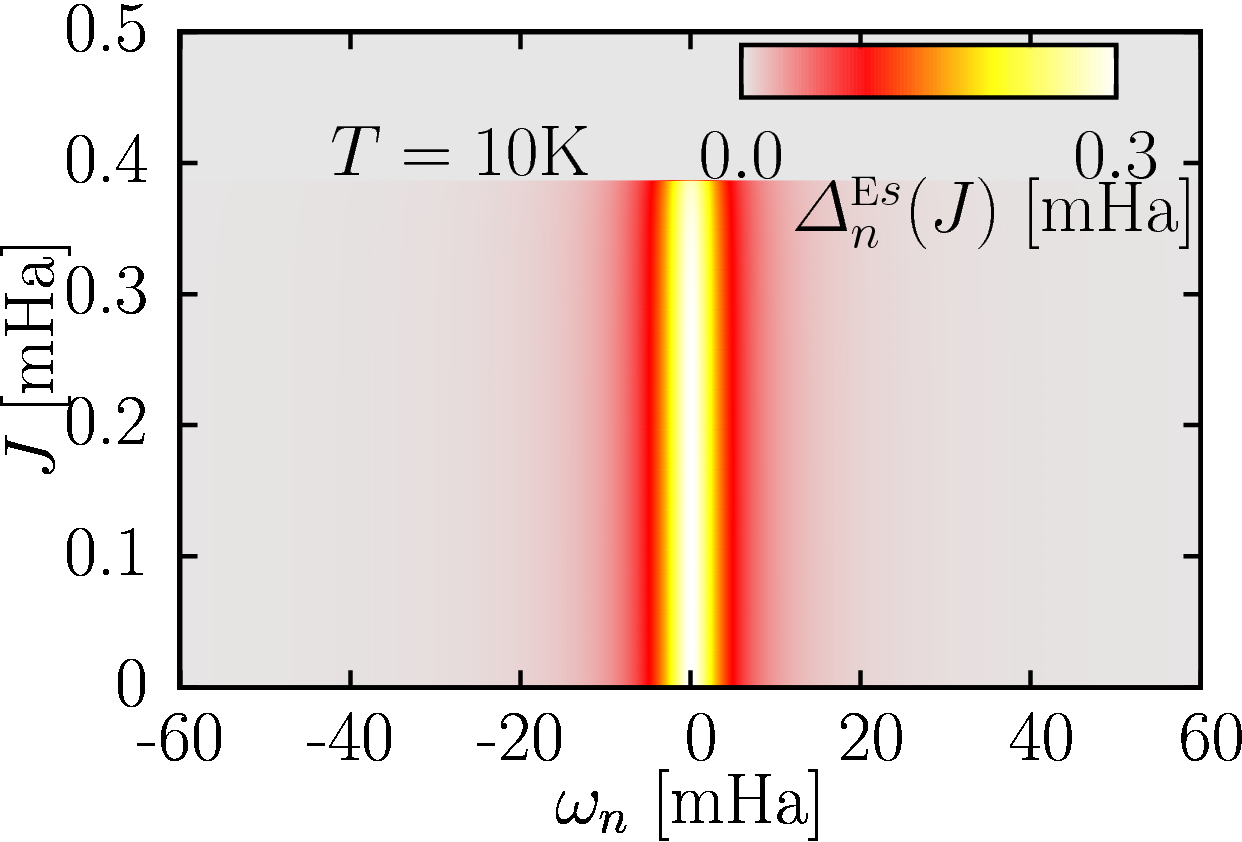}\\
a)
\par\end{center}%
\end{minipage}\nolinebreak%
\begin{minipage}[t]{0.5\columnwidth}%
\begin{center}
\includegraphics[width=1\textwidth]{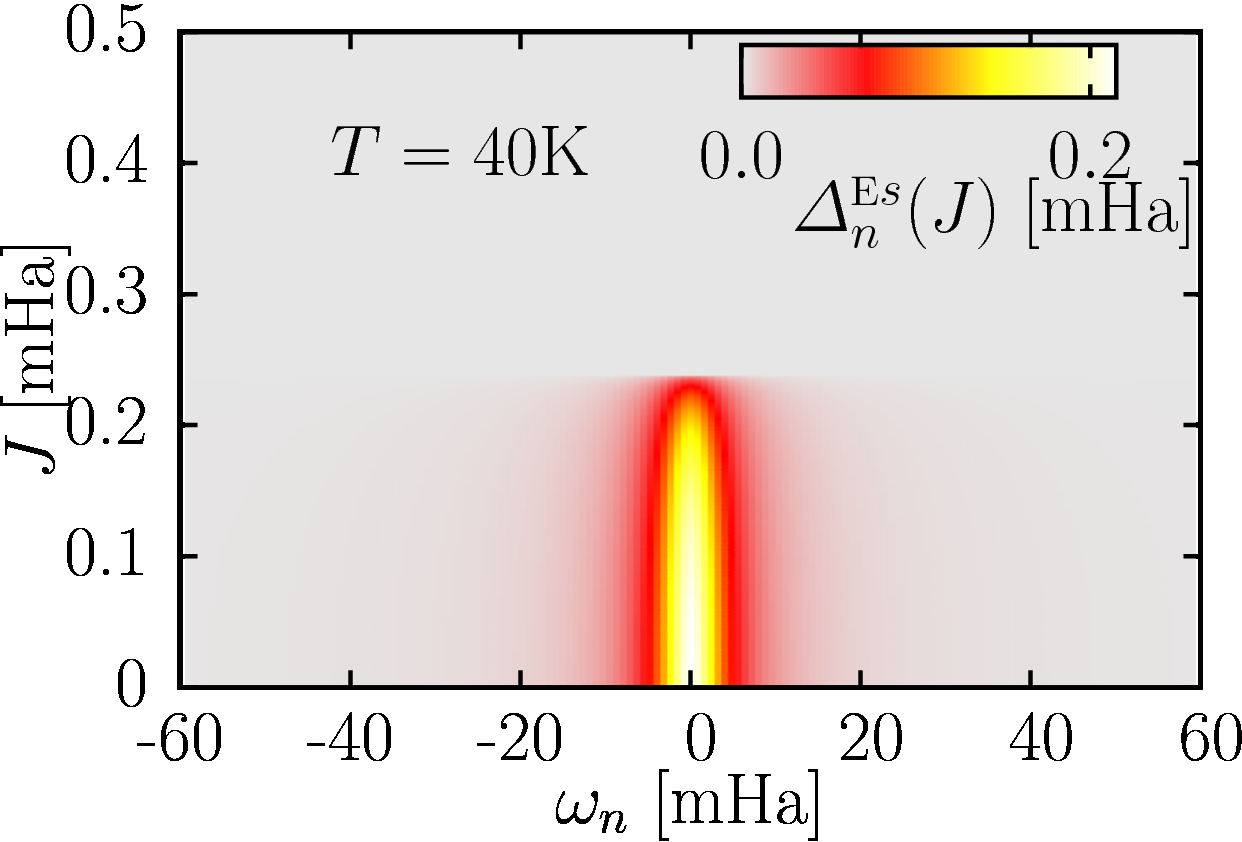}\\
b)
\par\end{center}%
\end{minipage}
\par\end{centering}

\caption{(color online) Solutions to the Eliashberg equations $\varDelta_{n}^{{\scriptscriptstyle {\rm E}}}(J)$
for $T=40{\rm K}$ a) and $T=10{\rm K}$ b) along the blue lines of
the left panel of Fig.~\ref{fig:PhaseDiagramEliashbergEquations}.,\label{fig:DeltaEliashberg40K10K}}
\end{figure}

We plot the $\varDelta_{n}^{{\scriptscriptstyle {\rm E}}}(J)$ at
$T=10{\rm K}$ in Fig.~\ref{fig:DeltaEliashberg40K10K} a) and $40{\rm K}$
in Fig.~\ref{fig:DeltaEliashberg40K10K} b) as a function of $J$
on the vertical axis. The corresponding equal temperature lines are
blue in Fig.~\ref{fig:PhaseDiagramEliashbergEquations}.
We find that the shape is largely independent on the splitting $J$ and the temperature $T$
except for a scale factor. Thus $\varDelta_{n=0}^{{\scriptscriptstyle {\rm E}}}$
is sufficient to investigate the behavior of the theory. For low temperatures
the down-scaling is much less pronounced and it is safe to say that
the pairing is almost unaffected by the presence of a splitting up
until the point where the SC phase is suppressed. For a high temperature,
instead, the down-scaling is more pronounced and the transition becomes
continuous above a certain temperature.

\section{Results of SpinSCDFT with the G0-Functional\label{sec:Results-of-SpinSCDFT}}

In this Section we discuss the numerical solution of the SpinSCDFT
gap equation (\NonLinearGapEq) using the $xc$-potential
derived in Sec.~\SubDerivationXCPotential. We refer to this functional
as the G0-functional.

In Eq.~(\NonLinearGapEq) of {\bf I}, we have derived the gap equation
of SpinSCDFT using the G0-functional. This equations (\NonLinearGapEq),
in turn, is derived from the Sham-Schl\"uter equation for a superconductor,
written in {\bf I} in the form
\begin{eqnarray}
\int \rm{d}\mathfrak{e}^{\prime}{S}_{\beta}[\varDelta_{{\rm {\scriptscriptstyle S}}}^{{\scriptscriptstyle {\rm s}}}](\mathfrak{e},\mathfrak{e}^{\prime})\varDelta_{{\rm {\scriptscriptstyle S}}}^{{\scriptscriptstyle {\rm s}}}(\mathfrak{e}^{\prime}) & = & 0\ . \label{eq:NonLinearSS}
\end{eqnarray}
From the previous discussion in the Secs.~\ref{sec:BCSwithJ}
and \ref{sec:Eliashberg}, a continuous transition is to be expected
for a small exchange field intensity $J$ as compared to the transition temperature.

For the point of the continuous transition
Eq.~\eqref{eq:NonLinearSS} can be linearized in $\varDelta_{{\rm {\scriptscriptstyle S}}}^{{\scriptscriptstyle {\rm s}}}$.
Similar to {\bf I} we use the notation with a breve
to indicate linearized entities $\breve{S}_{\beta}={S}_{\beta}[\varDelta_{{\rm {\scriptscriptstyle S}}}^{{\scriptscriptstyle {\rm s}}}=0]$.
Thus, in this case $T_{{\scriptscriptstyle {\rm c}}}(J)$
can be computed from the condition that $\breve{S}_{\beta}=\breve{S}_{\beta}^{{\scriptscriptstyle \mathfrak{C}}}+\breve{S}_{\beta}^{{\scriptscriptstyle \mathrm{M}}}+\breve{S}_{\beta}^{{\scriptscriptstyle \mathfrak{D}}}$
has a singular eigenvalue
\begin{eqnarray}
\rm{det}\breve{S}_{\beta} & = & 0\ .\label{eq:LinearSSEq}
\end{eqnarray}
The corresponding shape of the solution
$\varDelta_{{\rm {\scriptscriptstyle S}}}^{{\scriptscriptstyle {\rm s}}}/\vert\vert\varDelta_{{\rm {\scriptscriptstyle S}}}^{{\scriptscriptstyle {\rm s}}}\vert\vert$
is the right eigenfunction to such a singular eigenvalue.

$\breve{S}_{\beta}(\mathfrak{e},\mathfrak{e}^{\prime})$ is given in
Eq.~(\LinearSSOperator). To investigate the structure and properties of the
SpinSCDFT $xc$-potential is easier within the linearized form, since the matrix
$\breve{S}_{\beta}(\mathfrak{e},\mathfrak{e}^{\prime})=\breve{S}_{\beta}^{{\scriptscriptstyle \mathfrak{C}}}(\mathfrak{e},\mathfrak{e}^{\prime})+\breve{S}_{\beta}^{{\scriptscriptstyle \mathrm{M}}}(\mathfrak{e},\mathfrak{e}^{\prime})+\breve{S}_{\beta}^{{\scriptscriptstyle \mathfrak{D}}}(\mathfrak{e},\mathfrak{e}^{\prime})$
is independent of the potential $\varDelta_{{\rm {\scriptscriptstyle S}}}^{{\scriptscriptstyle {\rm s}}}$.
As discussed in detail in {\bf I}, $\breve{S}_{\beta}^{{\scriptscriptstyle \mathfrak{D}}}(\mathfrak{e},\mathfrak{e}^{\prime})$
($\breve{S}_{\beta}^{{\scriptscriptstyle \mathfrak{C}}}(\mathfrak{e},\mathfrak{e}^{\prime})$)
corresponds to the Nambu (off) diagonal self-energy contribution.
$\breve{S}_{\beta}^{{\scriptscriptstyle \mathrm{M}}}$ is due to the $v_{xc}$ part of the Sham-Schl\"uter equation.
In Sec.~\ref{sub:Linearized-Sham-Schluter-Equatio} we present and
discuss the shape of the contributions $\breve{S}_{\beta}^{{\scriptscriptstyle \mathrm{M}}}(\mathfrak{e},\mathfrak{e}^{\prime})$,
$\breve{S}_{\beta}^{{\scriptscriptstyle \mathfrak{D}}}(\mathfrak{e},\mathfrak{e}^{\prime})$,
$\breve{S}_{\beta}^{{\scriptscriptstyle \mathfrak{C}}}(\mathfrak{e},\mathfrak{e}^{\prime})$
and the $T_{{\scriptscriptstyle {\rm c}}}(J)$ curve from the linearized
$xc$-potential. 

Finally, the properties of the general non-linear gap equation, i.e.~%
the $J-T$ diagram of the solutions to Eq.~\eqref{eq:NonLinearSS}
with and without the Coulomb repulsion will
be presented in Sec.~\ref{sub:Non-Linear-Sham-Schluter-Equatio}.

\subsection{Linearized Sham-Schl\"uter Equation\label{sub:Linearized-Sham-Schluter-Equatio}}

As discussed before, in the part of the $J-T$ diagram for a relatively small applied field
(i.e.~low splitting $J$ and high $T$) we expect a second order phase transition.
This section deals with the corresponding continuous transition.
In Sec.~\ref{sub:Temperature-Dependence-ofS}, will show the shape
of $\breve{S}_{\beta}^{{\scriptscriptstyle \mathfrak{C}}}$, $\breve{S}_{\beta}^{{\scriptscriptstyle \mathrm{M}}}$,
and $\breve{S}_{\beta}^{{\scriptscriptstyle \mathfrak{D}}}$.
To determine the point of the transition according to Eq.~\eqref{eq:LinearSSEq},
in Sec.~\ref{sub:TcFromLinear} we investigate the spectrum of $\breve{S}_{\beta}$ as a function
of temperature and splitting and the corresponding solutions 
$\varDelta_{{\rm {\scriptscriptstyle S}}}^{{\scriptscriptstyle {\rm s}}}/\vert\vert\varDelta_{{\rm {\scriptscriptstyle S}}}^{{\scriptscriptstyle {\rm s}}}\vert\vert$.
Then we will discuss the shape of the
$T_{{\scriptscriptstyle {\rm c}}}(J)$ curve in Sec.~\ref{sub:TcVsSplitting} from this linear approach.

\subsubsection{Temperature Dependence of $\breve{S}_{\beta}$\label{sub:Temperature-Dependence-ofS}}

\begin{figure*}
\begin{minipage}[t]{0.5\textwidth}%
\begin{minipage}[t]{0.9\textwidth}%
\begin{center}
\begin{minipage}[t]{0.5\textwidth}%
\begin{center}
\includegraphics[width=0.9\columnwidth]{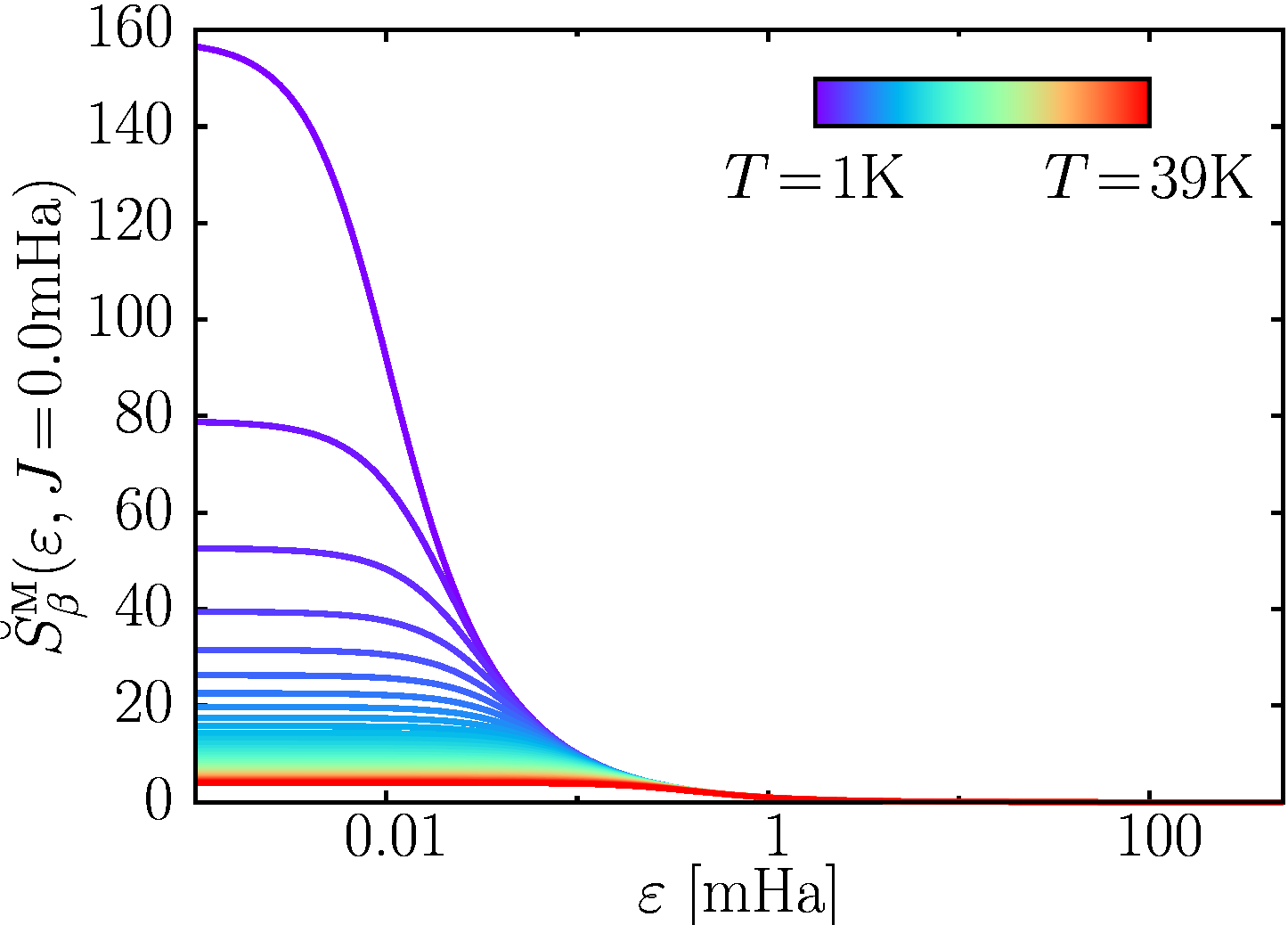}
\par\end{center}%
\end{minipage}\nolinebreak%
\begin{minipage}[t]{0.5\textwidth}%
\begin{center}
\includegraphics[width=0.9\columnwidth]{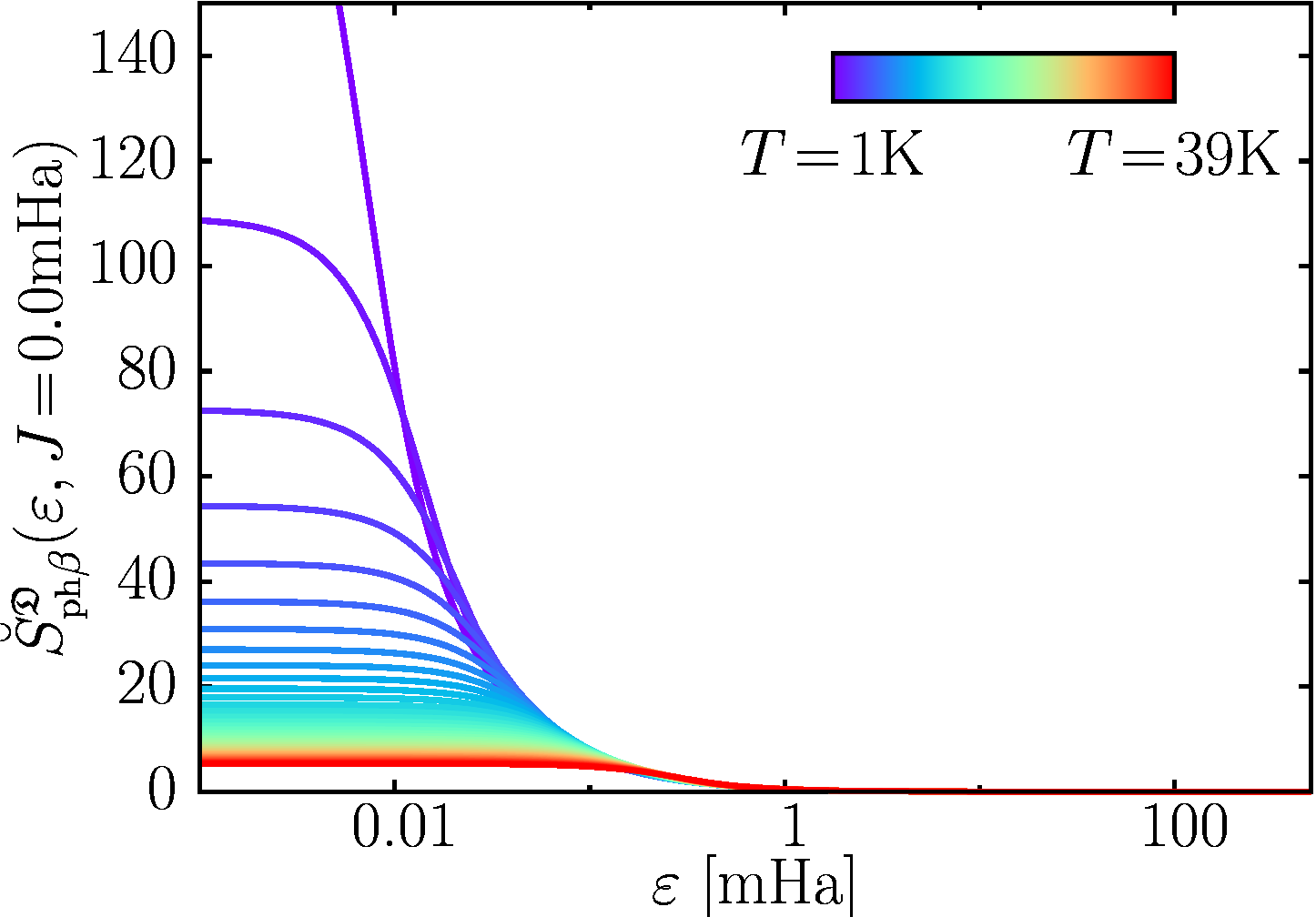}
\par\end{center}%
\end{minipage}\\
(a) $\breve{S}_{\beta}^{{\scriptscriptstyle \mathrm{M}}}(\mathfrak{e})$
(l.) and $\breve{S}_{\beta}^{{\scriptscriptstyle \mathfrak{D}}}(\mathfrak{e})$
(r.) for $J=0\mathrm{mHa}$\\
\mbox{ }
\par\end{center}%
\end{minipage}\\
\begin{minipage}[t]{0.9\textwidth}%
\begin{center}
\begin{minipage}[t]{0.333333\textwidth}%
\begin{center}
\includegraphics[width=1\columnwidth]{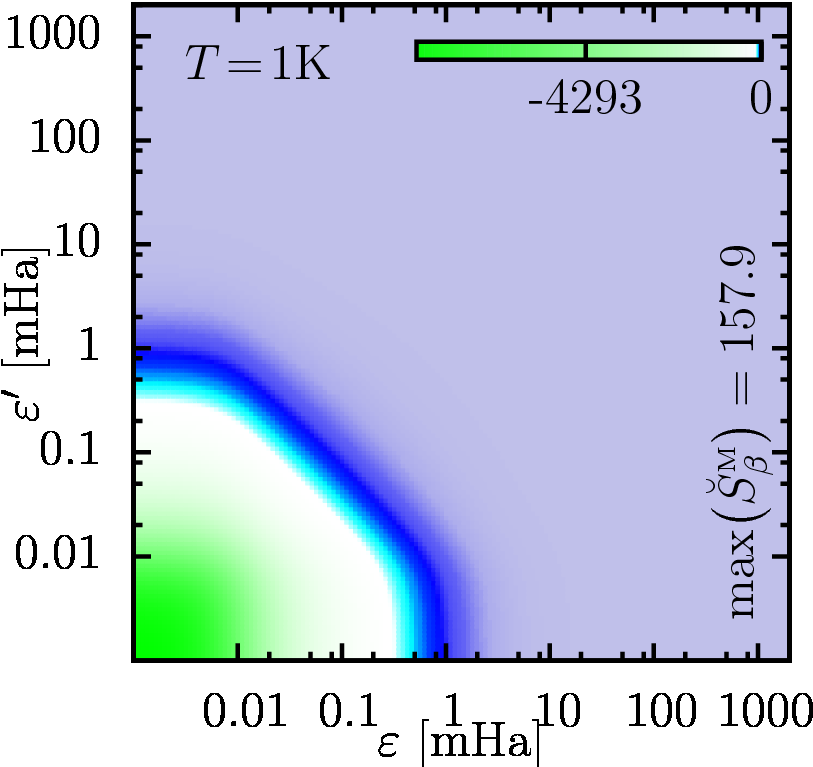}
\par\end{center}%
\end{minipage}\nolinebreak%
\begin{minipage}[t]{0.333333\textwidth}%
\begin{center}
\includegraphics[width=1\columnwidth]{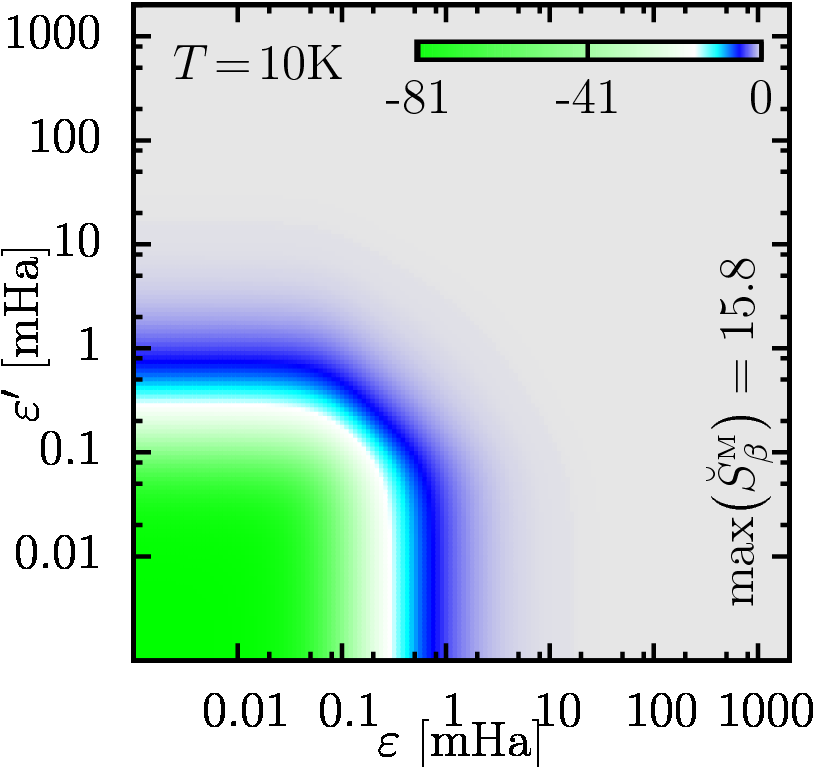}
\par\end{center}%
\end{minipage}\nolinebreak%
\begin{minipage}[t]{0.333333\textwidth}%
\begin{center}
\includegraphics[width=1\columnwidth]{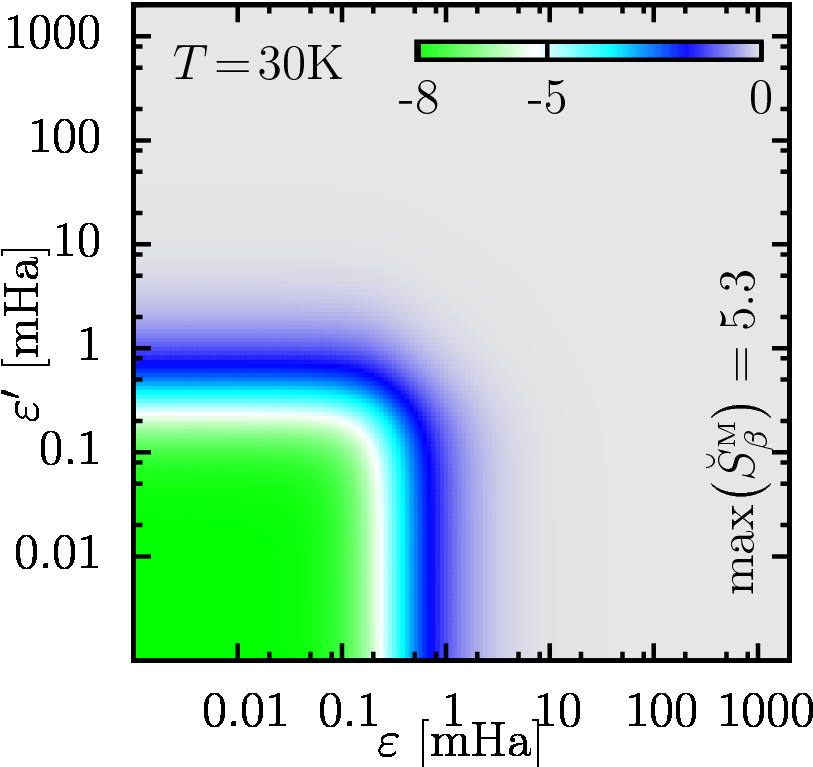}
\par\end{center}%
\end{minipage}\\
(c) $\breve{S}_{{\rm {\scriptscriptstyle ph}}\beta}^{{\scriptscriptstyle \mathfrak{C}}}(\mathfrak{e},\mathfrak{e}^{\prime})$
for $J=0.0\mathrm{mHa}$ at (l. to r.) $T=1{\rm K},\,10{\rm K},30{\rm K}$\\
\mbox{ }
\par\end{center}%
\end{minipage}\\
\begin{minipage}[t]{0.9\textwidth}%
\begin{center}
\begin{minipage}[t]{0.3333\columnwidth}%
\begin{center}
\includegraphics[width=1\columnwidth]{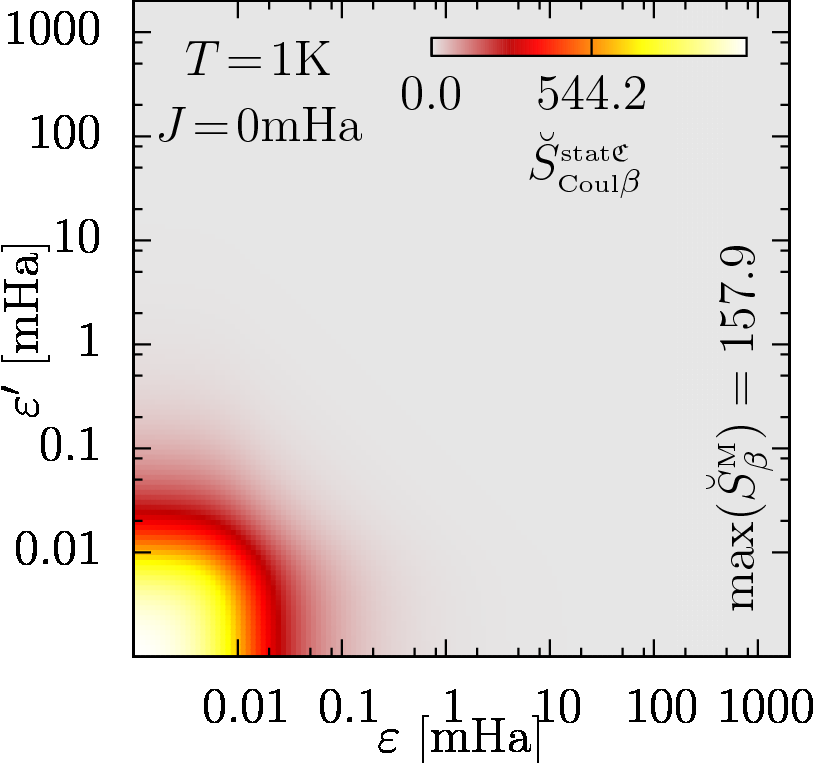}
\par\end{center}%
\end{minipage}\nolinebreak%
\begin{minipage}[t]{0.3333\columnwidth}%
\begin{center}
\includegraphics[width=1\columnwidth]{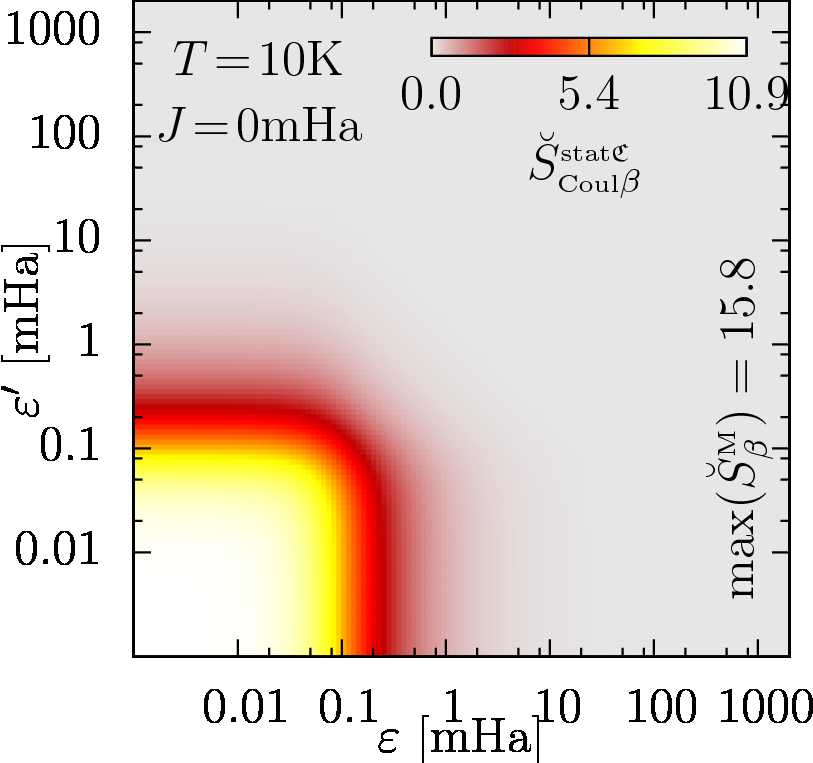}
\par\end{center}%
\end{minipage}\nolinebreak%
\begin{minipage}[t]{0.3333\columnwidth}%
\begin{center}
\includegraphics[width=1\columnwidth]{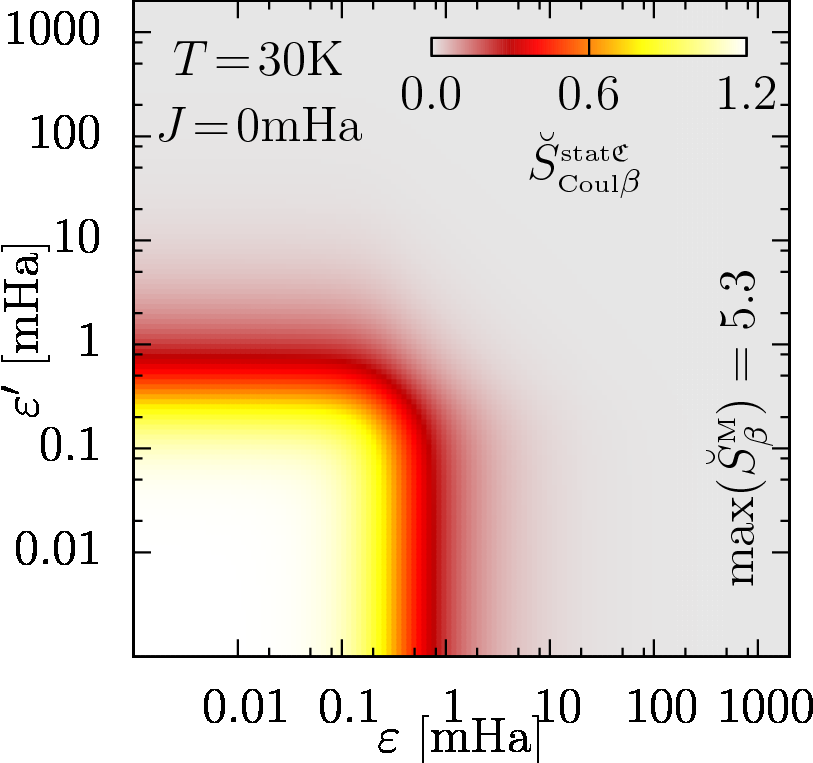}
\par\end{center}%
\end{minipage}\\
(d) $\breve{S}_{{\rm {\scriptscriptstyle C}}\beta}^{{\scriptscriptstyle \mathfrak{C}}}(\mathfrak{e},\mathfrak{e}^{\prime})$
for $J=0.0\mathrm{mHa}$ at (l. to r.) $T=1{\rm K},\,10{\rm K},30{\rm K}$\\
\mbox{ }
\par\end{center}%
\end{minipage}%
\end{minipage}\nolinebreak%
\begin{minipage}[t]{0.5\textwidth}%
\begin{minipage}[t]{0.9\textwidth}%
\begin{center}
\begin{minipage}[t]{0.5\textwidth}%
\begin{center}
\includegraphics[width=0.9\columnwidth]{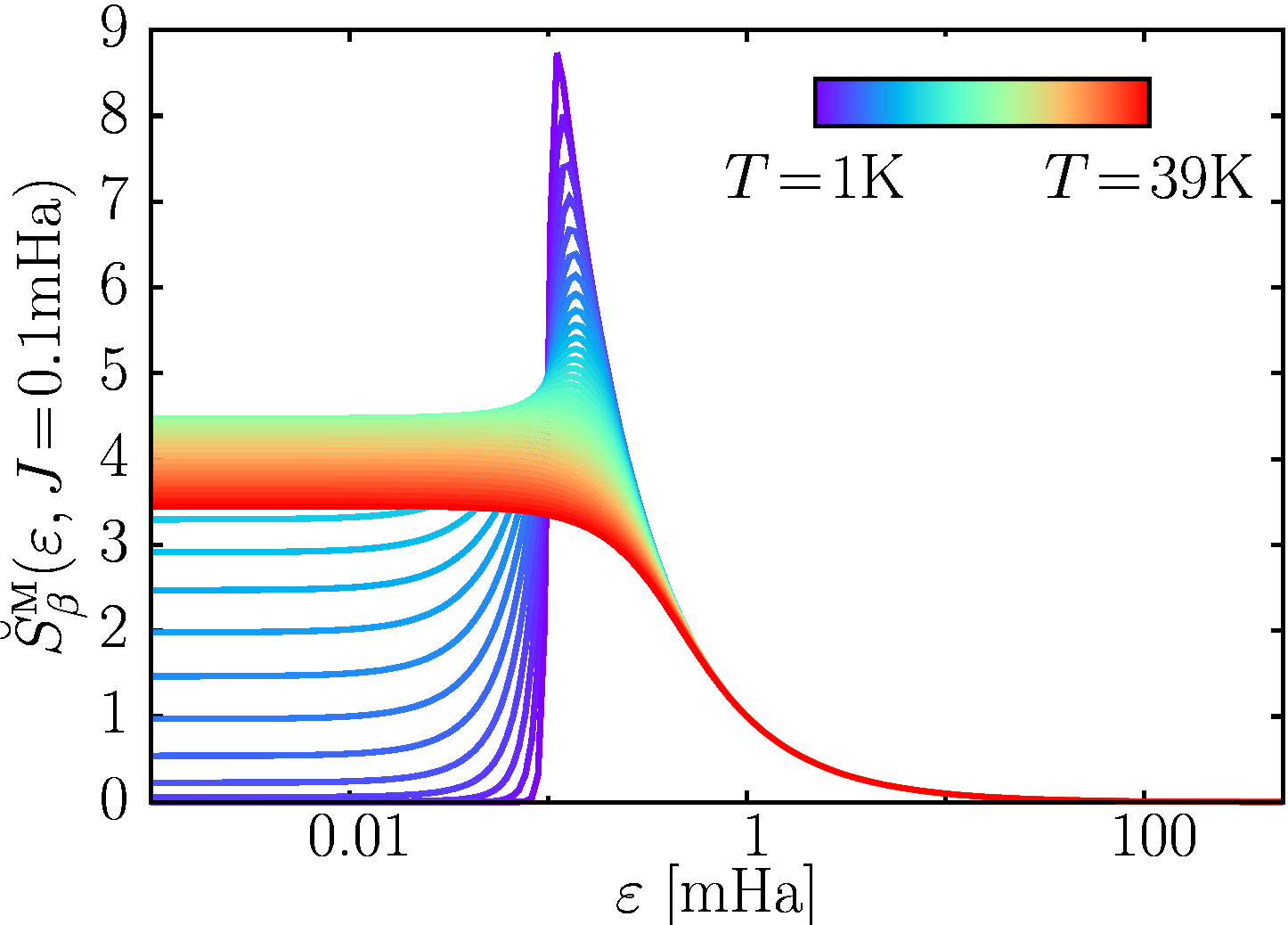}
\par\end{center}%
\end{minipage}\nolinebreak%
\begin{minipage}[t]{0.5\textwidth}%
\begin{center}
\includegraphics[width=0.9\columnwidth]{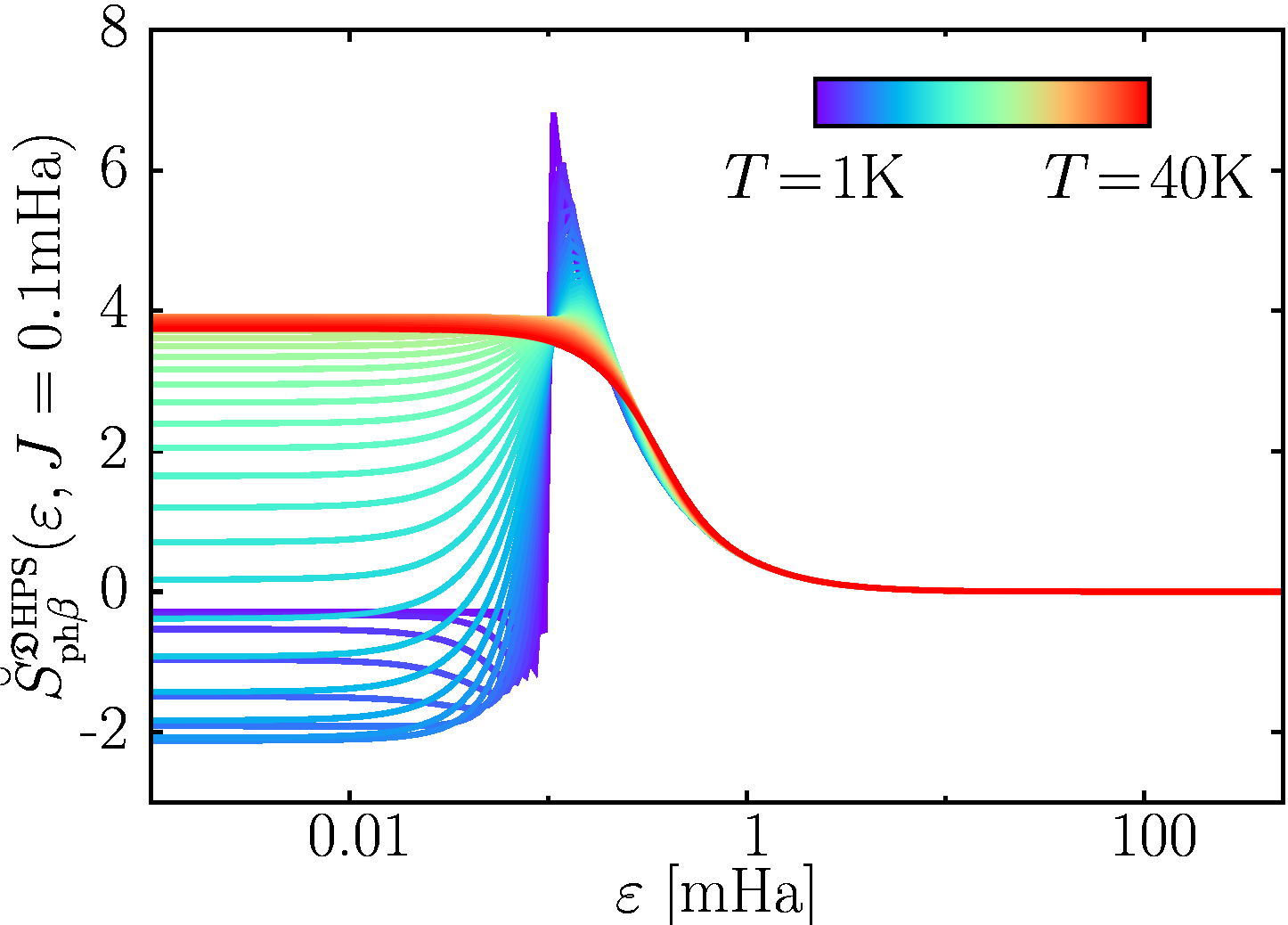}
\par\end{center}%
\end{minipage}\\
(b) $\breve{S}_{\beta}^{{\scriptscriptstyle \mathrm{M}}}(\mathfrak{e})$
(l.) and $\breve{S}_{\beta}^{{\scriptscriptstyle \mathfrak{D}}}(\mathfrak{e})$
(r.) for $J=0.1\mathrm{mHa}$.\\
\mbox{ }
\par\end{center}%
\end{minipage}\\
\begin{minipage}[t]{0.9\textwidth}%
\begin{center}
\begin{minipage}[t]{0.3333\textwidth}%
\begin{center}
\includegraphics[width=1\columnwidth]{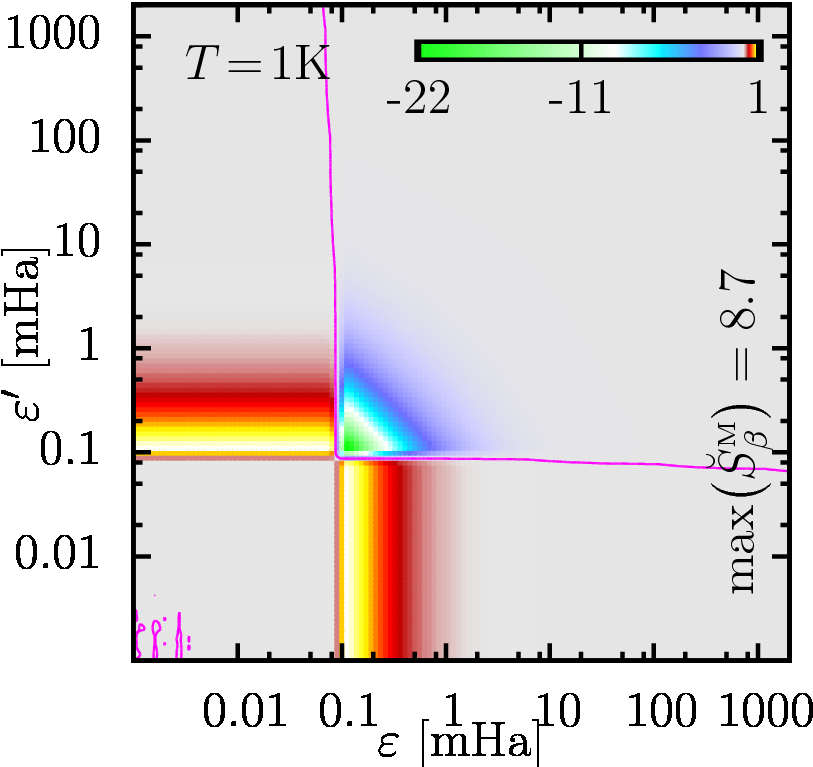}
\par\end{center}%
\end{minipage}\nolinebreak%
\begin{minipage}[t]{0.3333\textwidth}%
\begin{center}
\includegraphics[width=1\columnwidth]{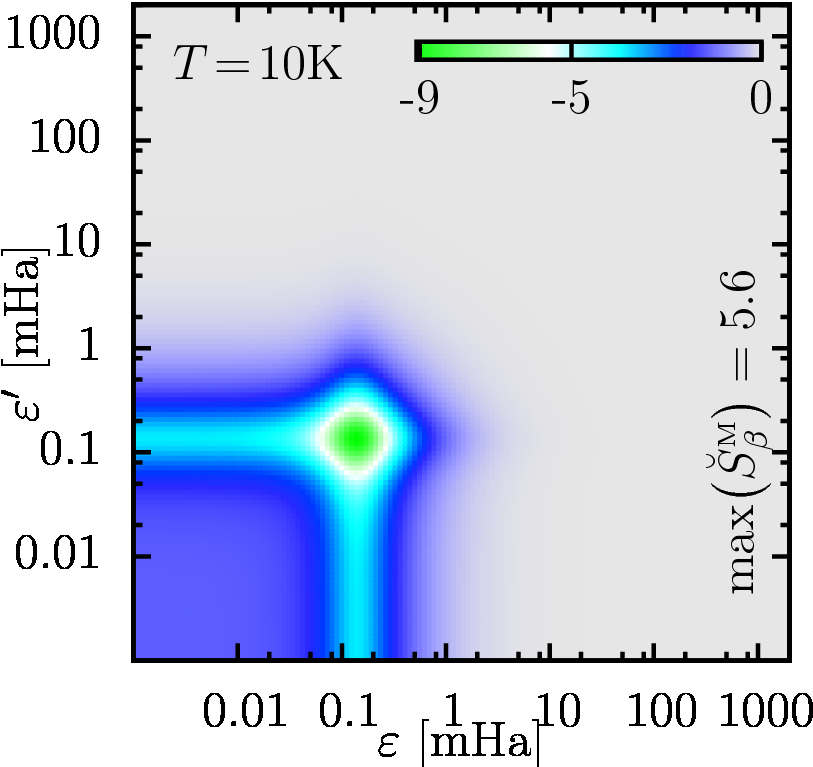}
\par\end{center}%
\end{minipage}\nolinebreak%
\begin{minipage}[t]{0.3333\textwidth}%
\begin{center}
\includegraphics[width=1\columnwidth]{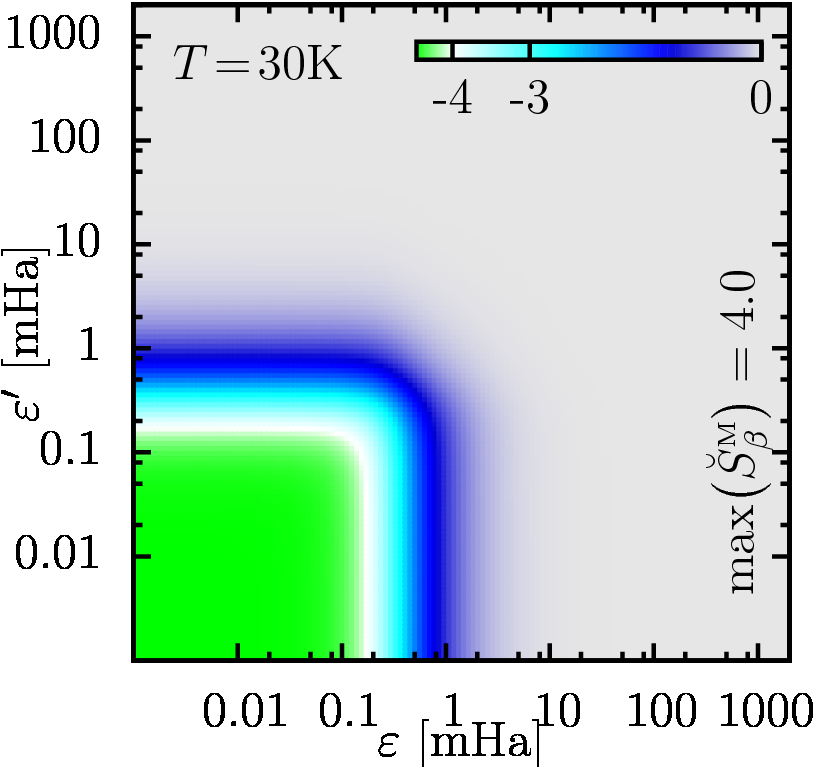}
\par\end{center}%
\end{minipage}\\
(d) $\breve{S}_{{\rm {\scriptscriptstyle ph}}\beta}^{{\scriptscriptstyle \mathfrak{C}}}(\mathfrak{e},\mathfrak{e}^{\prime})$
for $J=0.1\mathrm{mHa}$ at (l.~to r.) $T=1{\rm K},\,10{\rm K},30{\rm K}$\\
\mbox{ }
\par\end{center}%
\end{minipage}\\
\begin{minipage}[t]{0.9\textwidth}%
\begin{center}
\begin{minipage}[t]{0.333333\textwidth}%
\begin{center}
\includegraphics[width=1\columnwidth]{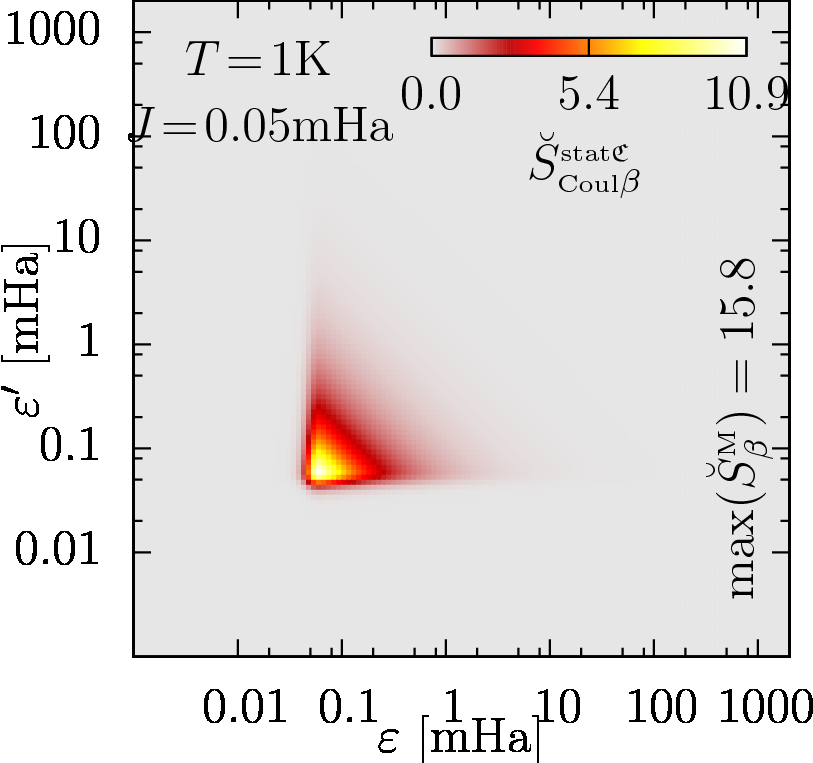}
\par\end{center}%
\end{minipage}\nolinebreak%
\begin{minipage}[t]{0.33333\textwidth}%
\begin{center}
\includegraphics[width=1\columnwidth]{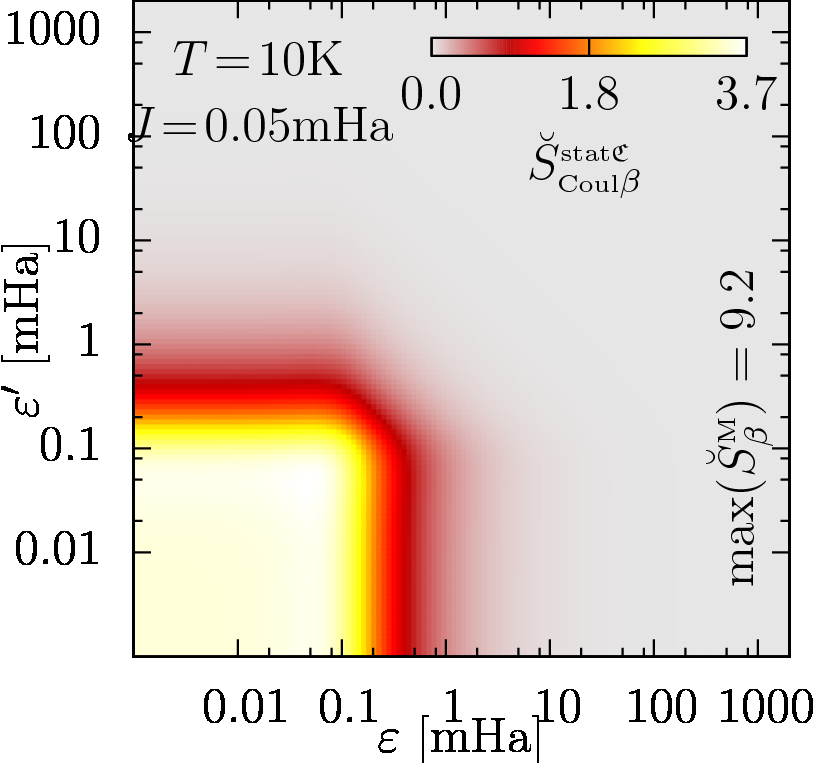}
\par\end{center}%
\end{minipage}\nolinebreak%
\begin{minipage}[t]{0.333333\textwidth}%
\begin{center}
\includegraphics[width=1\columnwidth]{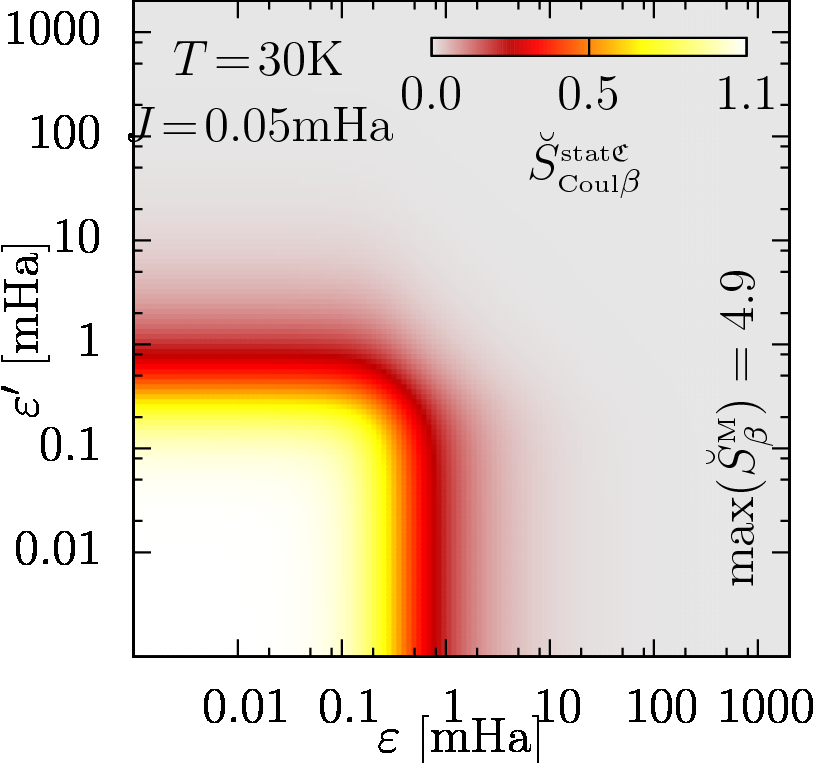}
\par\end{center}%
\end{minipage}\\
(f) $\breve{S}_{{\rm {\scriptscriptstyle C}}\beta}^{{\scriptscriptstyle \mathfrak{C}}}(\mathfrak{e},\mathfrak{e}^{\prime})$
for $J=0.05\mathrm{mHa}$ at (l. to r.) $T=1{\rm K},\,10{\rm K},30{\rm K}$\\
\mbox{ }
\par\end{center}%
\end{minipage}%
\end{minipage}

\caption{(color online) Contributions to the linearized Sham-Schl\"uter Eq.~\eqref{eq:NonLinearSS}.
In the top row we show the diagonal $\breve{S}_{\beta}^{{\scriptscriptstyle \mathrm{M}}}$ and
$\breve{S}_{\beta}^{{\scriptscriptstyle \mathfrak{D}}}$ that originate from the
$v_{xc}$ and Nambu diagonal self-energy in the Sham-Schl\"uter equation, respectively.
In the second (bottom) row, we show the contributions that originate from the Nambu off
phonon (Coulomb) self-energy.
The color scale of $\breve{S}_{\beta}^{{\scriptscriptstyle \mathfrak{C}}}(\mathfrak{e},\mathfrak{e}^{\prime})$
for negative values (decreasing: blue to white to green) is relative
to $\max(\breve{S}_{\beta}^{{\scriptscriptstyle \mathrm{M}}})$ (white).
Red to yellow to white indicates increasingly positive values. Note
that $\breve{S}_{\beta}^{{\scriptscriptstyle \mathfrak{D}}}$ and
$\breve{S}_{{\rm {\scriptscriptstyle ph}}\beta}^{{\scriptscriptstyle \mathfrak{C}}}$
switch sign at $\varepsilon\approx0$ for $J=0.1\mathrm{mHa}$ at
low $T$ as compared to $J=0\mathrm{mHa}$.\label{fig:TDepdenceKernel}}
\end{figure*}

The three contributions to $\breve{S}_{\beta}(\mathfrak{e},\mathfrak{e}^{\prime})$
are (see Sec.~\SubDerivationXCPotential, Eqs.~(\LinearSSOperatorM),(\LinearSSOperatorD),(\LinearSSOperatorCph) and (\LinearSSOperatorCc))
\begin{eqnarray}
\breve{S}_{\beta}(\mathfrak{e},\mathfrak{e}^{\prime}) & = & \bigl(\breve{S}_{\beta}^{{\scriptscriptstyle \mathfrak{D}}}(\mathfrak{e})+\breve{S}_{\beta}^{{\scriptscriptstyle \mathrm{M}}}(\mathfrak{e})\bigr)\updelta(\mathfrak{e}-\mathfrak{e}^{\prime})+\nonumber\\
 & + & \breve{S}_{{\rm {\scriptscriptstyle ph}}\beta}^{{\scriptscriptstyle \mathfrak{C}}}(\mathfrak{e},\mathfrak{e}^{\prime})+\breve{S}_{{\scriptscriptstyle \text{C}}\beta}^{{\scriptscriptstyle \mathfrak{C}}}(\mathfrak{e},\mathfrak{e}^{\prime}).
\end{eqnarray}
In this linear Sham-Schl\"uter form, $\breve{S}_{\beta}^{{\scriptscriptstyle \mathrm{M}}}(\mathfrak{e})$
and $\breve{S}_{\beta}^{{\scriptscriptstyle \mathfrak{D}}}(\mathfrak{e})$
multiply $\varDelta_{{\rm s}}^{{\rm {\scriptscriptstyle s}}}(\mathfrak{e})$
directly. They are shown for several $T$ for $J=0.0\mathrm{mHa}$
and $J=0.1\mathrm{mHa}$ in Fig.~\ref{fig:TDepdenceKernel} panel
a) and b), respectively. Note the logarithmic center of energy scale $\varepsilon$ in all
the plots in this section. The color scale (blue to red) indicates increasing
temperatures. All terms have features only in the close vicinity to
$\varepsilon=0$ and quickly decay to zero within a characteristic
energy width of the phonon coupling. This energy scale is the analog of
Debey frequency $\omega_{0}$ in Eq.~(\ref{eq:a2f}).
However the $\varepsilon$ dependence shown in Fig.~\ref{fig:TDepdenceKernel}
in the presence (panel b) and the absence (panel a) of an exchange splitting is very
different. In fact, in Fig.~\ref{fig:TDepdenceKernel} a) where
$J=0$ both $\breve{S}_{\beta}^{{\scriptscriptstyle \mathrm{M}}}(\mathfrak{e})$
and $\breve{S}_{\beta}^{{\scriptscriptstyle \mathfrak{D}}}(\mathfrak{e})$
are positive and monotonously decreasing as a function of $\vert\varepsilon\vert$.
In presence of a $J\neq0$ (Fig.~\ref{fig:TDepdenceKernel} b),
instead, they have the following complex temperature
and energy dependence: For small $T$ in the range $\vert\varepsilon\vert<J$,
$\breve{S}_{\beta}^{{\scriptscriptstyle \mathfrak{D}}}(\mathfrak{e})$
is negative and, in the limit $T\rightarrow0$, $\breve{S}_{\beta}^{{\scriptscriptstyle \mathrm{M}}}(\mathfrak{e})$
and $\breve{S}_{\beta}^{{\scriptscriptstyle \mathfrak{D}}}(\mathfrak{e})$
approach zero from opposite sides. At $\vert\varepsilon\vert\approx J$
both $\breve{S}_{\beta}^{{\scriptscriptstyle \mathrm{M}}}(\mathfrak{e})$
and $\breve{S}_{\beta}^{{\scriptscriptstyle \mathfrak{D}}}(\mathfrak{e})$
vary very rapidly. This behavior is smoothed out with increasing $T$
and at temperatures high enough with respect to $J$ the non-splitted
behavior is recovered.
The temperature and $J=0{\rm mHa}$ and $J=0.1{\rm mHa}$ dependence
of $\breve{S}_{{\scriptscriptstyle {\rm ph}}\beta}^{{\scriptscriptstyle \mathfrak{C}}}(\mathfrak{e},\mathfrak{e}^{\prime})$
is shown in Fig.~\ref{fig:TDepdenceKernel} c) and d), respectively.
$\breve{S}_{\beta}^{{\scriptscriptstyle \mathrm{M}}}$ serves as a
scale that other kernel contributions have to be compared with, so
we choose a color scale that is relative to the maximum of $\breve{S}_{\beta}^{{\scriptscriptstyle \mathrm{M}}}$,
indicated on the right of every plot.

For $J=0\mathrm{mHa}$ we note
that the size of $\breve{S}_{{\scriptscriptstyle {\rm ph}}\beta}^{{\scriptscriptstyle \mathfrak{C}}}(\mathfrak{e},\mathfrak{e}^{\prime})$
(Fig.~\ref{fig:TDepdenceKernel} c) decays faster with temperature the one of the diagonal $\breve{S}_{\beta}^{{\scriptscriptstyle \mathrm{M}}}$
and $\breve{S}_{\beta}^{{\scriptscriptstyle \mathfrak{D}}}$ (the
the position of "white" in the color scale of Fig.~\ref{fig:TDepdenceKernel} c) moves
to the left with increasing temperatures).
Furthermore, being both positive and diagonal,
$\breve{S}_{\beta}^{{\scriptscriptstyle \mathrm{M}}}$ and $\breve{S}_{\beta}^{{\scriptscriptstyle \mathfrak{D}}}$
have to be compared with the eigenvalues of $\breve{S}_{{\scriptscriptstyle {\rm ph}}\beta}^{{\scriptscriptstyle \mathfrak{C}}}$.
$\breve{S}_{\beta}^{{\scriptscriptstyle \mathrm{M}}}$ and $\breve{S}_{\beta}^{{\scriptscriptstyle \mathfrak{D}}}$ 
alone would result in a positive definite Sham-Schl\"uter matrix for  $J=0\mathrm{mHa}$ (compare Fig.~\ref{fig:TDepdenceKernel} a)
at all temperatures so there is not non-trivial solution to Eq.~\eqref{eq:NonLinearSS}.
Thus, technically, the phase transition from the SC to the non SC
regime with the singular eigenvalue is induced by this relative reduction
of $\breve{S}_{{\scriptscriptstyle {\rm ph}}\beta}^{{\scriptscriptstyle \mathfrak{C}}}$
as compared to $\breve{S}_{\beta}^{{\scriptscriptstyle \mathrm{M}}}$ plus $\breve{S}_{\beta}^{{\scriptscriptstyle \mathfrak{D}}}$.
We will turn to a systematic analysis of the eigenvalues of the linearized
Sham-Schl\"uter matrix $\breve{S}_{\beta}$ in Sec.~\ref{sub:TcFromLinear}.

The relative scale reduction is also found for the splitted $\breve{S}_{{\scriptscriptstyle {\rm ph}}\beta}^{{\scriptscriptstyle \mathfrak{C}}}$.
At $\varepsilon\approx0$, however, we stay much below the scale of
$\breve{S}_{\beta}^{{\scriptscriptstyle \mathrm{M}}}$ and exceed
it only for higher temperatures. Moreover, the sign change of {\small $\breve{S}_{\beta}^{{\scriptscriptstyle \mathfrak{D}}}$}
is effectively reducing the diagonal repulsion.

A purple line in Fig.~\ref{fig:TDepdenceKernel} d) indicates the zero contour and shows that for very low
$T$, $\breve{S}_{{\scriptscriptstyle {\rm ph}}\beta}^{{\scriptscriptstyle \mathfrak{C}}}(\mathfrak{e},\mathfrak{e}^{\prime})$
is positive for approximately the region where $\vert\varepsilon\vert<J$ or $\vert\varepsilon^{\prime}\vert<J^{\prime})$
and has a sharp negative spike at $\varepsilon=\varepsilon^{\prime}\approx J$.
Thus, as a curious fact, the phonon interaction is not "attractive" everywhere in this case.
We show the shape of the static Coulomb part $\breve{S}_{{\scriptscriptstyle \mathrm{C}}\beta}^{{\scriptscriptstyle \mathfrak{C}}}$
in Fig.~\ref{fig:TDepdenceKernel} row e) and f) for $J=0{\rm mHa}$
and $J=0.05{\rm mHa}$, respectively. Apart from the differences in
sign the overall behavior of the Coulomb term and phonon terms is
roughly similar with significant deviations in the fact that it does
not change sign for a low temperature and exchange splitting,
compare Fig.~\ref{fig:TDepdenceKernel} d) with f).

In summary we can say that we see relevant changes in the shape of
the contributions to $\breve{S}_{\beta}$ for a finite exchange splitting
for the low temperature limits in the region $\vert\varepsilon\vert<\vert J\vert$
as compared to the spin degenerate case. At higher temperatures the
splitting becomes less important.
We point out that we know from the earlier discussion that this is the region, where
we expect the linearization to be unjustified. From the form of the Bogoliubov
eigenvalues $E_{\alpha}^{\sigma}={\rm sign}(\sigma)J+{\rm sign}(\alpha)\sqrt{\varepsilon^{2}+\vert\varDelta_{{\rm s}}^{{\scriptscriptstyle \text{s}}}\vert^{2}}$
we expect that, whenever $\varDelta_{{\rm s}}^{{\scriptscriptstyle \text{s}}}$
is larger than $J$, will see a behavior more similar to the case
$J=0$ . The reason is that, then, only the $\alpha=+$ branch has positive
excitation energies $E_{+}^{\sigma}\geq0$, meaning that the ground
state does not correspond to some of the excitations $\hat{\gamma}_{k}$
being occupied (see the discussion in \ExcitationDiscussion
and by Ref.~\onlinecite{SarmaOnTheInfluenceOfAUniformExchangeFieldActingOnSC1963}).

\subsubsection{Critical Temperatures and the Shape of $\varDelta_{{\rm s}}^{{\scriptscriptstyle \text{s}}}$\label{sub:TcFromLinear}}

Since we compute the critical temperature from Eq.~\ref{eq:LinearSSEq}, i.e.~the occurrence of a
singular eigenvalue of $\breve{S}_{\beta}$, in this section, we will investigate
the full spectrum as a function of $T$ and $J$.
\begin{figure}
\centering{}%
\begin{minipage}[t]{0.5\columnwidth}%
\begin{center}
\begin{minipage}[t]{0.9\columnwidth}%
\begin{center}
\includegraphics[width=1\textwidth]{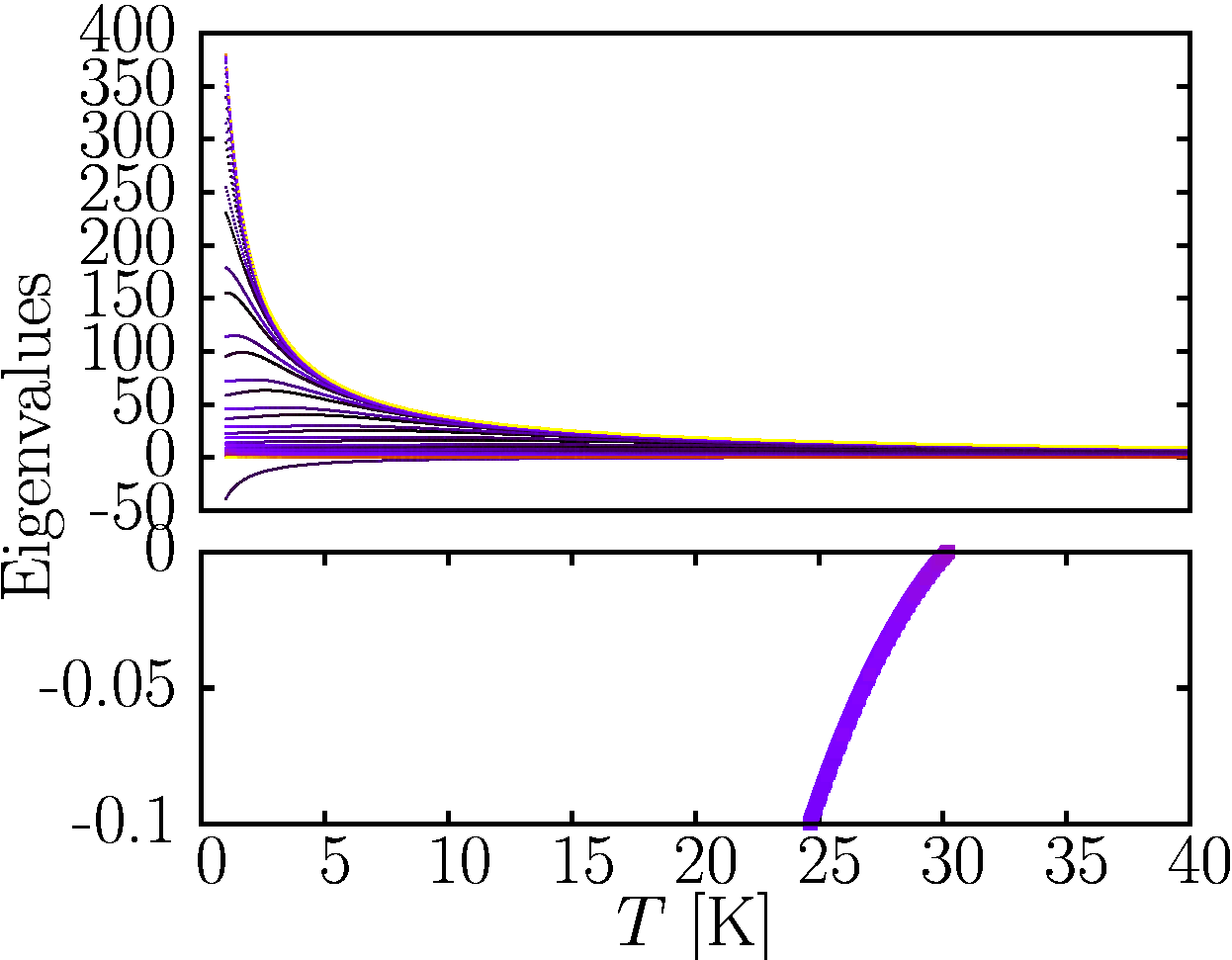}
\par\end{center}

\caption{(color online) Spectrum of $\breve{S}_{\beta}(J=0.0{\rm mHa})$ as a
function of $T$ with only one negative eigenvalue that leads
to a singular point.\label{fig:SpetraNonSPHDTerm}}
\end{minipage}
\par\end{center}%
\end{minipage}%
\begin{minipage}[t]{0.5\columnwidth}%
\begin{center}
\begin{minipage}[t]{0.9\columnwidth}%
\begin{center}
\includegraphics[width=1\textwidth]{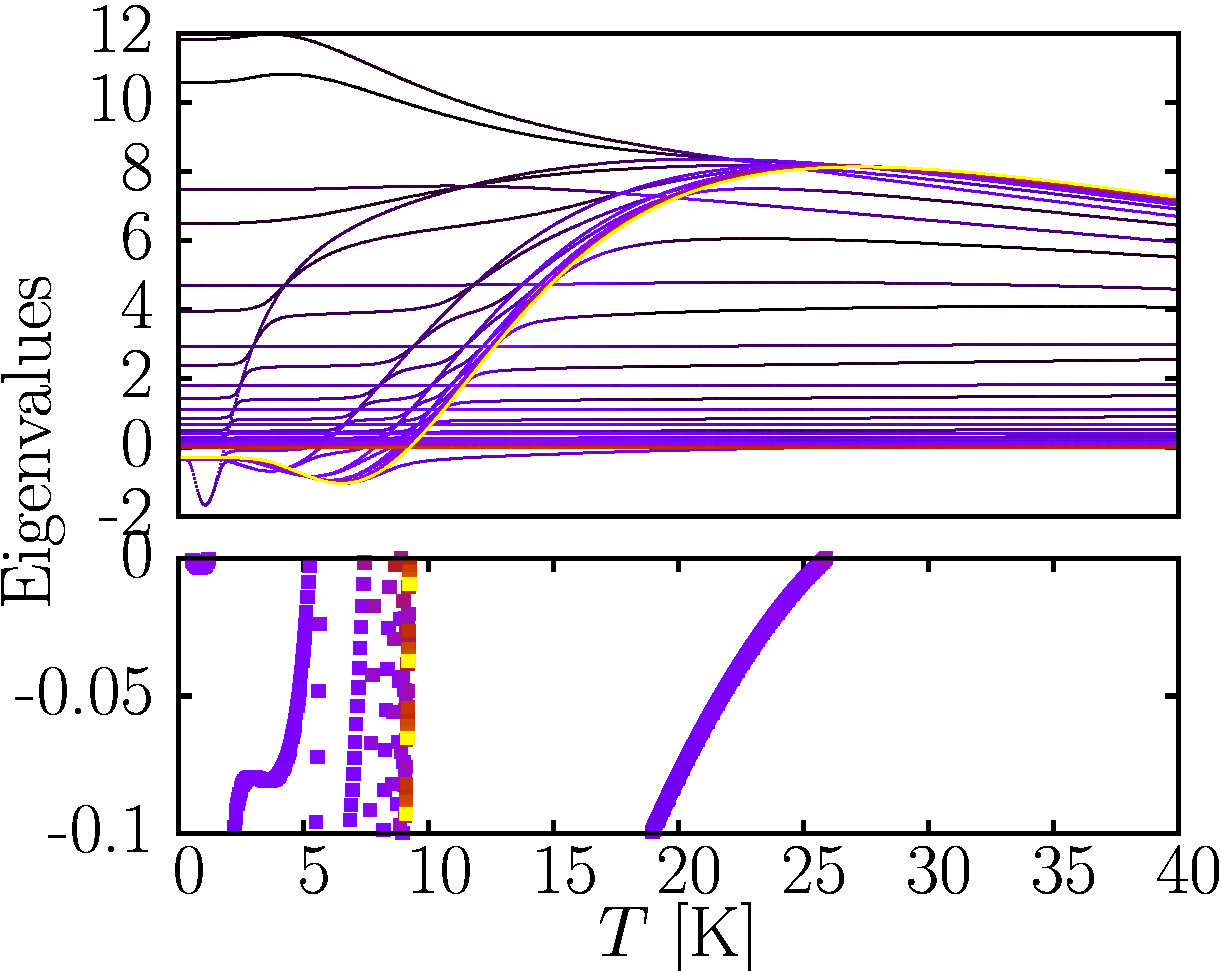}
\par\end{center}

\caption{(color online) Spectrum $\breve{S}_{\beta}(J=0.1{\rm mHa})$ as a
function of $T$ with many negative eigenvalues that cross zero
and lead to singular points.\label{fig:SpectrumSSEq}}
\end{minipage}
\par\end{center}%
\end{minipage}
\end{figure}
\begin{figure}
\centering{}%
\begin{minipage}[t]{0.5\columnwidth}%
\begin{center}
\begin{minipage}[t]{0.9\columnwidth}%
\includegraphics[width=1\columnwidth]{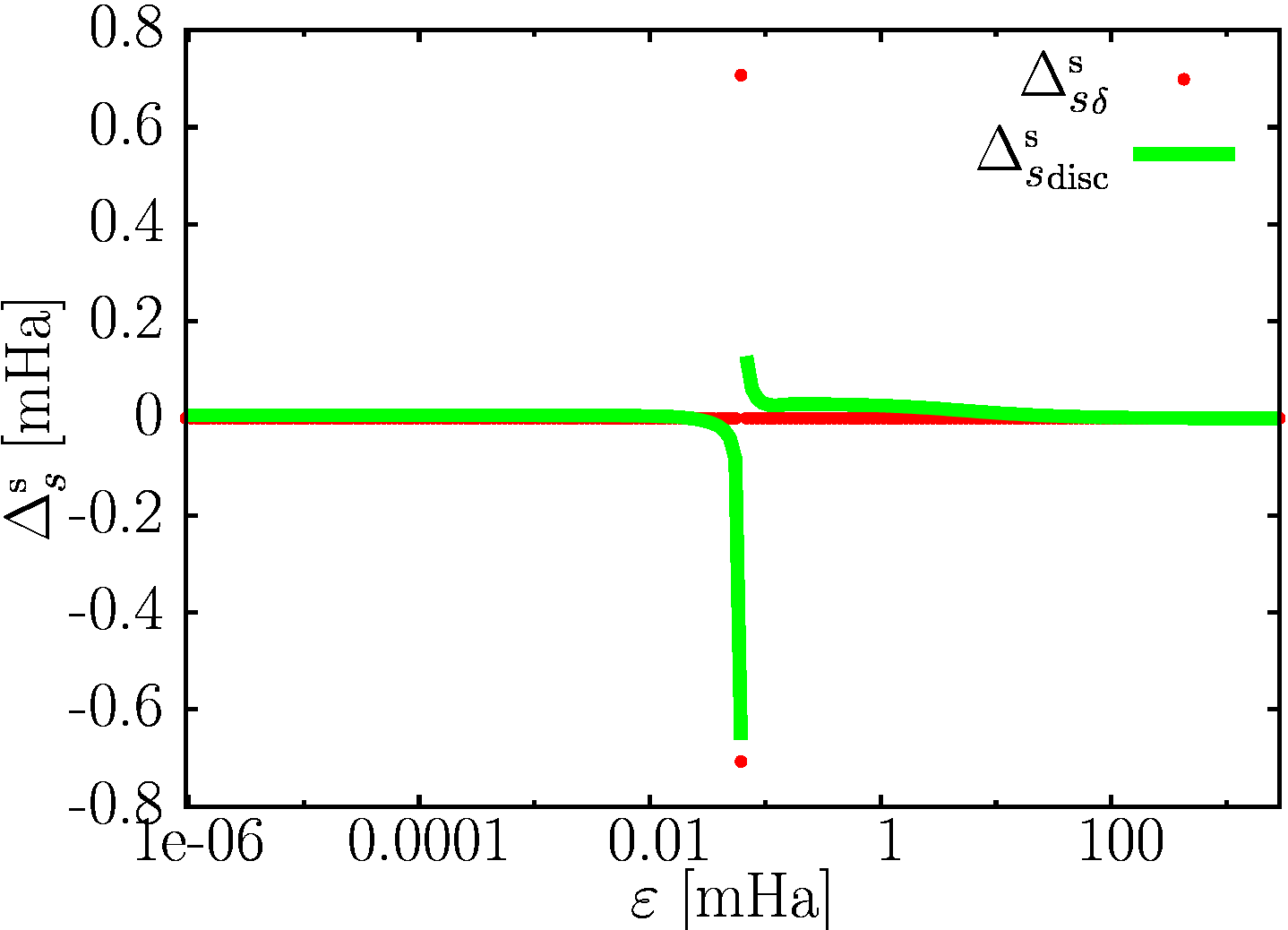}\caption{(color online) Eigenfunctions to a singular eigenvalue at $T\approx T_{{\scriptscriptstyle \text{cross}}}$.
All eigenfunctions except one are of either type.\label{fig:LinearGapDiscontinous}}
\end{minipage}
\par\end{center}%
\end{minipage}\nolinebreak%
\begin{minipage}[t]{0.5\columnwidth}%
\begin{center}
\begin{minipage}[t]{0.9\columnwidth}%
\includegraphics[width=1\columnwidth]{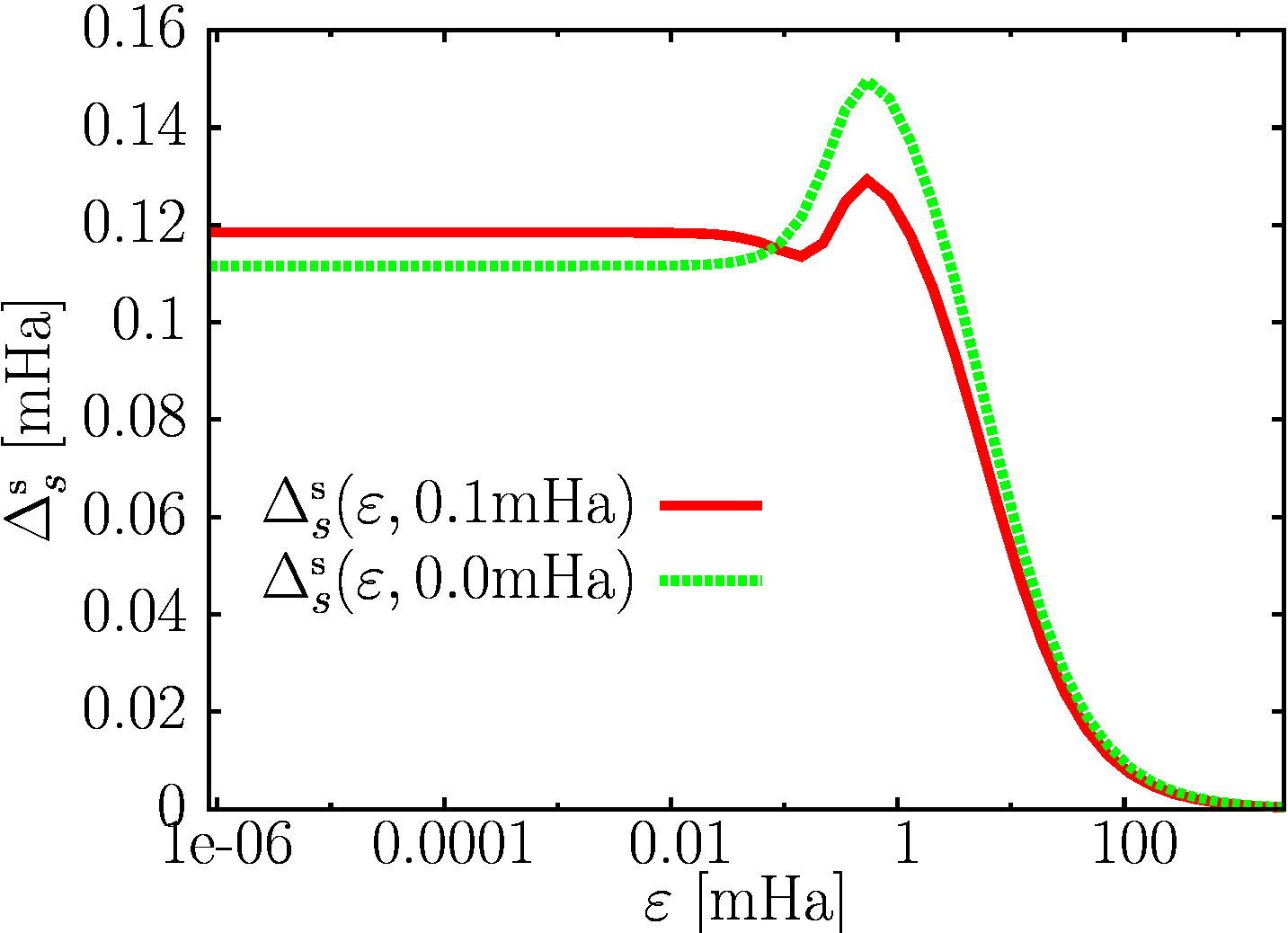}\caption{(color online) $\varDelta_{{\rm s}}^{{\scriptscriptstyle \text{s}}}(\mathfrak{e})$
at $T_{c}$ for $J=0.0{\rm mHa}\,,0.1{\rm mHa}$ without the Coulomb
interaction.\label{fig:LinearGapTc}}
\end{minipage}
\par\end{center}%
\end{minipage}
\end{figure}
 The KS potential $\varDelta_{{\rm s}}^{{\scriptscriptstyle \text{s}}}$
is proportional to the right eigenvector of $\breve{S}_{\beta}$ that
is associated to a singular eigenvalue. Thus, all eigenfunctions $\varDelta_{{\rm s}}^{{\scriptscriptstyle \text{s}}}$
we show are normalized to a common arbitrary value. In this Subsection,
we are not not considering the Coulomb contribution when we calculate
the spectrum of $\breve{S}_{\beta}$ as a function of temperature
in Fig.~\ref{fig:SpetraNonSPHDTerm}. In the spin-degenerate case
($J=0{\rm mHa}$) we see that the eigenvalues decrease in magnitude
with temperature in a monotonous way. At low temperature all eigenvalues
but one are positive valued; the negative eigenvalue crosses zero,
at the temperature $T_{{\rm {\scriptscriptstyle c}}}(J=0{\rm mHa})\approx30{\rm K}$
in the model present model, above which $\breve{S}_{\beta}$ becomes
positive definite. 

As compared to the $J=0{\rm mHa}$, the spectrum at finite splitting
$J=0.1{\rm mHa}$ is fundamentally different. For small $T$ we observe
many negative eigenvalues and, most interestingly, several solutions
$\det\bigl(\breve{S}_{\beta}\bigr)=0$ at low temperatures. There
is a temperature regime $T_{{\scriptscriptstyle \text{cross}}}\approx10{\rm K}$
(in this model) in which most negative eigenvalues cross zero and
become positive. Beyond $T_{{\scriptscriptstyle \text{cross}}}$ only
one negative eigenvalue remains, crossing later at $T_{{\rm {\scriptscriptstyle c}}}(J=0.1{\rm mHa})\approx25{\rm K}$.
Continuously reducing the splitting, this specific eigenvalue/eigenfunction
pair can be traced to the spin-degenerate limit, although we do not
show this here. Similarly, upon reducing the splitting, it is found
that the temperature range where the crossings appear goes to $T_{{\scriptscriptstyle \text{cross}}}\rightarrow0{\rm K}$
as $J\to0$. 

We analyze the eigenfunctions in Fig.~\ref{fig:LinearGapDiscontinous}
and \ref{fig:LinearGapTc} corresponding to these multiple solutions
and see that only the one at $T\approx25{\rm K}$ has a continuous
behavior. The other solutions are of two kinds and both show numerical
discontinuities (see Fig.~\ref{fig:LinearGapDiscontinous}). While
one (green in Fig.~\ref{fig:LinearGapDiscontinous}) has a $1/\varepsilon$-like
pole, there is a second kind (red in Fig.~\ref{fig:LinearGapDiscontinous})
which has a delta peak like structure, i.e.~the value at the pole
of the first kind is large while the rest is extremely small. Increasing
the number of sampling points increased the relative value at the
discontinuity so this lead us to the conclusion that we are numerically
sampling an unbound function. It has to be understood that an unbound
function cannot be the linearized solution to an originally non-linear
fixed-point problem. This because at the pole the function is not
small and a linearization cannot be justified. We expect that in the
non-linear equation these type of solutions will be suppressed. We
therefore ignore these other solutions in the following discussion
and always refer to the continuous, bound, high temperature solution. 

As a side remark we point out that comparing $T_{{\rm {\scriptscriptstyle c}}}^{{\rm {\scriptscriptstyle SCDFT}}}(J=0)\approx30{\rm K}$
with the solution the Eliashberg equations, the latter predicts a much
higher $T_{{\scriptscriptstyle {\rm c}}}^{{\rm {\scriptscriptstyle Eliash}}}(J=0)\approx50{\rm K}$.
For a detailed comparison, see Fig.~\ref{fig:PhaseDiagramEliashbergEquations}
b) where we show the linearized $T_{{\rm {\scriptscriptstyle c}}}^{{\rm {\scriptscriptstyle SCDFT}}}(J)$
in the phase diagram of the Eliashberg equations. We also observe
via the black curve of Fig.~\ref{fig:PhaseDiagramEliashbergEquations}
b), that the Eliashberg solutions predict a SC phase
that is less susceptible against a splitting. The reason for the lower
$T_{{\rm {\scriptscriptstyle c}}}$ prediction is that within the
$xc$-potential construction $\bar{G}$ was replaced with $\bar{G}^{{\scriptscriptstyle {\rm KS}}}$
which violates Migdal's theorem \cite{LuedersSCDFTI2005}. The solution
has recently presented by Sanna {\textit et al.}\cite{SannaMigdalFunctionalSCDFT2014}
using a corrected self-energy in the functional construction. We will
come back to this point and elaborate on the distinction in the Appendix \ref{sec:G0W0SC}.
As a curious result, the linearized
$T_{{\rm {\scriptscriptstyle c}}}^{{\rm {\scriptscriptstyle SCDFT}}}(J)$
curve bends upwards and starts an almost linear increase at the point
where the transition is expected to become of discontinuous type. We investigate
this issue in the next Subsection \ref{sub:TcVsSplitting}.

\subsubsection{Analysis of the $B_{\rm{\scriptscriptstyle{0}}}$ dependence of $T_{{\scriptscriptstyle {\rm c}}}$ \label{sub:TcVsSplitting}}

\begin{figure*}
\centering{}%
\begin{minipage}[t]{1\textwidth}%
\begin{center}
\begin{minipage}[t]{0.333333\textwidth}%
\begin{center}
\begin{minipage}[t]{0.9\textwidth}%
\begin{center}
\includegraphics[width=1\columnwidth]{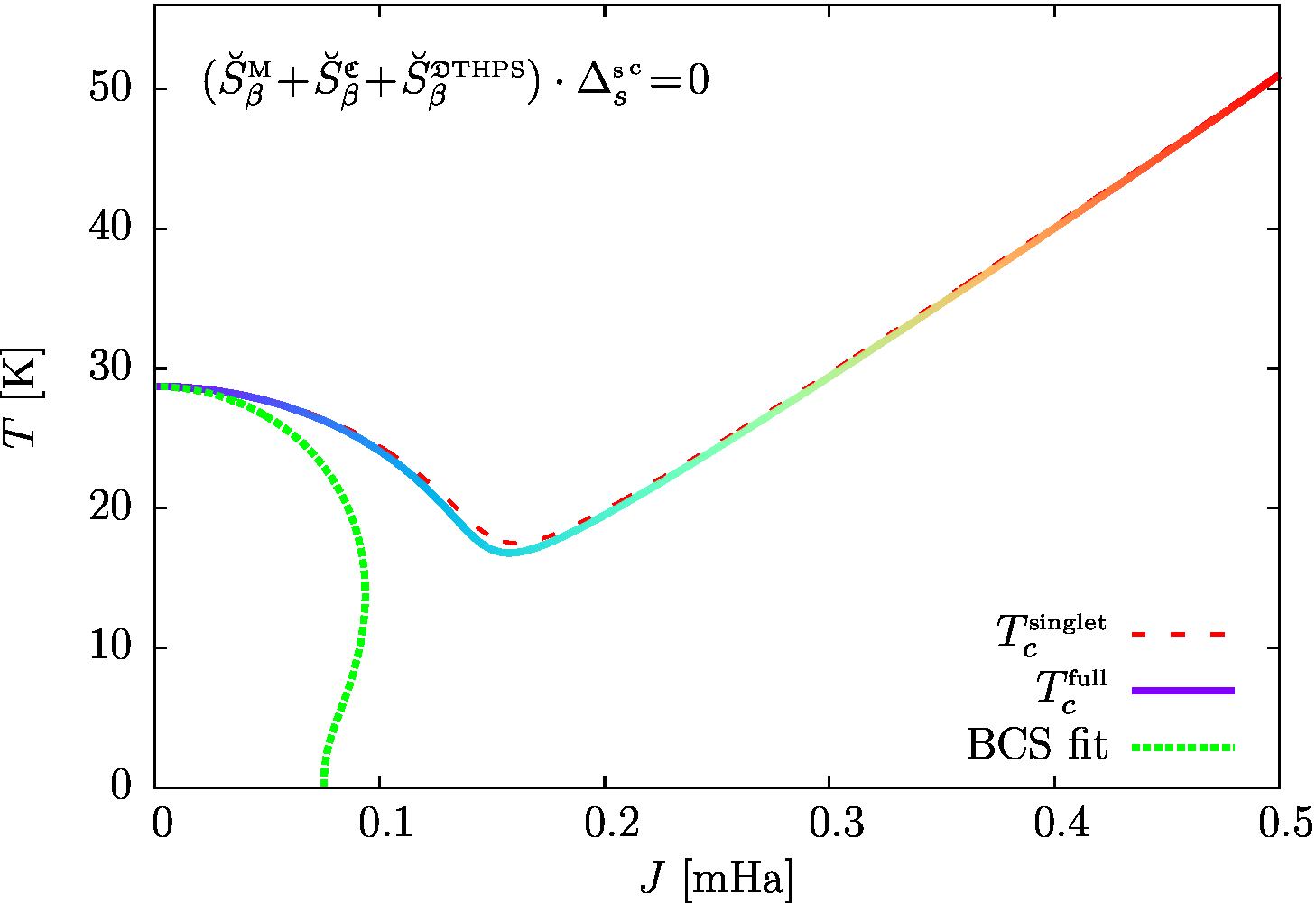}
\par\end{center}

\begin{flushleft}
{\small (a) $T_{{\scriptscriptstyle {\rm c}}}(J)$ excluding the Coulomb
coupling also removing triplet self-energy parts (singlet) or not
(full) plus a BCS fit to the same $T_{{\scriptscriptstyle {\rm c}}}(J=0)$.}
\par\end{flushleft}%
\end{minipage}
\par\end{center}%
\end{minipage}\nolinebreak%
\begin{minipage}[t]{0.333333\textwidth}%
\begin{center}
\begin{minipage}[t]{0.9\textwidth}%
\begin{center}
\includegraphics[width=1\textwidth]{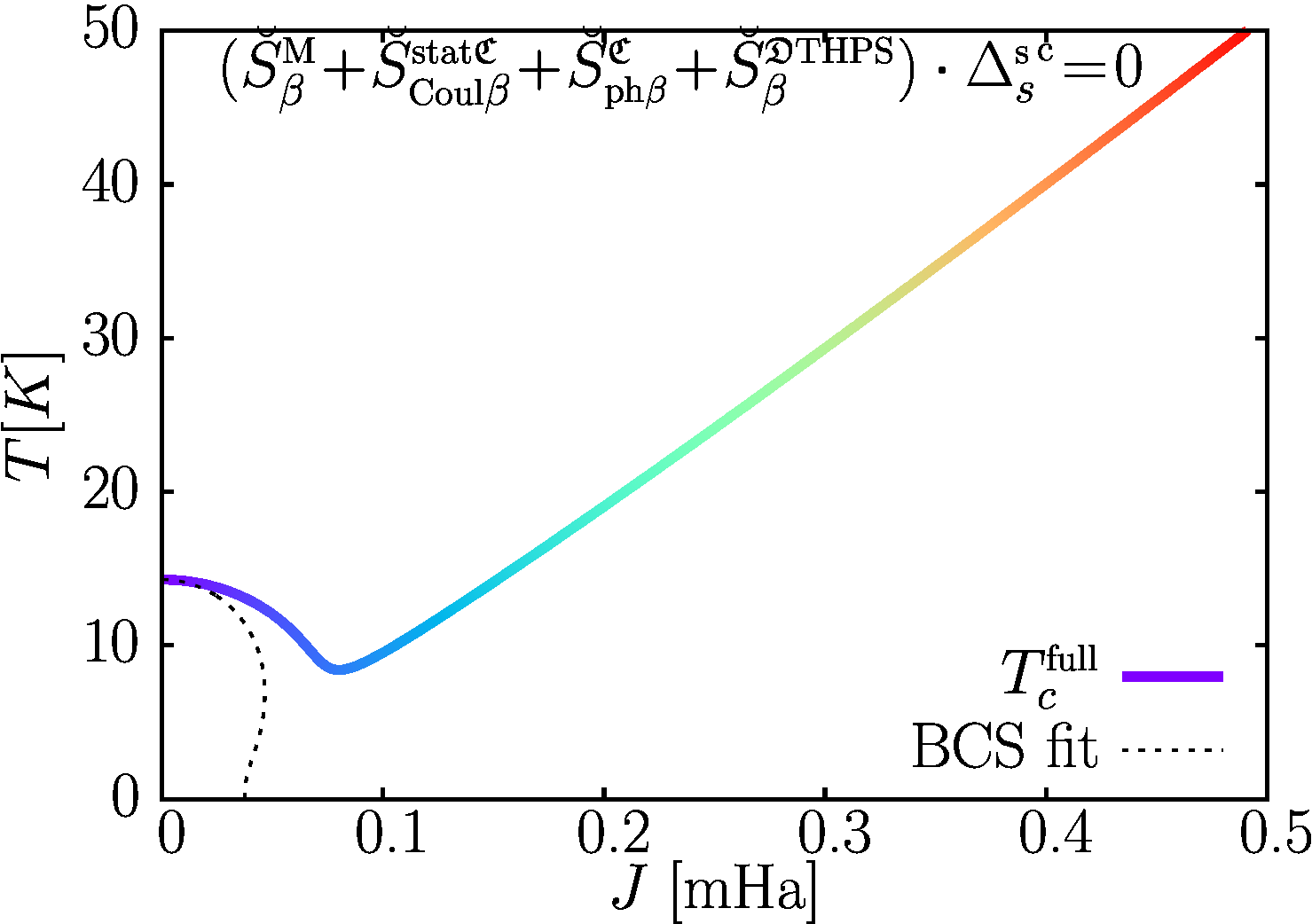}
\par\end{center}

\begin{flushleft}
{\small (c) $T_{{\scriptscriptstyle {\rm c}}}(J)$ including the Coulomb
coupling together with a BCS fit to the same $T_{{\scriptscriptstyle {\rm c}}}(J=0)$.}
\par\end{flushleft}%
\end{minipage}
\par\end{center}%
\end{minipage}\nolinebreak%
\begin{minipage}[t]{0.333333\textwidth}%
\begin{center}
\begin{minipage}[t]{0.9\textwidth}%
\begin{center}
\includegraphics[width=1\columnwidth]{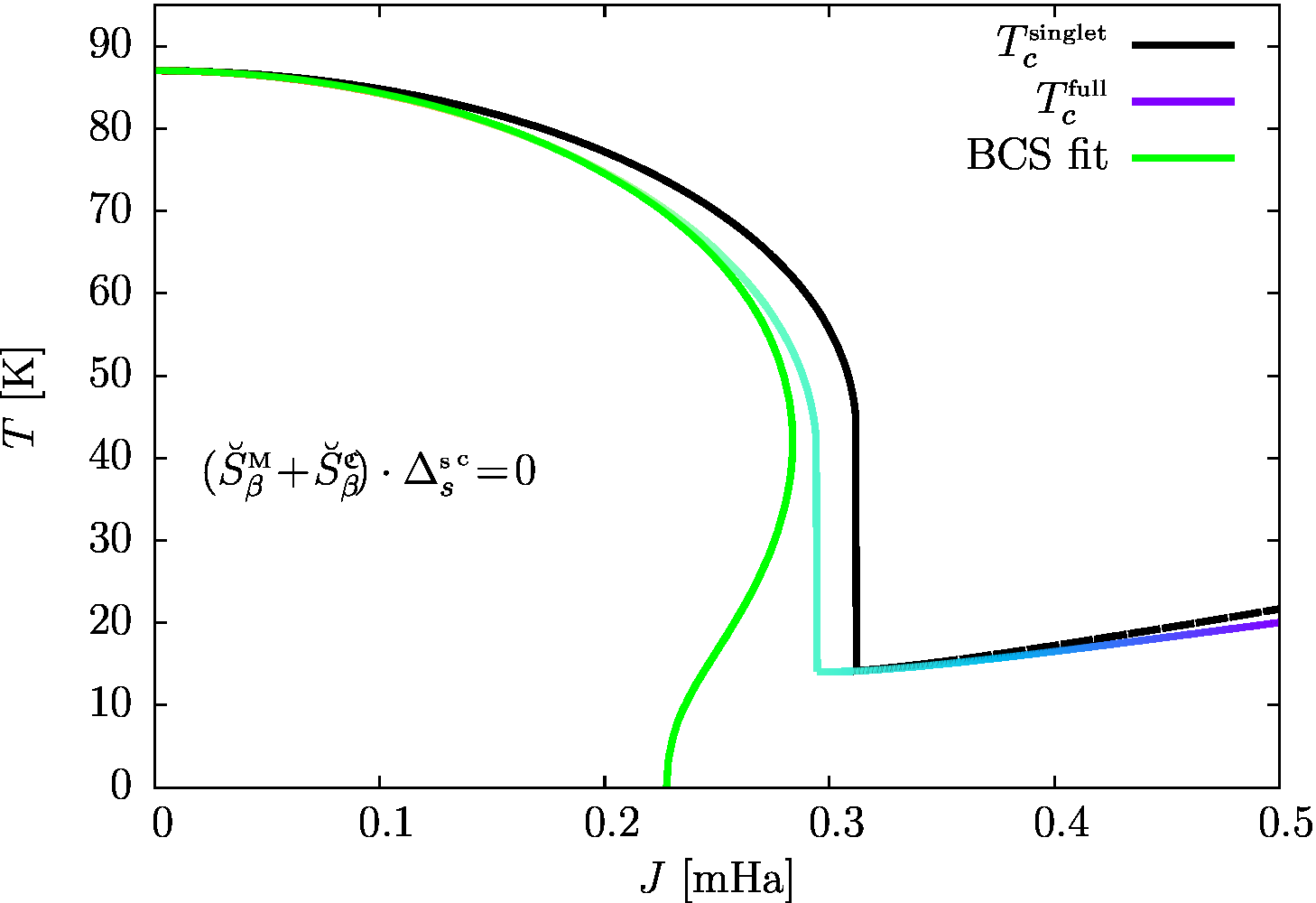}
\par\end{center}

\begin{flushleft}
{\small (e) $T_{{\scriptscriptstyle {\rm c}}}(J)$ excluding the Coulomb
coupling and $\breve{S}_{\beta}^{{\scriptscriptstyle \mathfrak{D}}}$
together with the linear BCS curve.}
\par\end{flushleft}%
\end{minipage}
\par\end{center}%
\end{minipage}\\
\begin{minipage}[t]{0.333333\textwidth}%
\begin{center}
\begin{minipage}[t]{0.9\textwidth}%
\begin{center}
\includegraphics[width=1\columnwidth]{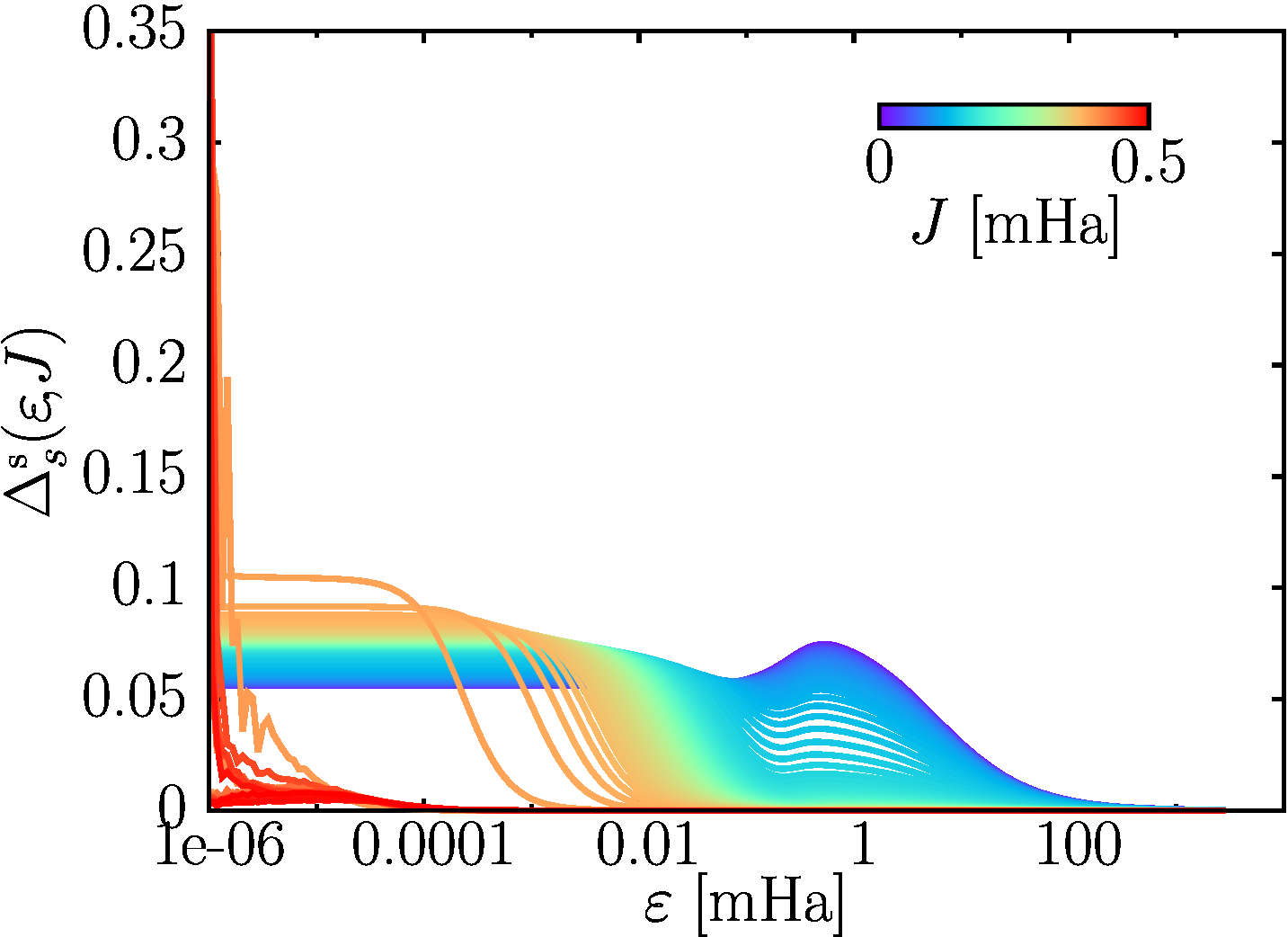}
\par\end{center}

\begin{flushleft}
{\small (b) Normalized $\varDelta_{{\rm s}}^{{\scriptscriptstyle \text{s}}}(\mathfrak{e})$
at $T_{{\scriptscriptstyle {\rm c}}}^{{\rm {\scriptscriptstyle full}}}(J)$
of (a).}
\par\end{flushleft}%
\end{minipage}
\par\end{center}%
\end{minipage}\nolinebreak%
\begin{minipage}[t]{0.333333\textwidth}%
\begin{center}
\begin{minipage}[t]{0.9\textwidth}%
\begin{center}
\includegraphics[width=1\textwidth]{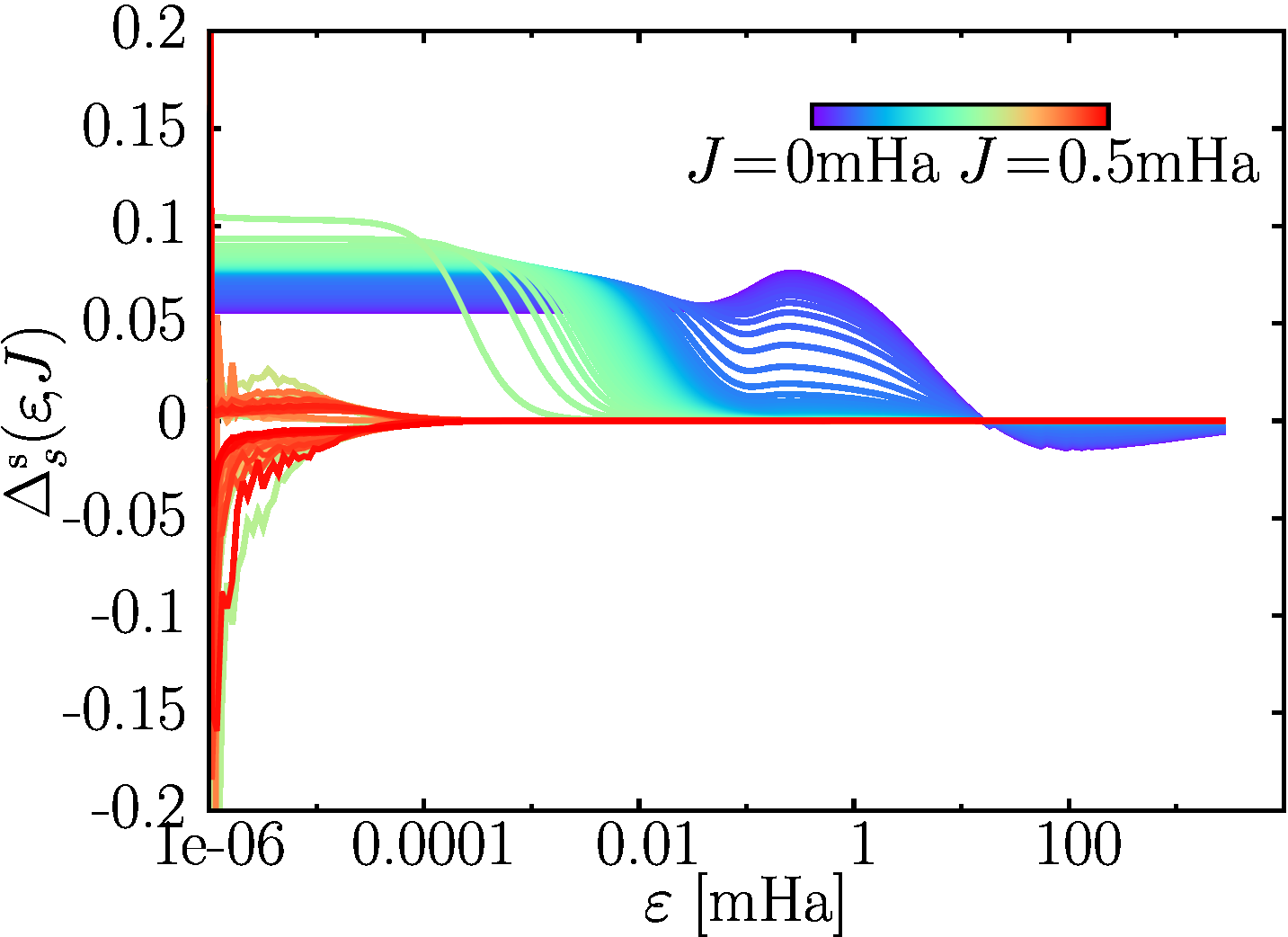}
\par\end{center}

\begin{flushleft}
{\small (d) Normalized $\varDelta_{{\rm s}}^{{\scriptscriptstyle \text{s}}}(\mathfrak{e})$
at $T_{{\scriptscriptstyle {\rm c}}}^{{\rm {\scriptscriptstyle full}}}(J)$
of (c).}
\par\end{flushleft}%
\end{minipage}
\par\end{center}%
\end{minipage}\nolinebreak%
\begin{minipage}[t]{0.333333\textwidth}%
\begin{center}
\begin{minipage}[t]{0.9\textwidth}%
\begin{center}
\includegraphics[width=1\columnwidth]{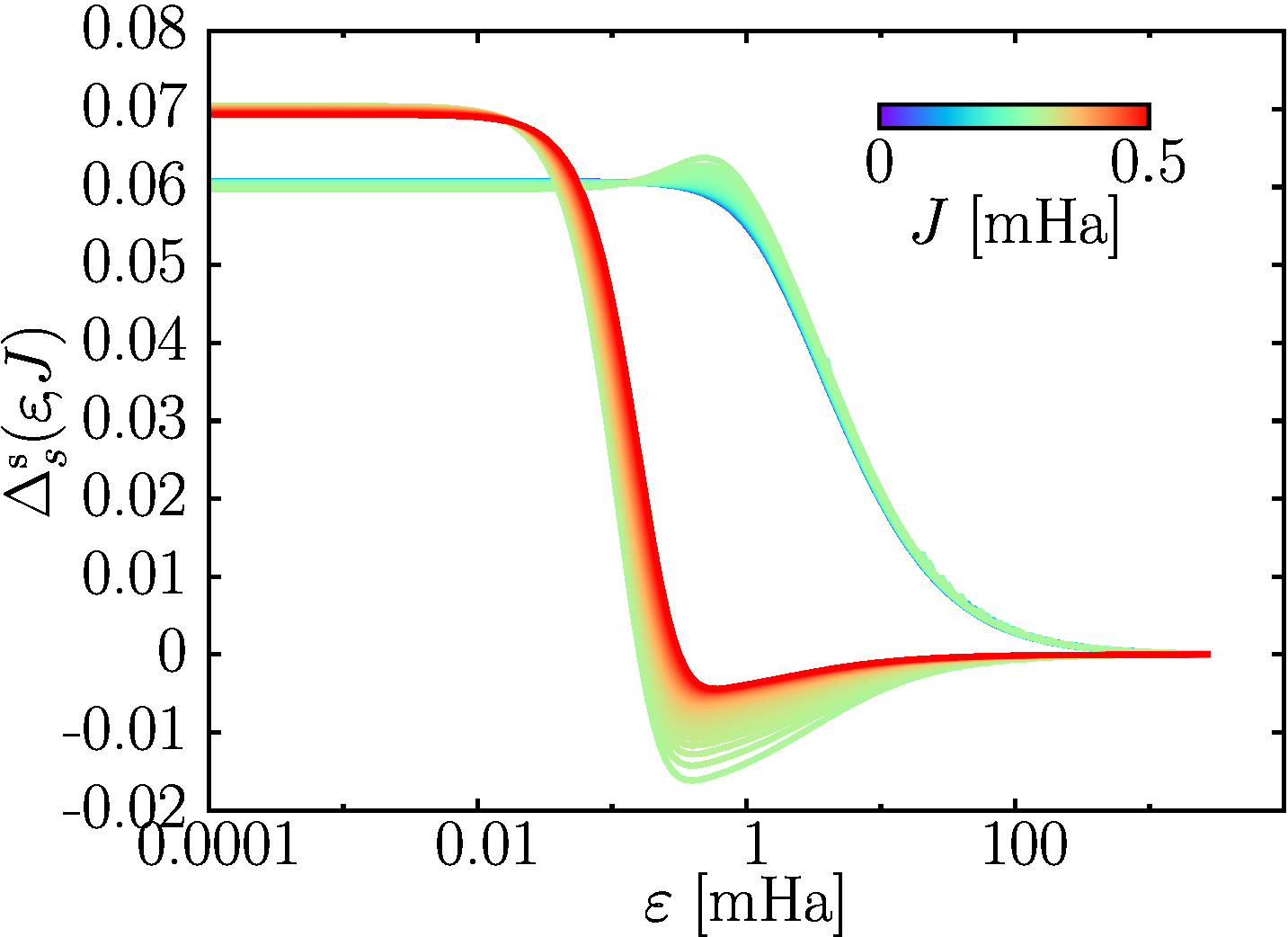}
\par\end{center}

\begin{flushleft}
{\small (f) The normalized $\varDelta_{{\rm s}}^{{\scriptscriptstyle {\rm s}}}(\mathfrak{e})$
from $\breve{S}_{\beta}-\breve{S}_{\beta}^{{\scriptscriptstyle \mathfrak{D}}}$
at $T_{{\scriptscriptstyle {\rm c}}}(J)$ of (e).}
\par\end{flushleft}%
\end{minipage}
\par\end{center}%
\end{minipage}
\par\end{center}

\caption{(color online) $T_{{\scriptscriptstyle {\rm c}}}(J)$ from SpinSCDFT
and the linear BCS curve with the same $T_{{\scriptscriptstyle {\rm c}}}(J=0)$
. We show the $T_{{\scriptscriptstyle {\rm c}}}(J)$ in - or excluding
the Coulomb coupling, in - or excluding contributions from triplet
self-energy parts. Beyond the range of the continuous transition,
the SpinSCDFT solutions start to increase at $J\approx0.15\mbox{mHa}$.
The effect of the Coulomb potential is to reduce the $T_{{\scriptscriptstyle {\rm c}}}(J=0)$
but the overall shape remains essentially unaltered. The normalized
eigenfunctions are shown with a color code indicated the respective
splitting. The solutions become numerically noisy at large splittings.\label{fig:TcVersusSplitting}}
\end{minipage}
\end{figure*}
Using the condition of $\breve{S}_{\beta}$ to be positive definite
we compute the $T_{{\scriptscriptstyle {\rm c}}}(J)$ curve of the
model (see Fig.~\ref{fig:TcVersusSplitting}). At a low field the
$T_{{\scriptscriptstyle {\rm c}}}(J)$ curve behaves as expected;
the critical temperature is slowly reducing with increasing $J$.
Similar to the Eliashberg results in Sec.~\ref{sec:Eliashberg},
the SpinSCDFT pair potential seems to be more resistant against a
splitting than the BCS approach predicts.

In the regime of a first order phase transition, where the conditions
for a linearization are not met, SpinSCDFT behaves differently as
compared to the linear BCS solution of Fig.~\ref{fig:BCSsolutions}.
While in neither case, BCS nor SpinSCDFT, a linearization can be expected
to yield sensible results for a discontinuous first order transition,
the behavior of the $T_{{\scriptscriptstyle {\rm c}}}(J)$ curve from
SpinSCDFT is certainly more unphysical. At high field, past $J\approx0.15{\rm mHa}$
the $T_{{\scriptscriptstyle {\rm c}}}(J)$ curve bends outwards and
starts an almost linearly increase with $J$.

In figure \ref{fig:TcVersusSplitting} b), the eigenfunctions to the
singular eigenvalues of $\breve{S}_{\beta}$ for increasing $J$ are
plotted. We can clearly observe that the upturn the $T_{{\scriptscriptstyle {\rm c}}}(J)$
curve is accompanied by an increasing localization of $\varDelta_{{\rm s}}^{{\scriptscriptstyle \text{s}}}$
at the Fermi level. The usual high energy tail gets more and more
suppressed. For very large splittings, $\varDelta_{{\rm s}}^{{\scriptscriptstyle \text{s}}}$
becomes numerically noisy. Reintroducing the Coulomb coupling to $\breve{S}_{\beta}$
we observe a similar behavior. In this case $\varDelta_{{\rm s}}^{{\scriptscriptstyle \text{s}}}$
shows a characteristic negative tail induced by the Coulomb renormalization
mechanism \cite{ScalapinoStrongCouplSC1966,MorelCalculationSCStateParametersWithRetardedElPhInteraction1962}
as it occurs within SCDFT \cite{LuedersSCDFTI2005,MarquesSCDFTIIMetals2005,MassiddaCoulombCaC6H2009}.
From the comparison between $T_{{\scriptscriptstyle {\rm c}}}^{{\rm {\scriptscriptstyle SpinSCDFT}}}(J)$
with the green dashed BCS curve in Fig.~\ref{fig:TcVersusSplitting},
we note that in the second order regime $T_{{\scriptscriptstyle {\rm c}}}^{{\rm {\scriptscriptstyle BCS}}}(J)$
scales down with $J$ faster. In order to make the strong coupling
SpinSCDFT theory more similar to the weak coupling BCS approach we
disregard $\breve{S}_{\beta}^{{\scriptscriptstyle \mathfrak{D}}}$
in Fig.~\ref{fig:TcVersusSplitting} e) and f). In this case we are
only considering the effectively attractive coupling among electrons
via phonons, similar to Fr\"ohlich \cite{HFroehlich1952} and BCS.
The effective Fr\"ohlich interaction requires the coupling to be small,
and moreover we neglect the phonon influence on the normal state (Nambu
diagonal) part of the self-energy entirely. Thus this approximation
is called the weak coupling limit. As expected, the resulting $T_{{\scriptscriptstyle {\rm c}}}^{{\rm {\scriptscriptstyle SpinSCDFT}}}(J)/T_{{\scriptscriptstyle {\rm c}}}^{{\rm {\scriptscriptstyle SpinSCDFT}}}(0)$
and $T_{{\scriptscriptstyle {\rm c}}}^{{\rm {\scriptscriptstyle BCS}}}(J)/T_{{\scriptscriptstyle {\rm c}}}^{{\rm {\scriptscriptstyle BCS}}}(0)$
behave very similarly. Here the $T_{{\scriptscriptstyle {\rm c}}}(J)$
curves shown in Fig.~\ref{fig:TcVersusSplitting} e) also feature
the linear increase for high splitting. Moreover we observe a discontinuous
jump of the critical temperature at a certain splitting $J_{c}$ which
is accompanied by the eigenfunction dramatically changing shape. After
the jump, the solution does not have a common sign convention but
shows positive and negative parts. Also here we find numerically noisy
solutions.

The BCS $T_{{\scriptscriptstyle {\rm c}}}(J)$ curve, fitted to the
same $T_{{\scriptscriptstyle {\rm c}}}(0\rm{mHa})$, matches the
weak coupling SpinSCDFT $T_{{\scriptscriptstyle {\rm c}}}(J)$ curve
Fig.~\ref{fig:TcVersusSplitting} e), not the strong coupling curve of 
Fig.~\ref{fig:TcVersusSplitting} a). This
points out that the strong coupling $\breve{S}_{\beta}^{{\scriptscriptstyle \mathfrak{D}}}$
term does not simply scale $T_{{\scriptscriptstyle {\rm c}}}(J)$
down equally on both, $T$ and $J$ axis. Instead, $\breve{S}_{\beta}^{{\scriptscriptstyle \mathfrak{D}}}$
leads to a larger $T_{{\scriptscriptstyle {\rm c}}}(J)$ reduction
of the temperature axis. Thus we conclude that strong coupling systems are less
effected by an exchange splitting relative to their $T_{{\scriptscriptstyle {\rm c}}}(0\rm{mHa})$.

\subsection{Non-Linear Sham-Schl\"uter Equation\label{sub:Non-Linear-Sham-Schluter-Equatio}}

\begin{figure*}
\begin{centering}
\begin{minipage}[t]{0.5\textwidth}%
\begin{center}
\begin{minipage}[t]{0.9\textwidth}%
\begin{center}
\includegraphics[width=0.9\textwidth]{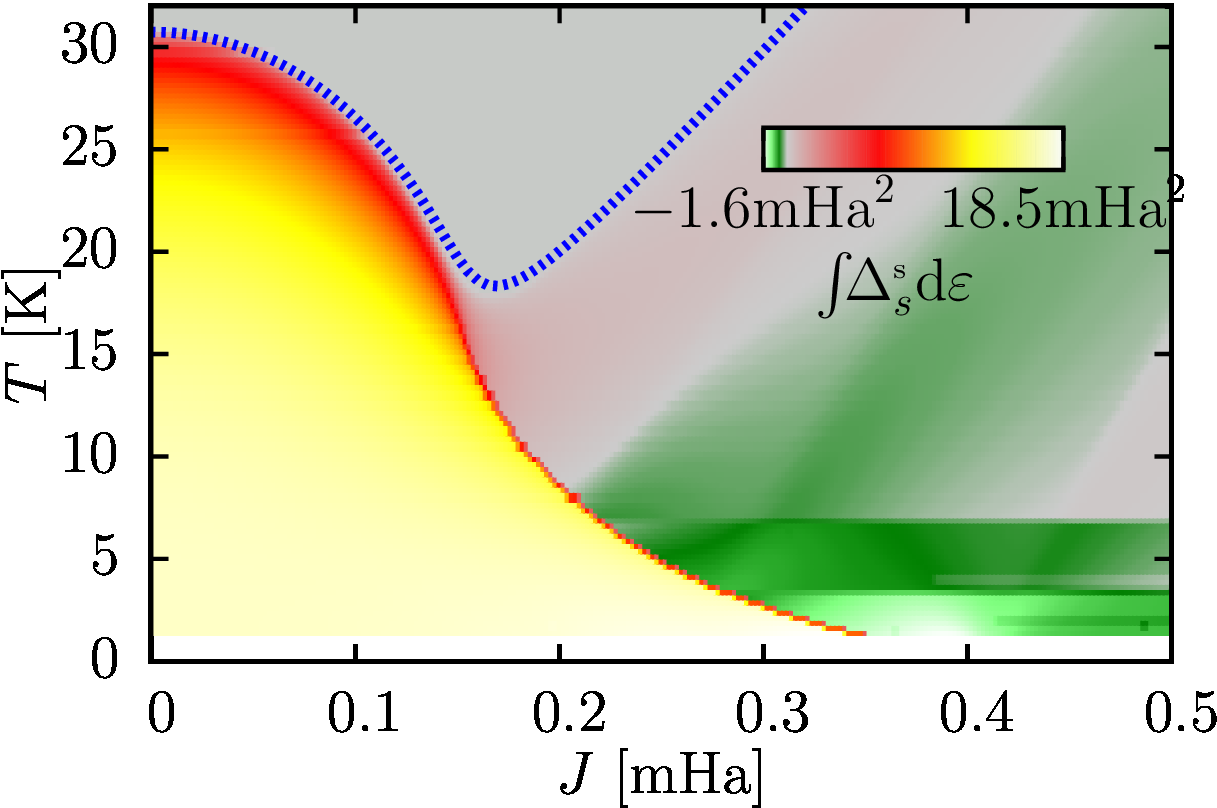}\caption{(color online) $J-T$ diagram of solutions to the non-linear gap equation.
We include the $T_{c}^{{\scriptscriptstyle {\rm full}}}(J)$ curve
(dashed blue) from the linearized functional of Fig.~\ref{fig:TcVersusSplitting}
a).\label{fig:PhaseDiagramNonLinear}}

\par\end{center}%
\end{minipage}
\par\end{center}%
\end{minipage}\nolinebreak%
\begin{minipage}[t]{0.5\textwidth}%
\begin{center}
\begin{minipage}[t]{0.9\textwidth}%
\begin{center}
\includegraphics[width=0.9\textwidth]{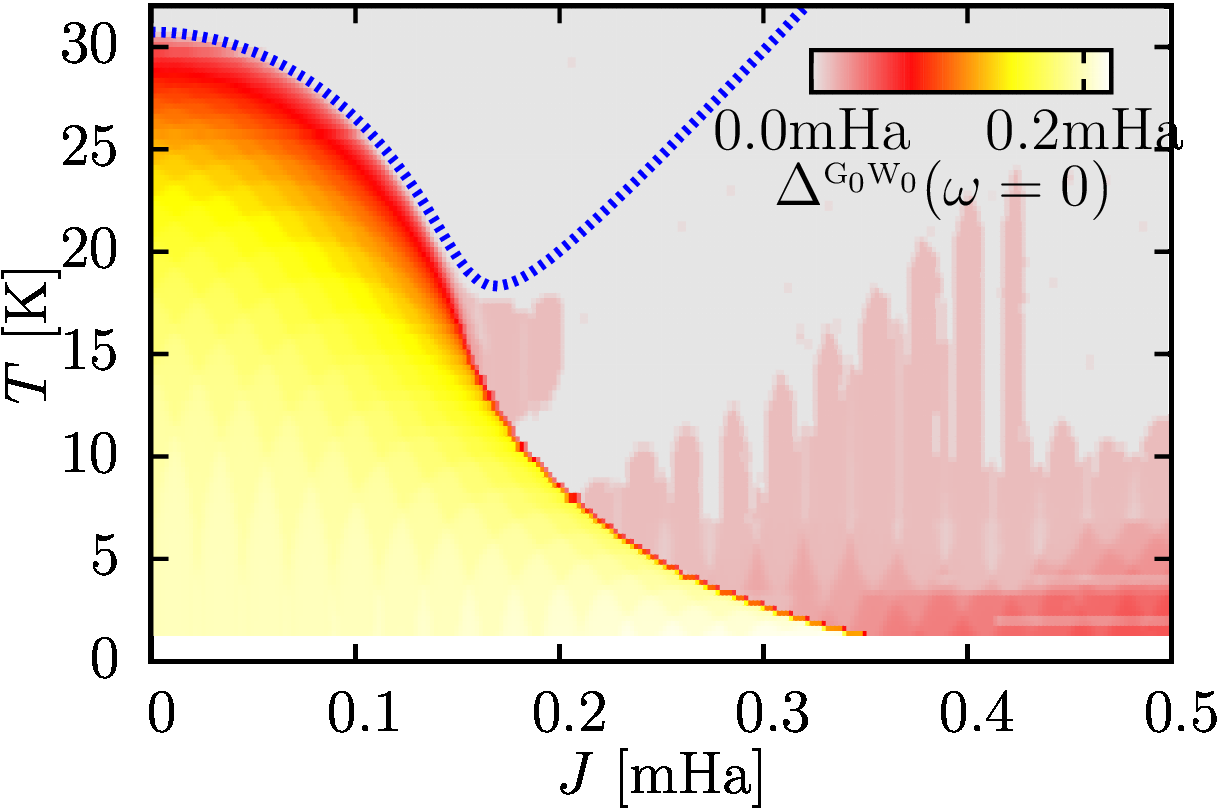}\caption{(color online) The SC gap in the SpinSCDFT $\rm{G}_0\rm{W}_0$ DOS. The dashed blue
line is the linear $T_{c}^{{\scriptscriptstyle {\rm full}}}(J)$ of
Fig.~\ref{fig:TcVersusSplitting} a).\label{fig:PhaseDiagramNonLinearG0W0}}

\par\end{center}%
\end{minipage}
\par\end{center}%
\end{minipage}\\

\par\end{centering}

\centering{}%
\begin{minipage}[t]{0.333\textwidth}%
\begin{center}
\begin{minipage}[t]{0.9\textwidth}%
\begin{center}
\includegraphics[width=1\textwidth]{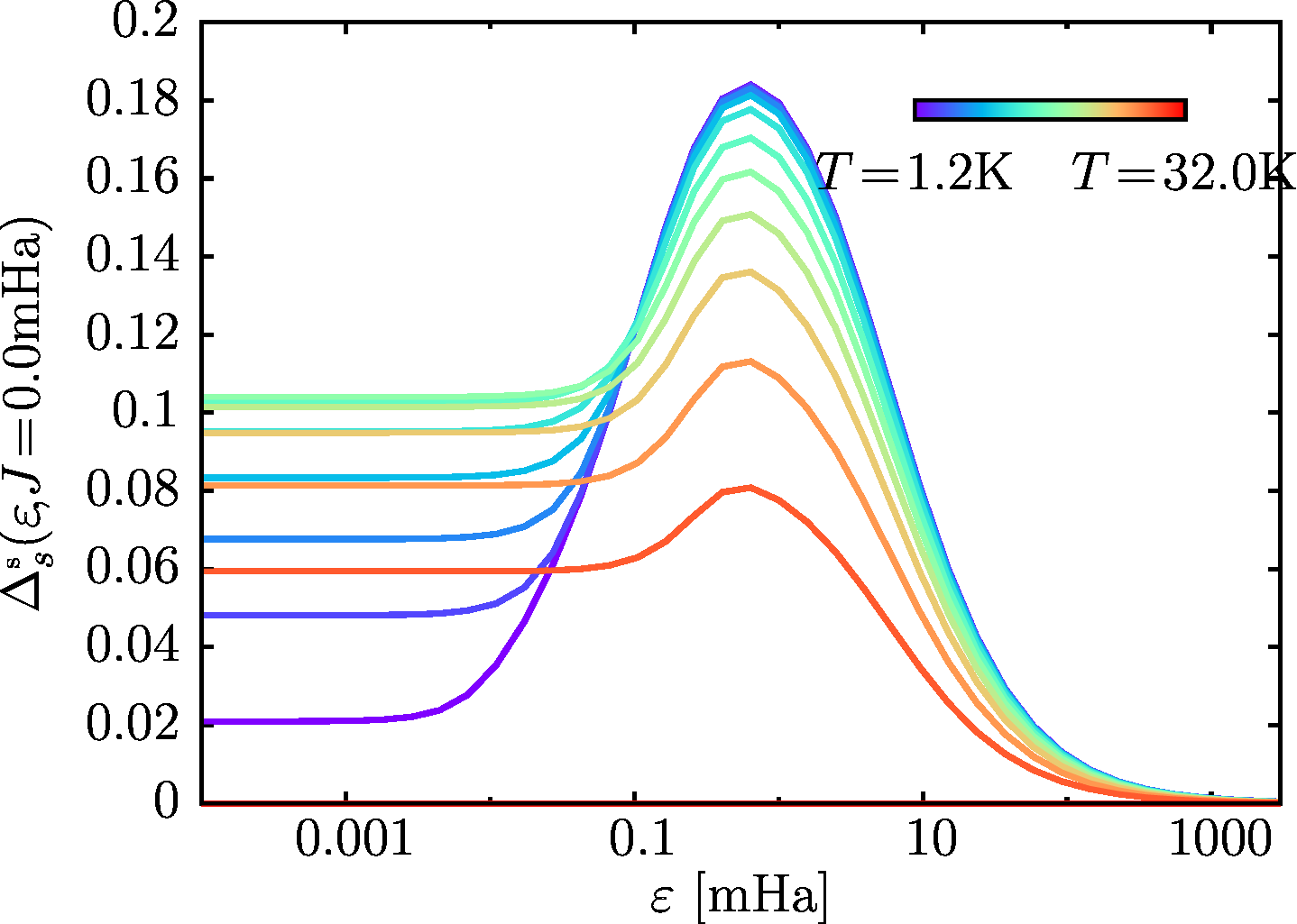}
\par\end{center}

\caption{(color online) $\varDelta_{{\rm s}}^{{\scriptscriptstyle \text{s}}}(\mathfrak{e})$
for $J=0.0{\rm mHa}$ as a function of $T$. For low $T$, $\varDelta_{{\rm s}}^{{\scriptscriptstyle \text{s}}}(\mathfrak{e})$
goes to zero at $\varepsilon\approx0$.\label{fig:NonLinearSolutionsIsoSplittingLineJZero}}
\end{minipage}
\par\end{center}%
\end{minipage}\nolinebreak%
\begin{minipage}[t]{0.333\textwidth}%
\begin{center}
\begin{minipage}[t]{0.9\textwidth}%
\begin{center}
\includegraphics[width=1\textwidth]{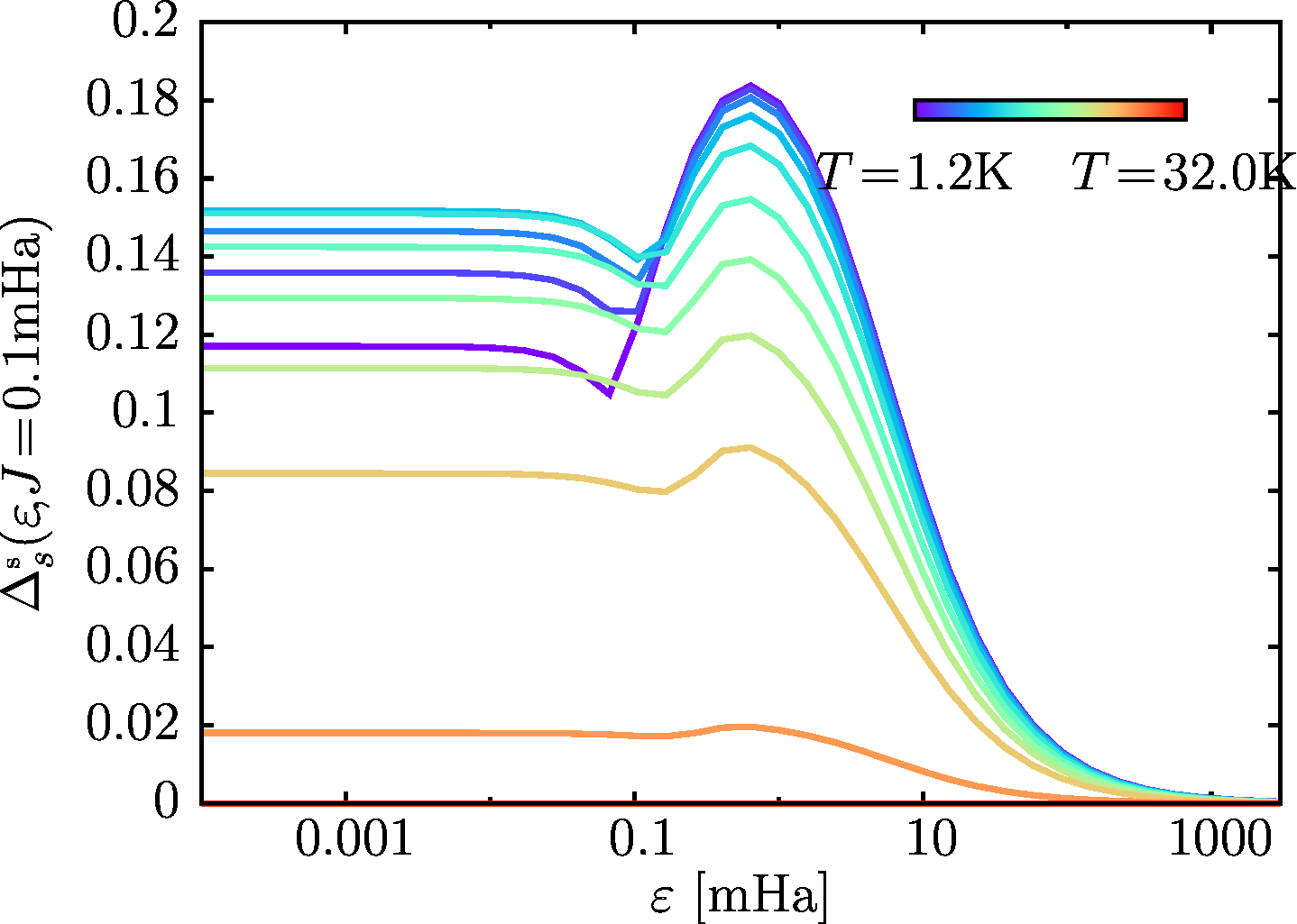}
\par\end{center}

\caption{(color online) $\varDelta_{{\rm s}}^{{\scriptscriptstyle \text{s}}}(\mathfrak{e})$
for $J=0.1{\rm mHa}$ as a function of $T$. For low $T$, $\varDelta_{{\rm s}}^{{\scriptscriptstyle \text{s}}}(\mathfrak{e})$
remains above $J$ at $\varepsilon\approx0$.\label{fig:NonLinearIsoTemperatureJ01}}
\end{minipage}
\par\end{center}%
\end{minipage}\nolinebreak%
\begin{minipage}[t]{0.333\textwidth}%
\begin{center}
\begin{minipage}[t]{0.9\textwidth}%
\begin{center}
\includegraphics[width=1\textwidth]{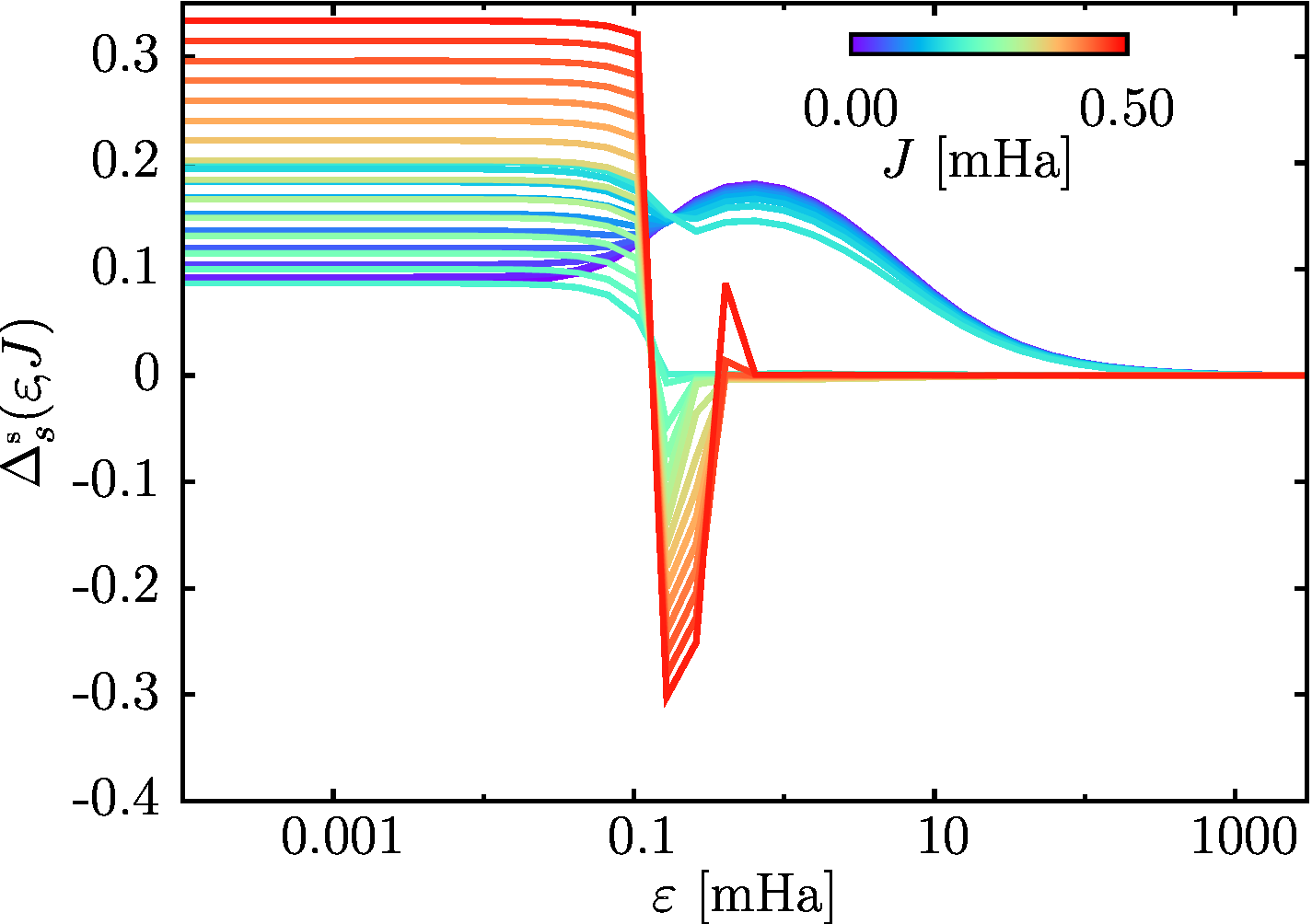}
\par\end{center}

\caption{(color online) $\varDelta_{{\rm s}}^{{\scriptscriptstyle \text{s}}}(\mathfrak{e})$
for $T=10{\rm K}$ as a function of $J$. At $J\approx0.17{\rm mHa}$
$\varDelta_{{\rm s}}^{{\scriptscriptstyle \text{s}}}(\mathfrak{e})$
dramatically change shape.\label{fig:IsoTemperatureLine}}
\end{minipage}
\par\end{center}%
\end{minipage}
\end{figure*}
\begin{figure*}
\begin{centering}
\begin{minipage}[t]{0.5\textwidth}%
\begin{center}
\begin{minipage}[t]{0.9\textwidth}%
\begin{center}
\includegraphics[width=0.9\textwidth]{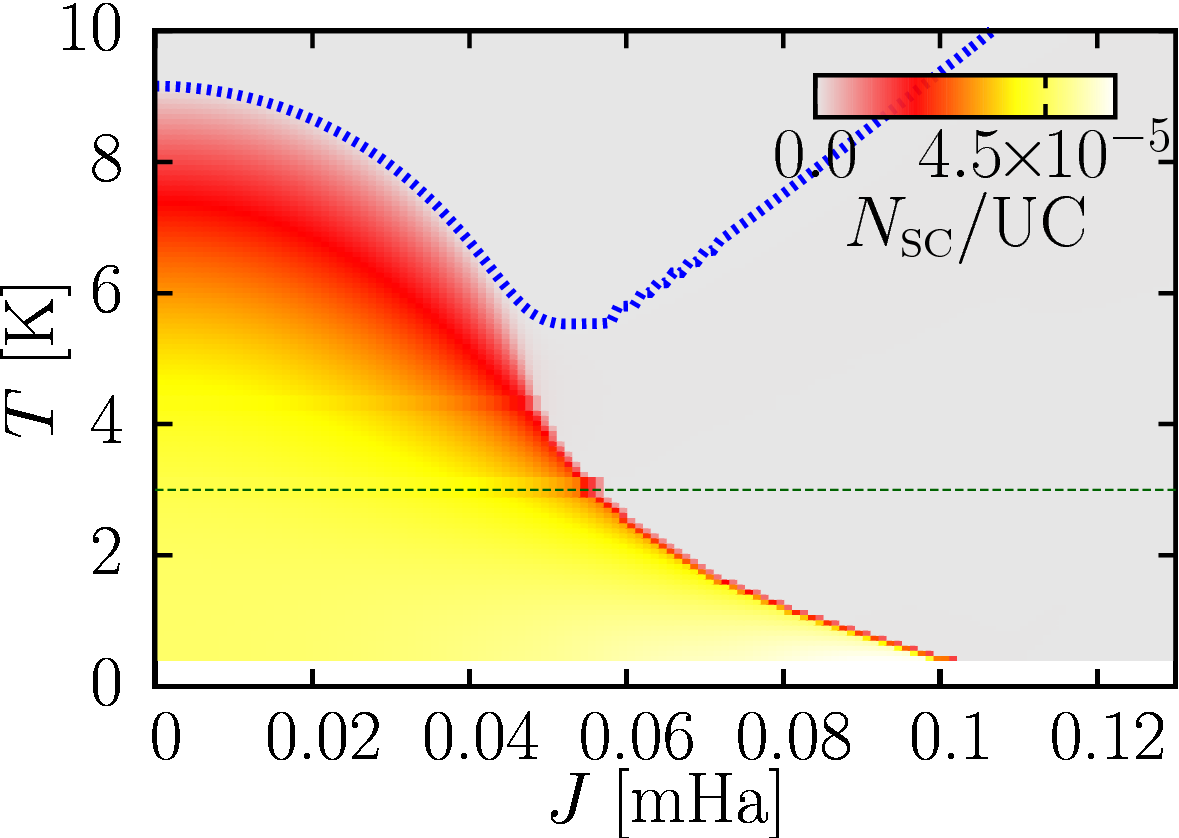}\caption{(color online) $N_{{\scriptscriptstyle {\rm SC}}}(T,J)$ and the linear
$T_{c}(J)$ (dashed blue) including the Coulomb repulsion. The $\varDelta_{s}^{{\scriptscriptstyle \mathrm{s}}}(\mathfrak{e})$
after the transition lead to almost no condensed electrons.\label{fig:CoulombResultsNsc}}

\par\end{center}%
\end{minipage}
\par\end{center}%
\end{minipage}\nolinebreak%
\begin{minipage}[t]{0.5\textwidth}%
\begin{center}
\begin{minipage}[t]{0.9\textwidth}%
\begin{center}
\includegraphics[width=0.9\textwidth]{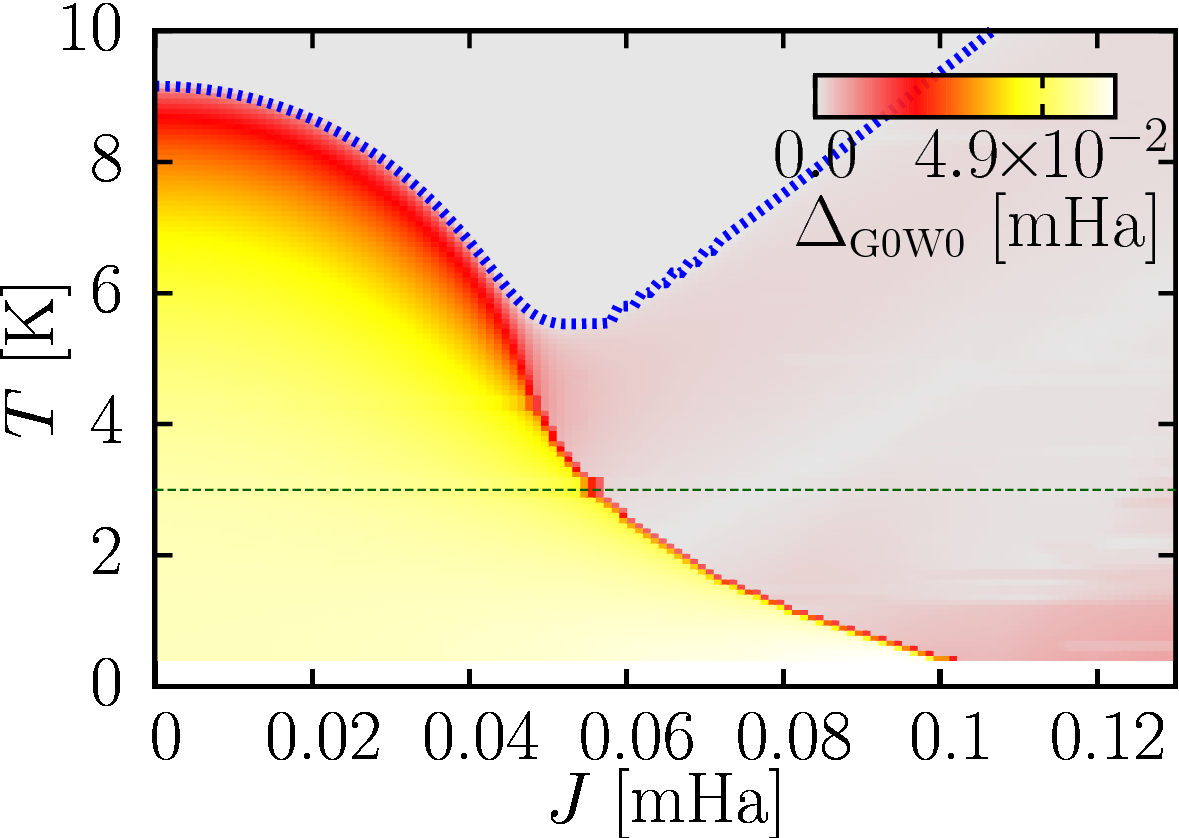}\caption{(color online) The SC gap in the SpinSCDFT $\rm{G}_0\rm{W}_0$ DOS. The dashed blue
line is the linear $T_{c}^{{\scriptscriptstyle {\rm full}}}(J)$ of
Fig.~\ref{fig:TcVersusSplitting} c).\label{fig:CoulombResultsG0W0Gap}}

\par\end{center}%
\end{minipage}
\par\end{center}%
\end{minipage}\\

\par\end{centering}

\centering{}%
\begin{minipage}[t]{0.5\textwidth}%
\begin{center}
\begin{minipage}[t]{0.9\textwidth}%
\begin{center}
\includegraphics[width=0.9\textwidth]{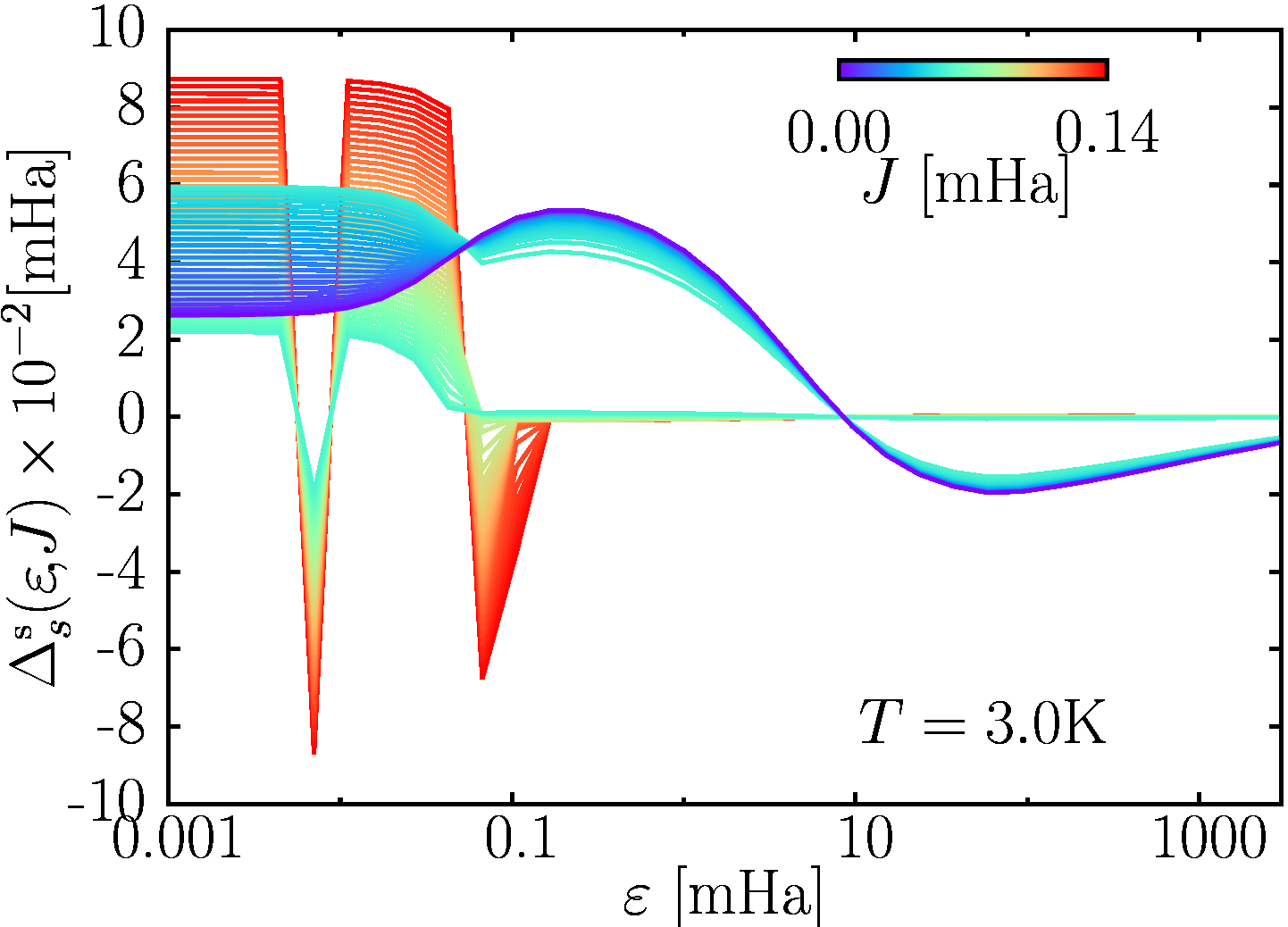}
\par\end{center}

\caption{(color online) $\varDelta_{s}^{{\scriptscriptstyle {\rm s}}}(\mathfrak{e})$
at $T=3{\rm K}$ for several $J$ along the dashed green line in Fig.~\ref{fig:CoulombResultsNsc}.
The $\varDelta_{s}^{{\scriptscriptstyle {\rm s}}}(\mathfrak{e})$
past the transition at $J=0.056\rm{mHa}$ are similarly confined to the Fermi level region as without the Coulomb repulsion.\label{fig:CoulombResultsT3KIsoline}}
\end{minipage}
\par\end{center}%
\end{minipage}\nolinebreak%
\begin{minipage}[t]{0.5\textwidth}%
\begin{center}
\begin{minipage}[t]{0.9\textwidth}%
\begin{center}
\includegraphics[width=0.9\textwidth]{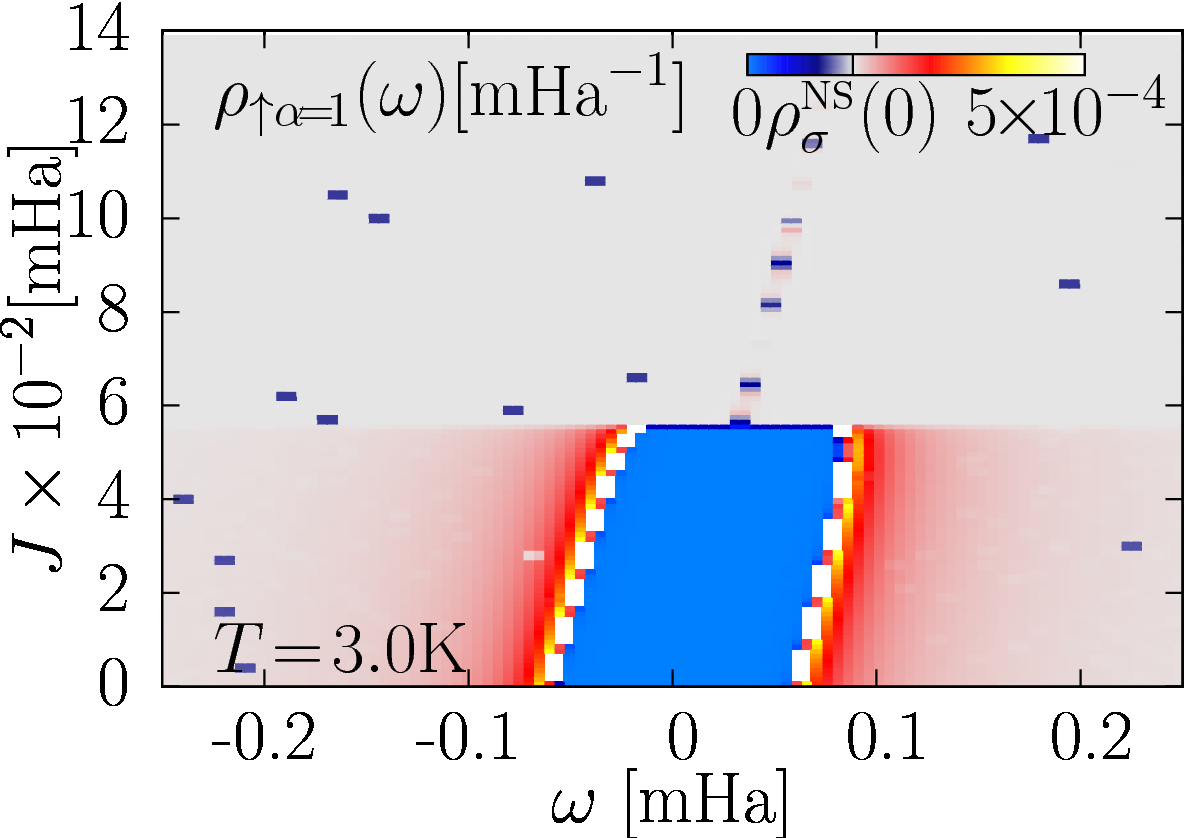}
\par\end{center}

\caption{(color online) Up-spin channel of the $\rm{G}_0\rm{W}_0$ DOS along the dashed green
line in Fig.~\ref{fig:CoulombResultsG0W0Gap} at $T=3{\rm K}$.
We see only small features from the unphysical, oscillatory solutions 
past the transition (light blue to red in Fig.~\ref{fig:CoulombResultsT3KIsoline}.\label{fig:CoulombResultsDOST3KIsoline})}
\end{minipage}
\par\end{center}%
\end{minipage}
\end{figure*}
\begin{figure*}
\centering{}%
\begin{minipage}[t]{1\textwidth}%
\begin{center}
\begin{minipage}[t]{0.333333\textwidth}%
\begin{center}
\includegraphics[width=0.9\textwidth]{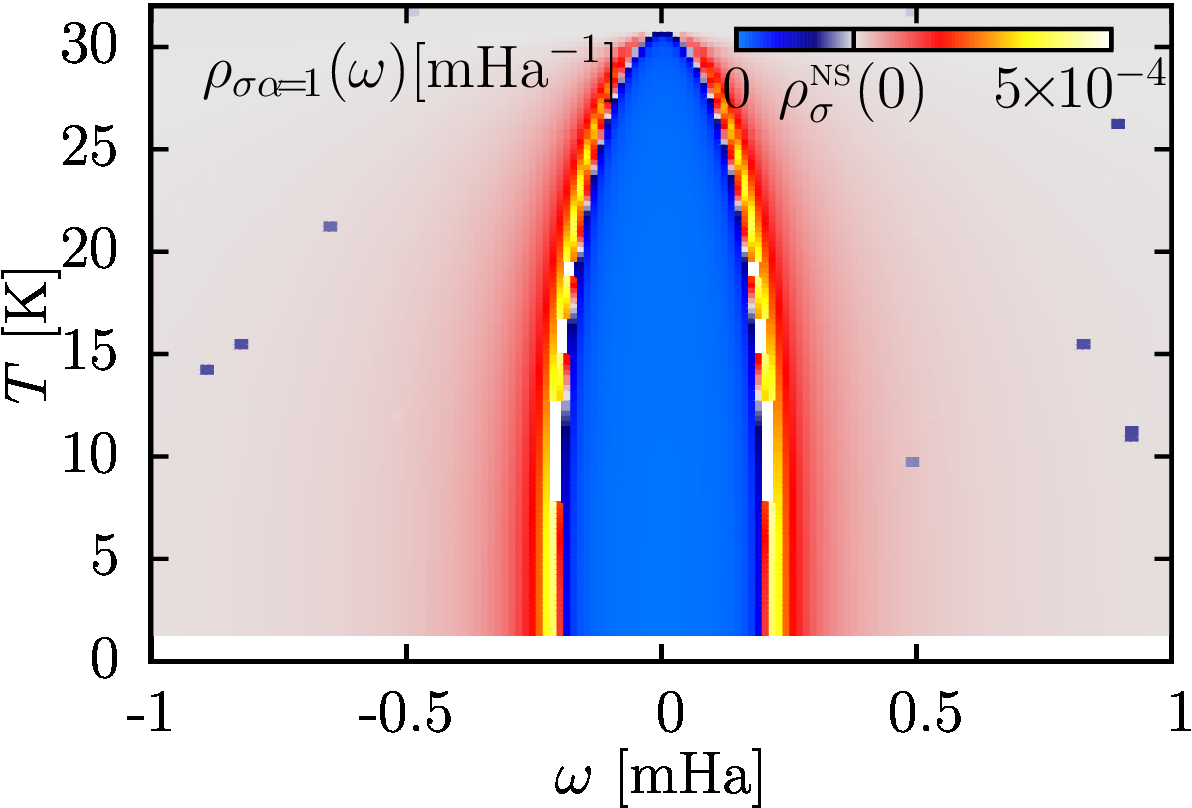}
\par\end{center}

\begin{center}
(a)
\par\end{center}%
\end{minipage}\nolinebreak%
\begin{minipage}[t]{0.333333\textwidth}%
\begin{center}
\includegraphics[width=0.9\textwidth]{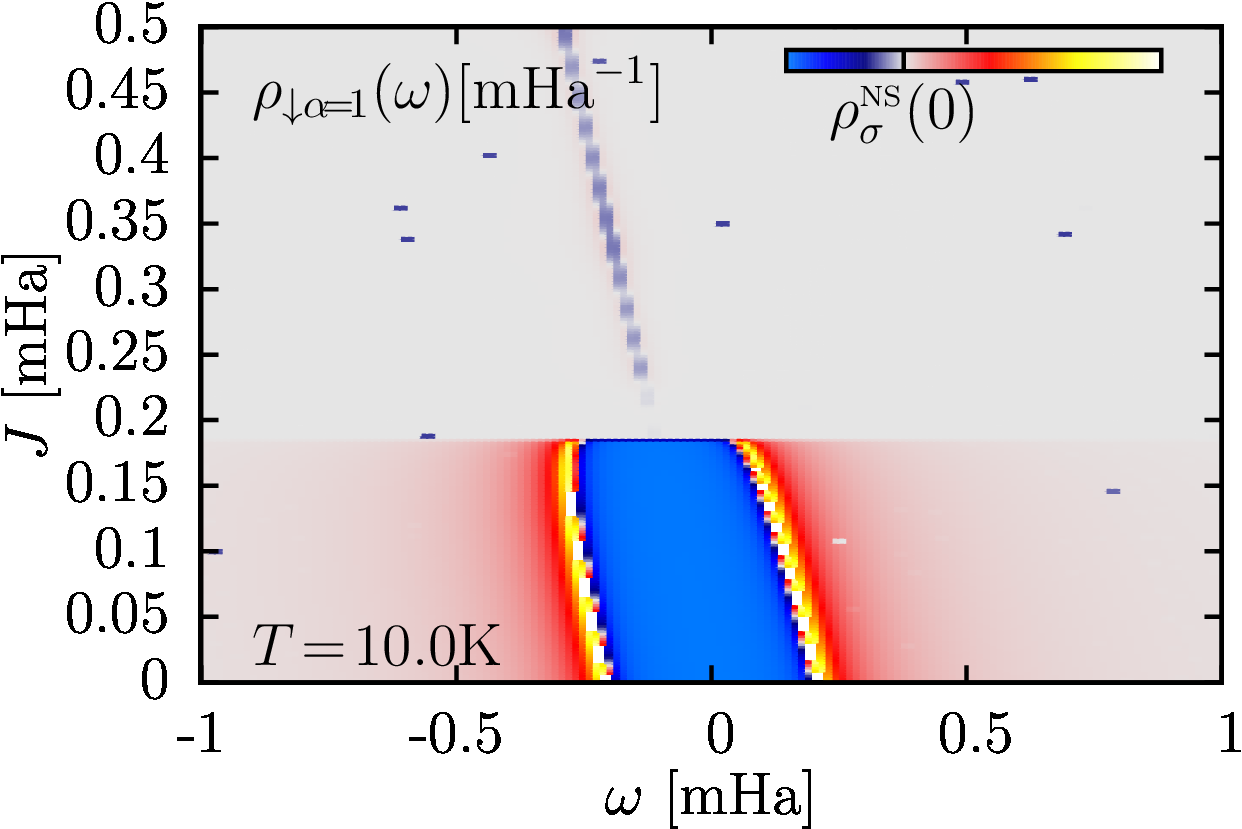}
\par\end{center}

\begin{center}
(b)
\par\end{center}%
\end{minipage}\nolinebreak%
\begin{minipage}[t]{0.333333\textwidth}%
\begin{center}
\includegraphics[width=0.9\textwidth]{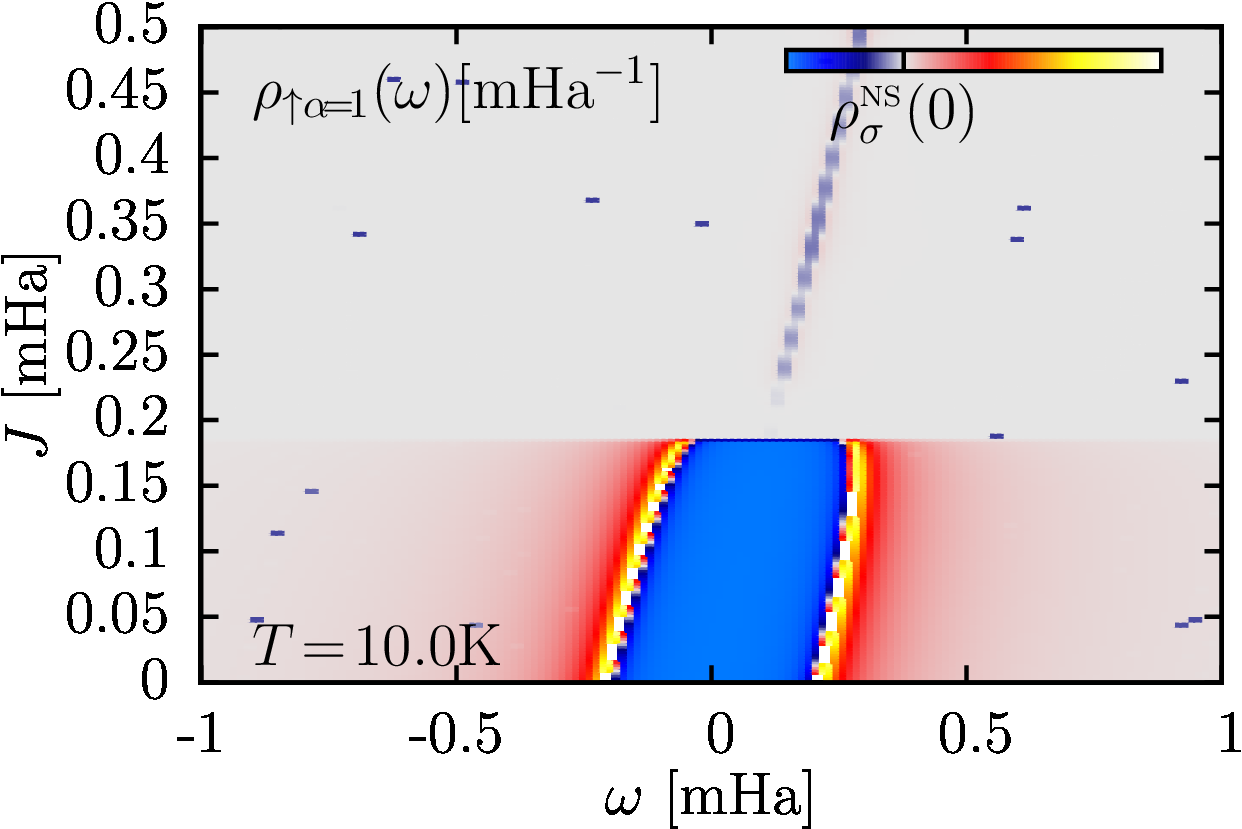}
\par\end{center}

\begin{center}
(c)
\par\end{center}%
\end{minipage}
\par\end{center}

\caption{(color online) DOS from the $\rm{G}_0\rm{W}_0$ GF. We show the DOS in (a) corresponding
to the SpinSCDFT results $\varDelta_{{\rm s}}^{{\scriptscriptstyle {\rm s}}}(\mathfrak{e})$
shown in Fig.~\ref{fig:NonLinearSolutionsIsoSplittingLineJZero}
for no splitting $J=0{\rm mHa}$. In (b) ($\sigma=\downarrow$) and
(c) ($\sigma=\uparrow$) we present the two different spin channels
of the DOS with the SpinSCDFT results for the $\varDelta_{{\rm s}}^{{\scriptscriptstyle {\rm s}}}(\mathfrak{e})$
along the iso-temperature line $T=10{\rm K}$ as shown in Fig.~\ref{fig:IsoTemperatureLine}.\label{fig:NonLinearMBDOS}}
\end{minipage}
\end{figure*}
The previous section has shown the importance to consider the fully
nonlinear Sham-Schl\"uter equation $S_{\beta}[\varDelta_{{\rm s}}^{{\scriptscriptstyle {\rm s}}}]\cdot\varDelta_{{\rm s}}^{{\scriptscriptstyle {\rm s}}}=0$ of Eq.~\eqref{eq:NonLinearSS}
when working in the limit of strong external field/large exchange
splitting $J$. We solve the fully non-linear Sham-Schl\"uter equation
\begin{eqnarray}
\varDelta_{{\rm s}}^{{\scriptscriptstyle {\rm s}}} & = &\mathcal{K}_{\mathcal{\mathcal{S}}}[\varDelta_{{\rm s}}^{{\scriptscriptstyle {\rm s}}}]\cdot\varDelta_{{\rm s}}^{{\scriptscriptstyle {\rm s}}}\\
\mathcal{K}_{\mathcal{S}} & = &\mathcal{S}^{-1}\cdot(S_{\beta}+\mathcal{S})\,,\label{eq:IsotropicShamSchlueterNull}
\end{eqnarray}
with the splitting matrix $\mathcal{\mathcal{S}}$ chosen to be $S_{\beta}^{{\scriptscriptstyle \text{M}}}(\varepsilon,J=0.0\,{\rm mHa})$
(more details on this procedure can be found in Sec.~\SubDerivationXCPotentialNonLinear).
In Fig.~\ref{fig:NonLinearSolutionsIsoSplittingLineJZero} we show
results, neglecting the Coulomb coupling along the iso-splitting line
$J=0.0{\rm mHa}$ as a function of temperature $T$. We obtain a $\varDelta_{{\rm s}}^{{\scriptscriptstyle \text{s}}}(\mathfrak{e})$
that goes to zero at the Fermi level for low temperatures (the purple
to blue lines in Fig.~\ref{fig:NonLinearSolutionsIsoSplittingLineJZero}).
This means the SC KS system is not gapped (still maintaining $\chi\neq0$)
and we cannot directly interpret the SC KS excitations as quasi particles.
In order to have the computationally convenient DFT scheme and a good
approximation to the quasi-particle structure at the same time we
introduce the one-cycle Dyson equation iteration for SC in the Appendix
\ref{sec:G0W0SC}. This approach is similar to the common $\rm{G}_0\rm{W}_0$ approximation
in band-structure theory \cite{HybertsenTheoryQuasiparticlesG0W01985}
and leads to excellent results in SpinSCDFT.

To complete the discussion of the $J$ and $T$ dependence of SpinSCDFT,
we need a characteristic number of a given $\varDelta_{{\rm s}}^{{\scriptscriptstyle \text{s}}}(\mathfrak{e})$
solution. As mentioned, $\varDelta_{{\rm s}}^{{\scriptscriptstyle \text{s}}}(\varepsilon=0,J)$
is not a sensible choice, because it neither corresponds to an excitation
gap nor is it a measure for the size of the potential $\varDelta_{{\rm s}}^{{\scriptscriptstyle \text{s}}}(\mathfrak{e})$.

Instead, we chose $\int\varDelta_{{\rm s}}^{{\scriptscriptstyle \text{s}}}(\mathfrak{e}){\rm d}\varepsilon$
and the resulting SpinSCDFT $J-T$ diagram of Fig.~\ref{fig:PhaseDiagramNonLinear}
shows a transition at a point where, from the shape of the non-linear
BCS and Eliashberg diagram the first order phase transition is to
be expected. However, following this discontinuous transition, the
solutions $\varDelta_{{\rm s}}^{{\scriptscriptstyle \text{s}}}(\mathfrak{e})$
do not vanish but have a different shape. In Fig.~\ref{fig:IsoTemperatureLine},
we show the $\varDelta_{{\rm s}}^{{\scriptscriptstyle \text{s}}}(\mathfrak{e})$
with increasing splitting on the equal-temperature line at $T=10{\rm K}$
and the transition is clearly seen. In general, while before a critical
splitting $J_{{\rm {\scriptscriptstyle c}}}(T)$ the potential is
little effected by the splitting, past $J_{{\rm {\scriptscriptstyle c}}}(T)$
the solutions $\varDelta_{{\rm s}}^{{\scriptscriptstyle \text{s}}}(\mathfrak{e})$
localize at the $\varDelta_{n}^{{\scriptscriptstyle {\rm E}}}(J)$
Fermi level and show positive as well as negative regions. This behavior
is similar to the shape of the potential from the linearized $\breve{S}_{\beta}$
as given in Fig.~\ref{fig:TcVersusSplitting} b). We show the $T_{{\rm {\scriptscriptstyle c}}}(J)$
curve from the linear equation as a dashed blue line in Fig.~\ref{fig:PhaseDiagramNonLinear}
and see that it marks the border of the appearance of the curious
solutions in the non-linear equation past the range in $J$ of the
second order phase transition.

Due to the Coulomb renormalization, including the Coulomb repulsion,
$\int\varDelta_{{\rm s}}^{{\scriptscriptstyle \text{s}}}(\mathfrak{e}){\rm d}\varepsilon$
is predominantly negative. Thus, as a physical property, we compute
the number of condensed electrons $N_{{\rm {\scriptscriptstyle SC}}}=\int{\rm d}\boldsymbol{r}\int{\rm d}\boldsymbol{r}^{\prime}\vert\boldsymbol{\chi}(\boldsymbol{r},\boldsymbol{r}^{\prime})\vert^{2}$
instead. We show the SpinSCDFT $J-T$ diagram including the Coulomb
coupling in Fig.~\ref{fig:CoulombResultsNsc}. The region past the
transition has essentially no condensed electrons, while, still, the
$\varDelta_{{\rm s}}^{{\scriptscriptstyle \text{s}}}(\mathfrak{e})$
is not zero (compare Fig.~\ref{fig:CoulombResultsT3KIsoline}). The
SC $\rm{G}_0\rm{W}_0$ gap is shown in Fig.~\ref{fig:CoulombResultsG0W0Gap} and
is very similar to the results without the Coulomb interaction of
Fig.~\ref{fig:PhaseDiagramNonLinearG0W0}. Again, we find only small
features past the transition (compare also Fig.~\ref{fig:CoulombResultsDOST3KIsoline}).

\subsection{Numerical calculation of the DOS from the $\rm{G}_0\rm{W}_0$ GF}

We compute the GF according to the SC $\rm{G}_0\rm{W}_0$ scheme derived in the
Appendix \ref{sec:G0W0SC}. In detail,
we solve the Eq.~(\ref{eq:ManyBodyDOSKSSelfEnergy}) using the
Eqs.~(\ref{eq:G0W0SelfEnergyImSigmaOmega}) to (\ref{eq:G0W0SelfEnergyReSigmaEpsilon})
together with the Eqs.~(\ref{eq:ManyBodyWithKSSelfEnergyG11}) to
(\ref{eq:ManyBodyWithKSSelfEnergyG21}) for the model and couplings
described in Section \ref{sec:The-model-system}. We exclude the Coulomb
potential at this point for a better comparison with Eliashberg theory
although there is no difficulty to include it.

In Fig.~\ref{fig:PhaseDiagramNonLinearG0W0} we compute the $\rm{G}_0\rm{W}_0$
corrected DOS at every point in $J$ and $T$ and extract the SC excitation
gap. We find that the curious solutions past the transition $J_{{\rm {\scriptscriptstyle c}}}(T)$
lead to almost no excitation gap. The reason is that for the self-energy
in the calculation of the SC DOS in the Appendix \ref{sec:G0W0SC}
$\varDelta_{{\rm s}}^{{\scriptscriptstyle \text{s}}}(\mathfrak{e})$
is integrated in $\varepsilon$. If the high $\varepsilon$ region, away form the Fermi level
are strongly suppressed, as in the KS potential past the $J_{{\rm {\scriptscriptstyle c}}}(T)$,
the effect on the excitation gap is negligible.

Comparing with SpinSCDFT $\rm{G}_0\rm{W}_0$ gap of Fig.~\ref{fig:PhaseDiagramNonLinearG0W0}
with the BCS (Fig.~\ref{fig:BCSsolutions}) and the Eliashberg $J-T$
diagram (\ref{fig:PhaseDiagramEliashbergEquations}) we conclude that
the point of the transition can be clearly identified. Moreover
this one-cycle correction sheds light onto the appearance of the Fermi-level
localized solutions past the critical field $J_{{\rm {\scriptscriptstyle c}}}(T)$.
We have seen that for small $T$ and $J=0$ the non-linear $\varDelta_{{\rm s}}^{{\scriptscriptstyle \text{s}}}(\mathfrak{e})$
go to zero at the Fermi level (compare Fig.~\ref{fig:NonLinearSolutionsIsoSplittingLineJZero})
while the analogue of $\rm{G}_0\rm{W}_0$ GF, the excitation gap of Fig.~\ref{fig:PhaseDiagramNonLinearG0W0},
takes its largest value at $T=0$ and shows the expected monotonous
decay with temperature to $T_{{\rm {\scriptscriptstyle c}}}$.

This implies a significant difference in the quasi particle states
if a splitting occurs with such a $\varDelta_{{\rm s}}^{{\scriptscriptstyle \text{s}}}(\mathfrak{e})$.
While the KS particle with the dispersion $E_{\sigma}^{\alpha}={\rm sign}(\sigma)J+\alpha\sqrt{\varepsilon^{2}+{\varDelta_{{\rm s}}^{{\scriptscriptstyle \text{s}}}(\mathfrak{e})}^{2}}$
is strongly altered by the splitting because the Bogoliubov branches
change their order (compare the earlier discussion in the conclusion of
Sec.~\ref{sub:Temperature-Dependence-ofS} and in {\bf I}) this is not the case
in the true quasi particle structure. In fact, from Fig.~\ref{fig:NonLinearIsoTemperatureJ01},
we see that the SC solutions $\varDelta_{{\rm s}}^{{\scriptscriptstyle \text{s}}}(\mathfrak{e})$
if $J>0$ do not go to zero and, instead, rise with $J$ to prevent
this situation. On the other hand, after the discontinuous transition
we find $\varDelta_{{\rm s}}^{{\scriptscriptstyle \text{s}}}(0,J)<J$.

In the functional construction, the replacement $\bar{G}\rightarrow\bar{G}^{{\rm {\scriptscriptstyle KS}}}$
is thus a strong suspect for the occurrence of this curious solutions
past the SC transition. This is because $\bar{G}$ and $\bar{G}^{{\rm {\scriptscriptstyle KS}}}$
deviate in that the latter can be non-gapped while still corresponding
to a SC solution.

\subsection{Triplet components}

The present implementation of SpinSCDFT assumes the spin decoupling
approximation, i.e.~assumes the pairing to be
of spin singlet type (compare Sec.~\SpinSCDFTSDA). However, it was also shown in {\bf I} that a magnetic
splitting creates triplet components in the pairing potential, even
for a purely singlet order parameter density.
Triplet components appear as an intermediate step, in the self-energy
that leads to the G0-functional in {\bf I} since the Nambu off diagonal
upspin and downspin components are in general not equal and of opposite sign.
They can be intermediate since such triplet self-energy contributions lead to triplet as well
as singlet order parameter contributions. The intermediate triplet
self-energy that leads to singlet order parameter contributions can
be included in the spin decoupling approximation functional without difficulties.
From the theoretical side, this is an
unpleasant signature of formal inconsistency. We have, in fact, computed
the critical temperature and KS gaps with and without these intermediate
triplet self-energy terms. In Fig.~\ref{sub:TcFromLinear}
and Fig.~\ref{fig:TcVersusSplitting}, we observe that their effect
is negligibly small. The possibility of a triplet condensation, i.e.~non-vanishing
triplet order parameter contributions, in not investigated further
in this work.

\subsection{Extension to real materials}

In this work, properties of the free electron gas with a phonon
and Coulomb coupling subject to an homogeneous exchange splitting 
have been calculated.
To compute real materials without the use of adjustable parameters,
the electron-phonon coupling and the Coulomb potential has to be calculated
from first principles. Then, according to the equations (\aSqareFDiagonal),(\aSqareFOffDiagonal)
and (\CoulombCoupl) these couplings, as well as the computed single
particle states $\varepsilon_{k\sigma}$ may well have a distribution
in $J$ different from the homogeneous $J_{{\scriptscriptstyle 0}}=-\mu_{{\rm {\scriptscriptstyle B}}}B_{{\rm {\scriptscriptstyle 0}}}$
that we are considering here. Also, sometimes, several regions in
the Brillouin zone (or: in $k$) have different couplings and a different
SC pairing as in the well known case of ${\rm MgB}_{2}$ \cite{FlorisMgB2SCDFTStudy2005}.
The isotropic formulation does not have to be given up, often it is enough to group
this regions which we refer to as multi-band SC \cite{An2001,FlorisMgB2SCDFTStudy2005}.
We extend notation $\mathfrak{e}=(\varepsilon,J,b)\quad\int{\rm d}\mathfrak{e}=\int{\rm d}\varepsilon\int{\rm d}J\sum_{b}$
where $b$ labels the groups of quantum numbers $\{k\}$ sharing similar
pairing.

\section{Summary and Conclusion}

In this work, we have presented parameter free ab-initio calculations
of a superconductor in presence of an homogeneous exchange splitting
as for example the result of an external magnetic field. We have used
two approaches: A generalization of the Eliashberg approach and SpinSCDFT.
SCDFT allows the direct inclusion of Coulomb interactions
in a straightforward way, while its direct inclusion remains to be
problematic within Eliashberg where one has to rely on the $\mu^\star$ approach\cite{Scalapino1966,McMillanTransitionTemperatureStrongCoupledSuperconductors1968}.
The Eliashberg equations, on the other hand, provide the reference
for the phononic self-energy, allowing to understand and develop functionals
for SpinSCDFT.

We have implemented a code that solves the SpinSCDFT equations with
a linear and non-linear $xc$-potential and the non-linear Eliashberg
equations derived in {\bf I}. The $xc$ functional is derived in {\bf I}
from the Sham-Schl\"uter equation based on the replacement of the
interacting with the SC KS GF. We have investigated the behavior of
the $xc$-potential on a model of a free electron gas with a tunable,
homogeneous exchange splitting $J$, a phonon coupling that resembles
to the one of ${\rm MgB}_{2}$ and, optional, a static Coulomb interaction
in the Thomas-Fermi approximation. We compute the SC properties of
this system and find that in the regime of a second order phase transition
in the $T$ vs $J$ diagram, SpinSCDFT results in a curve that compares
similar in shape to the Eliashberg solutions. Removing the contributions
in the functional that arise from the normal state (Nambu diagonal)
part of the self-energy we arrive at a shape that is very similar
to the BCS behavior. Including the Coulomb interaction reduces the
critical temperature but otherwise does not largely effect the shape
of the $J-T$ diagram.

In agreement with BCS and Eliashberg, SpinSCDFT predicts a discontinuous
transition in $\varDelta_{{\rm s}}^{{\scriptscriptstyle {\rm s}}}(\mathfrak{e})$
for large $J$ except that the $\varDelta_{{\rm s}}^{{\scriptscriptstyle {\rm s}}}(\mathfrak{e})$
past the transition are not zero but have a curious shape that has
positive and negative values. Furthermore, the solutions $\varDelta_{{\rm s}}^{{\scriptscriptstyle {\rm s}}}(\mathfrak{e})$
increasingly adopt non-vanishing values more or less only directly at the Fermi level $\varepsilon\approx0$.
In addition, we find that the non-linear SpinSCDFT solutions go to
zero at $\varepsilon\approx0$ for $T\rightarrow0$ and thus the SC
KS GF is not gapped while the interacting and $\rm{G}_0\rm{W}_0$ GF is. Since we have noted in
{\bf I} that the low center of energy range is where $\varDelta_{{\rm s}}^{{\scriptscriptstyle {\rm s}}}(\mathfrak{e})$
compared with $J$ because the Bogoliubov eigenvalues at $\varepsilon=0$
read $E_{\sigma}^{\alpha}(0,J)={\rm sign}(\sigma)J+\alpha\vert\varDelta_{{\rm s}}^{{\scriptscriptstyle {\rm s}}}(0,J)\vert$
we believe that this range is crucial. In contrast to the ones before
the transition, the curious $\varDelta_{{\rm s}}^{{\scriptscriptstyle {\rm s}}}(\mathfrak{e})$
past $J_{{\scriptscriptstyle {\rm c}}}$ have $\vert J\vert>\vert\varDelta_{{\rm s}}^{{\scriptscriptstyle {\rm s}}}(0,J)\vert$.

We perform a $\rm{G}_0\rm{W}_0$ like correction to the GF where we solve the Dyson
equation with the same self-energy that we used originally for the
$xc$-potential construction. The resulting excitation spectrum (here
in the isotropic case the DOS) is gapped and behaves as one would expect
for a SC. From this result we conclude that a fitting technique of
the self-energy similar to A.~Sanna \textit{et al.}\cite{SannaMigdalFunctionalSCDFT2014} will allow us to reproduce
the $J-T$ diagram of Eliashberg while keeping the possibility to include
the Coulomb potential in addition to a numerically simple form where
the Matzubara summations can be computed analytically.

\appendix

\section{Quasi-particle Excitations from the One-cycle Interaction Green's
Function\label{sec:G0W0SC}}

The theoretical definition of SC is the existence of a non vanishing
order parameter $\chi$ (Eq.~\OrderParameterDefinition), while experimentally
SC are usually characterized by the properties of their excitation
spectrum, namely the single particle gap at the Fermi level \cite{Damascelli2003}. This
can be rather directly extracted from the solution to the Eliashberg
equations on the imaginary axis since $\varDelta_{n=0}^{{\scriptscriptstyle {\rm E}}}(J)$
is closely related this excitation gap itself \cite{CarbottePropertiesOfBosonExchangeSC1990}
and we use it in Fig.~\ref{fig:DeltaEliashberg40K10K}
for the $J-T$ diagram.

The SC KS system of SpinSCDFT is designed to reproduce the densities
of the interacting system not the quasi particle spectrum. On the
other hand, for a normal metal the KS particles are often in good
agreement with experiment so that the resulting KS excitation spectrum
is used as an approximation to the interacting quasi particle spectrum.
With the potential $\varDelta_{{\rm s}}^{{\scriptscriptstyle {\rm s}}}(\mathfrak{e})$
of Fig.~\ref{fig:NonLinearSolutionsIsoSplittingLineJZero} it turns
out in SCDFT, also for the zero field case\cite{SannaMigdalFunctionalSCDFT2014},
this is not always the case, since e.g.~for $T\rightarrow0$
the SC KS system is not gapped.

To predict a proper excitation spectrum
without having to solve the Many-Body problem self-consistently we
introduce the $\rm{G}_0\rm{W}_0$ approximation in the context of SC. This means
to solve the Dyson equation once while replacing the interacting
GF with the SC KS GF in the self-energy. Here we use the same approximations for the
self energy made to arrive at the functional in {\bf I} which means we
use $\bar{\varSigma}^{{\scriptscriptstyle {\rm KS}}}=\bar{\varSigma}[\bar{G}^{\scriptscriptstyle {\rm KS}}]$
instead of the true self-energy $\bar{\varSigma}[\bar{G}]$.

In this Section we work in the isotropic formulation but note that
the approach is easily generalized to the anisotropic case. We use
the notation $\mathfrak{e}=\varepsilon,J,b$ and the isotropic Dyson
equation
\begin{equation}
\bar{G}_{n}(\mathfrak{e})=\Bigl(\bigl(\bar{G}_{n}^{{\scriptscriptstyle {\rm KS}}}(\mathfrak{e})\bigr)^{-1}+\bar{\varSigma}_{n}^{{\scriptscriptstyle {\rm KS}}}(\mathfrak{e})\Bigr)^{-1}\,,\label{eq:IsotropicDyson}
\end{equation}
that follows form the assumption that the couplings depend on $k$
via the center of energy $\frac{\varepsilon_{k\uparrow}-\varepsilon_{-k,\downarrow}}{2}\rightarrow\varepsilon$
and the splitting $\frac{\varepsilon_{k\uparrow}-\varepsilon_{-k,\downarrow}}{2}\rightarrow J$
and the isotropic bands $b$ (that is a set of quantum numbers $\{k\}$).
We introduce the notation $\bar{G}_{n\sigma}^{{\scriptscriptstyle \alpha,\alpha^{\prime}}}(\mathfrak{e})=\hat{I}_{k\sigma}(\mathfrak{e})\bar{G}_{k\sigma,\pm k\pm\sigma}^{{\scriptscriptstyle \alpha,\alpha^{\prime}}}(\omega_{n})$.
The averaging procedure $\hat{I}_{k\sigma}(\mathfrak{e})$ on equal splitting and equal center of energy surfaces is defined in
Eq.~\AvaragingProcedure\ .
We refer to the non-vanishing matrix elements with a spin label that refers to the first index of $\bar{G}_{k\sigma,\pm k\pm\sigma}^{{\scriptscriptstyle \alpha,\alpha^{\prime}}}(\omega_{n})$
and similar for the self-energy.

\subsection{Imaginary Axis Formulation}

The inversion of the Dyson Eq.~\eqref{eq:IsotropicDyson} to compute the
GF explicitly is very analogous to the derivation of the Eliashberg equations
in {\bf I}, Sec.~\EliashbergCalc\ . We compute $\bar{G}_{n}(\mathfrak{e})$
via Eq.~\eqref{eq:IsotropicDyson} and the non-vanishing components are found to be (suppressing the arguments of
$\mathfrak{F}_{n\sigma}(\mathfrak{e}),\varSigma_{n}^{\omega}(\mathfrak{e}),\varSigma_{n}^{J}(\mathfrak{e}),A_{n}^{\omega z}(\mathfrak{e}),\varSigma_{n}^{{\scriptscriptstyle \Re\Delta}}(\mathfrak{e}),\varSigma_{n}^{{\scriptscriptstyle \Im\Delta}}(\mathfrak{e})$
and $\varSigma_{n}^{{\scriptscriptstyle t\pm}}(\mathfrak{e})$) 
\begin{eqnarray}
\bar{G}_{n\sigma}^{{\scriptscriptstyle 1,1}} & \!\!\!= & \!\!\frac{1}{2\mathfrak{F}_{n\sigma}}\sum_{\alpha}\frac{\mathfrak{F}_{n\sigma}\!+\!\alpha\bigl(\varepsilon\!+\!\varSigma_{n}^{\varepsilon}\!+\!\mbox{sign}(\sigma)A_{n}^{\omega z}\bigr)}{\text{i}\omega_{n}\!-\!\varSigma_{n}^{\omega}\!-\!\mbox{sign}(\sigma)\bigl(J\!+\!\varSigma_{n}^{J}\bigr)\!-\!\alpha\mathfrak{F}_{n\sigma}}\nonumber \\
\label{eq:ManyBodyWithKSSelfEnergyG11}\\
\bar{G}_{n\sigma}^{{\scriptscriptstyle -1,-1}} & \!\!\!= & \!\!\frac{1}{2\mathfrak{F}_{n-\sigma}}\sum_{\alpha}\frac{\mathfrak{F}_{n-\sigma}\!+\!\alpha\bigl(\varepsilon\!+\!\varSigma_{n}^{\varepsilon}\!-\!\mbox{sign}(\sigma)A_{n}^{\omega z}\bigr)}{\text{i}\omega_{n}\!-\!\varSigma_{n}^{\omega}\!+\!\mbox{sign}(\sigma)\bigl(J\!+\!\varSigma_{n}^{J}\bigr)\!+\!\alpha\mathfrak{F}_{n-\sigma}}\nonumber \\
\\
\bar{G}_{n\sigma}^{{\scriptscriptstyle 1,-1}} & \!\!\!= & \!\!\frac{\mbox{sign}(\sigma)}{2\mathfrak{F}_{n\sigma}}\sum_{\alpha}\!\!\frac{\alpha\Bigl(\!\varSigma_{n}^{{\scriptscriptstyle \Re\Delta}}\!+\!\mbox{i}\varSigma_{n}^{{\scriptscriptstyle \Im\Delta}}\!+\!\mbox{sign}(\sigma)\bigl(\varSigma_{n}^{{\scriptscriptstyle t-}}\!+\!\varSigma_{n}^{{\scriptscriptstyle t+}}\bigr)\Bigr)}{\text{i}\omega_{n}\!-\!\varSigma_{n}^{\omega}\!-\!\mbox{sign}(\sigma)\bigl(J\!+\!\varSigma_{n}^{J}\bigr)\!-\!\alpha\mathfrak{F}_{n\sigma}}\nonumber \\
\\
\bar{G}_{n\sigma}^{{\scriptscriptstyle -1,1}} & \!\!\!= & \!\!\frac{\mbox{sign}(\sigma)}{2\mathfrak{F}_{n-\sigma}}\sum_{\alpha}\!\!\frac{\alpha\Bigl(\!\varSigma_{n}^{{\scriptscriptstyle \Re\Delta}}\!-\!\mbox{i}\varSigma_{n}^{{\scriptscriptstyle \Im\Delta}}\!+\!\mbox{sign}(\sigma)\bigl(\varSigma_{n}^{{\scriptscriptstyle t-}}\!-\!\varSigma_{n}^{{\scriptscriptstyle t+}}\bigr)\Bigr)}{\text{i}\omega_{n}\!-\!\varSigma_{n}^{\omega}\!+\!\mbox{sign}(\sigma)\bigl(J\!+\!\varSigma_{n}^{J}\bigr)\!+\!\alpha\mathfrak{F}_{n-\sigma}}\nonumber \\
\label{eq:ManyBodyWithKSSelfEnergyG21}
\end{eqnarray}
with
\begin{eqnarray}
\mathfrak{F}_{n\sigma}(\mathfrak{e}) & = & \Bigl(\bigl(\varepsilon+\varSigma_{n}^{\varepsilon}+\mbox{sign}(\sigma)A_{n}^{\omega z}\bigr)^{2}+\nonumber \\
 &  & +\bigl(\varSigma_{n}^{{\scriptscriptstyle \Re\Delta}}+\mbox{i}\varSigma_{n}^{{\scriptscriptstyle \Im\Delta}}+\mbox{sign}(\sigma)(\varSigma_{n}^{{\scriptscriptstyle t+}}+\varSigma_{n}^{{\scriptscriptstyle t-}})\bigr)\times\nonumber \\
 &  & \times\bigl(\varSigma_{n}^{{\scriptscriptstyle \Re\Delta}}-\mbox{i}\varSigma_{n}^{{\scriptscriptstyle \Im\Delta}}+\mbox{sign}(\sigma)(\varSigma_{n}^{{\scriptscriptstyle t+}}-\varSigma_{n}^{{\scriptscriptstyle t-}})\bigr)\Bigr)^{\frac{1}{2}}\nonumber \\
\label{eq:MBPhiTerm}
\end{eqnarray}
where the self-energy parts are constructed similar to the Eliashberg
theory with the result
\begin{eqnarray}
\varSigma_{n}^{\omega} & = & \frac{1}{4}\sum_{\sigma}\bigl(\bar{\varSigma}_{\sigma n}^{{\scriptscriptstyle {\rm KS}}1,1}+\bar{\varSigma}_{\sigma n}^{{\scriptscriptstyle {\rm KS}}-1,-1}\bigr)\label{eq:KSSelfEnergyOmega}\\
A_{n}^{\omega z} & = & \frac{1}{4}\sum_{\sigma}\text{sign}(\sigma)\bigl(\bar{\varSigma}_{\sigma n}^{{\scriptscriptstyle {\rm KS}}1,1}+\bar{\varSigma}_{\sigma n}^{{\scriptscriptstyle {\rm KS}}-1,-1}\bigr)\\
\varSigma_{n}^{\varepsilon} & = & \frac{1}{4}\sum_{\sigma}\bigl(\bar{\varSigma}_{\sigma n}^{{\scriptscriptstyle {\rm KS}}1,1}-\bar{\varSigma}_{\sigma n}^{{\scriptscriptstyle {\rm KS}}-1,-1}\bigr)\\
\varSigma_{n}^{J} & = & \frac{1}{4}\sum_{\sigma}\text{sign}(\sigma)\bigl(\bar{\varSigma}_{\sigma n}^{{\scriptscriptstyle {\rm KS}}1,1}-\bar{\varSigma}_{\sigma n}^{{\scriptscriptstyle {\rm KS}}-1,-1}\bigr)\\
\varSigma_{n}^{{\scriptscriptstyle t+}} & = & \frac{1}{4}\sum_{\sigma}\bigl(\bar{\varSigma}_{\sigma n}^{{\scriptscriptstyle {\rm KS}}1,-1}+\bar{\varSigma}_{\sigma n}^{{\scriptscriptstyle {\rm KS}}-1,1}\bigr)\\
\varSigma_{n}^{{\scriptscriptstyle t-}} & = & \frac{1}{4}\sum_{\sigma}\bigl(\bar{\varSigma}_{\sigma n}^{{\scriptscriptstyle {\rm KS}}1,-1}-\bar{\varSigma}_{\sigma n}^{{\scriptscriptstyle {\rm KS}}-1,1}\bigr)\\
\varSigma_{n}^{{\scriptscriptstyle \Im\Delta}} & = & \frac{-\mbox{i}}{4}\sum_{\sigma}\text{sign}(\sigma)\bigl(\bar{\varSigma}_{\sigma n}^{{\scriptscriptstyle {\rm KS}}1,-1}+\bar{\varSigma}_{\sigma n}^{{\scriptscriptstyle {\rm KS}}-1,1}\bigr)\\
\varSigma_{n}^{{\scriptscriptstyle \Re\Delta}} & = & \frac{1}{4}\sum_{\sigma}\text{sign}(\sigma)\bigl(\bar{\varSigma}_{\sigma n}^{{\scriptscriptstyle {\rm KS}}1,-1}-\bar{\varSigma}_{\sigma n}^{{\scriptscriptstyle {\rm KS}}-1,1}\bigr)\,.\label{eq:KSSelfEnergyRealDelta}
\end{eqnarray}
Note, however, that $\bar{\varSigma}_{\sigma n}^{{\scriptscriptstyle {\rm KS}}1,-1}$
contains a triplet contribution that is generated by the coupling
imbalance of the spin channels. The isotropic variants of the Eqs.~(\KSSEPhOneOne)
to (\KSSEPhMinusOneMinusOne) are given by
\begin{eqnarray}
{\varSigma_{{\scriptscriptstyle \text{ph}}}^{{\scriptscriptstyle {\rm KS}}}}_{\sigma n}^{{\scriptscriptstyle 1,1}} & = & \int\hspace{-0.18cm}\mbox{d}\varOmega\int\hspace{-0.18cm}\mbox{d}\mathfrak{e}^{\prime}\,\alpha^{\!2}\! F_{\sigma}^{{\scriptscriptstyle {\rm D}}}(\mathfrak{e},\mathfrak{e}^{\prime},\varOmega)\times\nonumber \\
 &  & \hspace{-0.18cm}\times\sum_{\alpha}\frac{\alpha\varepsilon^{\prime}+F^{\prime}}{2F^{\prime}}M_{{\scriptscriptstyle {\rm ph}}}(\varOmega,{E_{\sigma}^{\alpha}}^{\prime},\omega_{n})\\
{\varSigma_{{\scriptscriptstyle \text{ph}}}^{{\scriptscriptstyle {\rm KS}}}}_{\sigma n}^{{\scriptscriptstyle -1,-1}} & = & \int\hspace{-0.18cm}\mbox{d}\varOmega\int\hspace{-0.18cm}\mbox{d}\mathfrak{e}^{\prime}\,\alpha^{\!2}\! F_{\sigma}^{{\scriptscriptstyle {\rm D}}}(\mathfrak{e},\mathfrak{e}^{\prime},\varOmega)\times\nonumber \\
 &  & \hspace{-0.18cm}\times\sum_{\alpha}\frac{\alpha\varepsilon^{\prime}+F^{\prime}}{2F^{\prime}}M_{{\scriptscriptstyle {\rm ph}}}(\varOmega,-{E_{\sigma}^{\alpha}}^{\prime},\omega_{n})\\
{\varSigma_{{\scriptscriptstyle \text{ph}}}^{{\scriptscriptstyle {\rm KS}}}}_{\sigma n}^{{\scriptscriptstyle 1,-1}} & = & -\mbox{sign}(\sigma)\int\hspace{-0.18cm}\mbox{d}\varOmega\int\hspace{-0.18cm}\mbox{d}\mathfrak{e}^{\prime}\,\alpha^{\!2}\! F(\mathfrak{e},\mathfrak{e}^{\prime},\varOmega)\times\nonumber \\
 &  & \hspace{-0.18cm}\times\sum_{\alpha}\frac{\alpha\varDelta_{{\rm s}}^{{\scriptscriptstyle {\rm s}}\prime}}{2F^{\prime}}M_{{\scriptscriptstyle {\rm ph}}}(\varOmega,{E_{\sigma}^{\alpha}}^{\prime},\omega_{n})\label{eq:EliashbergNambuOffDiagonalSelfEnergy}\\
{\varSigma_{{\scriptscriptstyle \text{ph}}}^{{\scriptscriptstyle {\rm KS}}}}_{\sigma n}^{{\scriptscriptstyle -1,1}} & = & -\mbox{sign}(\sigma)\int\hspace{-0.18cm}\mbox{d}\varOmega\int\hspace{-0.18cm}\mbox{d}\mathfrak{e}^{\prime}\,\alpha^{\!2}\! F(\mathfrak{e},\mathfrak{e}^{\prime},\varOmega)\times\nonumber \\
 &  & \hspace{-0.18cm}\times\sum_{\alpha}\frac{\alpha{\varDelta_{{\rm s}}^{{\scriptscriptstyle {\rm s}}\prime}}^{\ast}}{2F^{\prime}}M_{{\scriptscriptstyle {\rm ph}}}(\varOmega,-{E_{\sigma}^{\alpha}}^{\prime},\omega_{n})\label{eq:EliashbergNambuOffDiagonalSelfEnergy21}
\end{eqnarray}
with $\varDelta_{{\rm s}}^{{\scriptscriptstyle {\rm s}}\prime}$ short
hand for $\varDelta_{{\rm s}}^{{\scriptscriptstyle {\rm s}}}(\mathfrak{e}^{\prime})$,
the averaged ${\varDelta_{{\rm s}}^{{\scriptscriptstyle {\rm s}}}}_{k}$
and $F^{\prime}=\sqrt{{\varepsilon^{\prime}}^{2}+{\varDelta_{{\rm s}}^{{\scriptscriptstyle {\rm s}}\prime}}^{2}}$.
Furthermore ${E_{\sigma}^{\alpha}}^{\prime}=\mbox{sign}(\sigma)J+\text{sign}(\alpha)F^{\prime}$
and similarly the Eqs.~(\KSSECoulOneMinusOne) and ~(\KSSECoulMinusOneOne)
become
\begin{eqnarray}
{\varSigma_{{\rm {\scriptscriptstyle C}}}^{{\scriptscriptstyle {\rm KS}}}}_{\sigma n}^{{\scriptscriptstyle 1,-1}} & \hspace{-0.18cm}= & \!\!-\mbox{sign}(\sigma)\!\sum_{\alpha}\int\hspace{-0.18cm}\mbox{d}\mathfrak{e}^{\prime}\frac{\alpha\varDelta_{s}^{{\scriptscriptstyle \text{s}}\prime}}{2F^{\prime}}C^{{\scriptscriptstyle \text{stat}}}(\mathfrak{e},\mathfrak{e}^{\prime})f_{\beta}({E_{\sigma}^{\alpha}}^{\prime})\nonumber \\
\\
{\varSigma_{{\rm {\scriptscriptstyle C}}}^{{\scriptscriptstyle {\rm KS}}}}_{\sigma n}^{{\scriptscriptstyle -1,1}} & \hspace{-0.18cm}= & \!\!-\mbox{sign}(\sigma)\!\sum_{\alpha}\int\hspace{-0.18cm}\mbox{d}\mathfrak{e}^{\prime}\frac{\alpha{\varDelta_{s}^{{\scriptscriptstyle \text{s}}\prime}}^{\ast}}{2F^{\prime}}C^{{\scriptscriptstyle \text{stat}}}(\mathfrak{e},\mathfrak{e}^{\prime})f_{\beta}(-{E_{\sigma}^{\alpha}}^{\prime})\,.\nonumber \\
\end{eqnarray}
With these equations we can compute the $\rm{G}_0\rm{W}_0$ GF from the results of
a converged SpinSCDFT calculation.

\subsection{Real Axis Formulation\label{sub:MBRealAxisFormulation}}

To obtain the (L)DOS from the temperature GF we substitute 
\begin{equation}
\mbox{i}\omega_{n}\rightarrow\omega+\mbox{i}\eta
\end{equation}
where $\eta$ is a real positive infinitesimal \cite{fetterWaleckaQuantumTheoryOfManyParticles1971}.
The expression Eqs.~(\ref{eq:ManyBodyWithKSSelfEnergyG11}) to (\ref{eq:ManyBodyWithKSSelfEnergyG21})
remain essentially unchanged on the real axis, except that we have
to insert the SE parts Eq.~(\ref{eq:KSSelfEnergyOmega}) to (\ref{eq:KSSelfEnergyRealDelta})
on the real axis and write $\text{i}\eta+\omega$ instead of the Matsubara
frequency. Here we have two options, first we may compute the SE parts
on the imaginary axis and use a numerical analytic continuation to
the real axis, or we can compute analytic formulas for the real axis
and use them. We choose the latter because this avoids the sometimes
unstable analytical continuation.

We will see that the SE parts, e.g.~$\varSigma_{n}^{{\scriptscriptstyle \Re\Delta}}(\mathfrak{e})$,
on the real axis have to be computed via independent calculations
of imaginary and real part. The dependence on the Matsubara index
of the SE is only via the function $M_{{\scriptscriptstyle {\rm ph}}}$
of Eq.~(\MatsubaraIntegralMph), i.e.~the results of the first Matsubara
summation in the SE. Thus on the real axis
\begin{eqnarray}
M_{{\scriptscriptstyle {\rm ph}}}(\varOmega,E,\omega) & = & \hat{{\rm P}}\frac{n_{\beta}(\varOmega)+f_{\beta}(E)}{\varOmega-E+\omega}-\hat{{\rm P}}\frac{n_{\beta}(\varOmega)+f_{\beta}(-E)}{\varOmega+E-\omega}\nonumber \\
 &  & \hspace{-0.5cm}\hspace{-0.5cm}-\text{i}\pi\Bigl(\bigl(n_{\beta}(\varOmega)+f_{\beta}(E)\bigr)\updelta(\varOmega-E+\omega)\nonumber \\
 &  & \hspace{-0.5cm}\hspace{-0.5cm}+\bigl(n_{\beta}(\varOmega)+f_{\beta}(-E)\bigr)\updelta(\varOmega+E-\omega)\Bigr)\,.
\end{eqnarray}
Here $\hat{{\rm P}}$ is the principle value operator. Because of
the very different nature of the imaginary and real part of the SE
we compute both parts independently. Then we obtain\begin{widetext}
\begin{eqnarray}
\Im\varSigma^{\omega}(\mathfrak{e},\omega) & = & -\pi\int\hspace{-0.18cm}\mbox{d}\mathfrak{e}^{\prime}\sum_{\mu\alpha}\frac{\alpha\varepsilon^{\prime}+F^{\prime}}{8F^{\prime}}\Bigl(\bigl(n_{\beta}({E_{\mu}^{\alpha}}^{\prime}-\omega)+f_{\beta}({E_{\mu}^{\alpha}}^{\prime})\bigr)\bigl(\alpha^{\!2}\! F_{\mu}^{{\scriptscriptstyle {\rm D}}}(\mathfrak{e},\mathfrak{e}^{\prime},{E_{\mu}^{\alpha}}^{\prime}-\omega)-\alpha^{\!2}\! F_{\mu}^{{\scriptscriptstyle {\rm D}}}(\mathfrak{e},\mathfrak{e}^{\prime},\omega-{E_{\mu}^{\alpha}}^{\prime})\bigr)+\nonumber \\
 &  & +\bigl(n_{\beta}({E_{\mu}^{\alpha}}^{\prime}+\omega)+f_{\beta}({E_{\mu}^{\alpha}}^{\prime})\bigr)\bigl(\alpha^{\!2}\! F_{\mu}^{{\scriptscriptstyle {\rm D}}}(\mathfrak{e},\mathfrak{e}^{\prime},{E_{\mu}^{\alpha}}^{\prime}+\omega)-\alpha^{\!2}\! F_{\mu}^{{\scriptscriptstyle {\rm D}}}(\mathfrak{e},\mathfrak{e}^{\prime},-{E_{\mu}^{\alpha}}^{\prime}-\omega)\bigr)\Bigr)\label{eq:G0W0SelfEnergyImSigmaOmega}\\
\Re\varSigma^{\omega}(\mathfrak{e},\omega) & = & \int\hspace{-0.18cm}\mbox{d}\varOmega\int\hspace{-0.18cm}\mbox{d}\mathfrak{e}^{\prime}\sum_{\mu\alpha}\frac{\alpha\varepsilon^{\prime}+F^{\prime}}{8F^{\prime}}\alpha^{\!2}\! F_{\mu}^{{\scriptscriptstyle {\rm D}}}(\mathfrak{e}\!,\!\mathfrak{e}^{\prime},\varOmega)\Bigl(\hat{{\rm P}}\frac{n_{\beta}(\varOmega)+f_{\beta}({E_{\mu}^{\alpha}}^{\prime})}{\varOmega-{E_{\mu}^{\alpha}}^{\prime}+\omega}\nonumber \\
 &  & -\hat{{\rm P}}\frac{n_{\beta}(\varOmega)+f_{\beta}({E_{\mu}^{\alpha}}^{\prime})}{\varOmega-{E_{\mu}^{\alpha}}^{\prime}-\omega}-\hat{{\rm P}}\frac{n_{\beta}(\varOmega)+f_{\beta}(-{E_{\mu}^{\alpha}}^{\prime})}{\varOmega+{E_{\mu}^{\alpha}}^{\prime}-\omega}+\hat{{\rm P}}\frac{n_{\beta}(\varOmega)+f_{\beta}(-{E_{\mu}^{\alpha}}^{\prime})}{\varOmega+{E_{\mu}^{\alpha}}^{\prime}+\omega}\Bigr)\\
\Im\varSigma^{\varepsilon}(\mathfrak{e},\omega) & = & -\pi\int\hspace{-0.18cm}\mbox{d}\mathfrak{e}^{\prime}\sum_{\mu\alpha}\frac{\alpha\varepsilon^{\prime}+F^{\prime}}{8F^{\prime}}\Bigl(\bigl(n_{\beta}({E_{\mu}^{\alpha}}^{\prime}-\omega)+f_{\beta}({E_{\mu}^{\alpha}}^{\prime})\bigr)\bigl(\alpha^{\!2}\! F_{\mu}^{{\scriptscriptstyle {\rm D}}}(\mathfrak{e},\mathfrak{e}^{\prime},{E_{\mu}^{\alpha}}^{\prime}-\omega)-\alpha^{\!2}\! F_{\mu}^{{\scriptscriptstyle {\rm D}}}(\mathfrak{e},\mathfrak{e}^{\prime},\omega-{E_{\mu}^{\alpha}}^{\prime})\bigr)\nonumber \\
 &  & -\bigl(n_{\beta}({E_{\mu}^{\alpha}}^{\prime}+\omega)+f_{\beta}({E_{\mu}^{\alpha}}^{\prime})\bigr)\bigl(\alpha^{\!2}\! F_{\mu}^{{\scriptscriptstyle {\rm D}}}(\mathfrak{e},\mathfrak{e}^{\prime},{E_{\mu}^{\alpha}}^{\prime}+\omega)-\alpha^{\!2}\! F_{\mu}^{{\scriptscriptstyle {\rm D}}}(\mathfrak{e},\mathfrak{e}^{\prime},-{E_{\mu}^{\alpha}}^{\prime}-\omega)\bigr)\Bigr)\\
\Re\varSigma^{\varepsilon}(\mathfrak{e},\omega) & = & \int\hspace{-0.18cm}\mbox{d}\varOmega\int\hspace{-0.18cm}\mbox{d}\mathfrak{e}^{\prime}\sum_{\mu\alpha}\frac{\alpha\varepsilon^{\prime}+F^{\prime}}{8F^{\prime}}\alpha^{\!2}\! F_{\mu}^{{\scriptscriptstyle {\rm D}}}(\mathfrak{e},\mathfrak{e}^{\prime},\varOmega)\Bigl(\hat{{\rm P}}\frac{n_{\beta}(\varOmega)+f_{\beta}({E_{\mu}^{\alpha}}^{\prime})}{\varOmega-{E_{\mu}^{\alpha}}^{\prime}+\omega}\nonumber \\
 &  & +\hat{{\rm P}}\frac{n_{\beta}(\varOmega)+f_{\beta}({E_{\mu}^{\alpha}}^{\prime})}{\varOmega-{E_{\mu}^{\alpha}}^{\prime}-\omega}-\hat{{\rm P}}\frac{n_{\beta}(\varOmega)+f_{\beta}(-{E_{\mu}^{\alpha}}^{\prime})}{\varOmega+{E_{\mu}^{\alpha}}^{\prime}-\omega}-\hat{{\rm P}}\frac{n_{\beta}(\varOmega)+f_{\beta}(-{E_{\mu}^{\alpha}}^{\prime})}{\varOmega+{E_{\mu}^{\alpha}}^{\prime}+\omega}\Bigr)\label{eq:G0W0SelfEnergyReSigmaEpsilon}
\end{eqnarray}
\end{widetext}and very similar for $A^{\omega z}(\mathfrak{e}\omega)$
that only differs from $\varSigma^{\omega}$ by putting a $\text{sign}(\mu)$
into the spin sums. We also obtain $\varSigma^{J}(\mathfrak{e}\omega)$
from the relation for $\varSigma^{\varepsilon}(\mathfrak{e},\omega)$
in the same way, i.e.~we put a $\text{sign}(\mu)$ into the spin
sum. The above equation again points out the problem in the $\varepsilon^{\prime}$
integral if the energy dependence of $\alpha^{\!2}\! F_{\mu}^{{\scriptscriptstyle {\rm D}}}(\mathfrak{e}\!,\!\mathfrak{e}^{\prime},\varOmega)$
is neglected. Here ${E_{\mu}^{\alpha}}^{\prime}\rightarrow\alpha\vert\varepsilon^{\prime}\vert$
for large $\vert\varepsilon^{\prime}\vert$ so there are parts in the integral
that behave as $\frac{1}{\varepsilon^{\prime}}$ leading to logarithmic
divergence. Thus we see explicitly that we cannot compute the energy
renormalization without considering the influence of the interaction
on the full energy spectrum and quasi-particle occupations as was already
discussed in {\bf I} and Ref.~\onlinecite{LuedersSCDFTI2005}.

We define the integrand
\begin{eqnarray}
 &  & \hspace{-0.25cm}\Im\mathcal{B}_{\pm}(\mathfrak{e},\mathfrak{e}^{\prime},\omega)=\nonumber \\
 &  & \pi\sum_{\mu\alpha}\mbox{sign}(\mu)^{\frac{1\pm1}{2}}\frac{\text{sign}(\alpha)}{2F^{\prime}}\bigl(n_{\beta}({E_{\mu}^{\alpha}}^{\prime}-\omega)+f_{\beta}({E_{\mu}^{\alpha}}^{\prime})\bigr)\times\nonumber \\
 &  & \times\bigl(\alpha^{\!2}\! F_{\mu}^{{\scriptscriptstyle {\rm D}}}(\mathfrak{e},\mathfrak{e}^{\prime},{E_{\mu}^{\alpha}}^{\prime}-\omega)\!-\!\alpha^{\!2}\! F_{\mu}^{{\scriptscriptstyle {\rm D}}}(\mathfrak{e},\mathfrak{e}^{\prime},\omega-{E_{\mu}^{\alpha}}^{\prime})\bigr)\\
 &  & \hspace{-0.25cm}\Re\mathcal{B}_{\pm}(\mathfrak{e},\mathfrak{e}^{\prime},\omega)\nonumber \\
 &  & =-\sum_{\mu\alpha}\mbox{sign}(\mu)^{\frac{1\pm1}{2}}\frac{\text{sign}(\alpha)}{4F^{\prime}}\biggl(\int\hspace{-0.18cm}\mbox{d}\varOmega\alpha^{\!2}\! F_{\mu}^{{\scriptscriptstyle {\rm D}}}(\mathfrak{e},\mathfrak{e}^{\prime},\varOmega)\times\nonumber \\
 &  & \times\Bigl(\hat{{\rm P}}\frac{n_{\beta}(\varOmega)+f_{\beta}({E_{\mu}^{+}}^{\prime})}{\varOmega-{E_{\mu}^{+}}^{\prime}+\omega}-\hat{{\rm P}}\frac{n_{\beta}(\varOmega)+f_{\beta}(-{E_{\mu}^{\alpha}}^{\prime})}{\varOmega+{E_{\mu}^{\alpha}}^{\prime}-\omega}\Bigr)\!+\nonumber \\
 &  & +f_{\beta}({E_{\mu}^{\alpha}}^{\prime})C^{{\scriptscriptstyle \text{stat}}}(\mathfrak{e},\mathfrak{e}^{\prime})\biggr)
\end{eqnarray}
and further introducing
\begin{eqnarray}
B_{n}^{s}(\mathfrak{e}) & \equiv & \varSigma_{n}^{{\scriptscriptstyle \Re\Delta}}(\mathfrak{e})+\mbox{i}\varSigma_{n}^{{\scriptscriptstyle \Im\Delta}}(\mathfrak{e})\\
B_{n}^{s\star}(\mathfrak{e}) & \equiv & \varSigma_{n}^{{\scriptscriptstyle \Re\Delta}}(\mathfrak{e})-\mbox{i}\varSigma_{n}^{{\scriptscriptstyle \Im\Delta}}(\mathfrak{e})\\
B_{n}^{t}(\mathfrak{e}) & \equiv & \varSigma_{n}^{{\scriptscriptstyle t+}}(\mathfrak{e})+\varSigma_{n}^{{\scriptscriptstyle t-}}(\mathfrak{e})\\
B_{n}^{t\star}(\mathfrak{e}) & \equiv & \varSigma_{n}^{{\scriptscriptstyle t+}}(\mathfrak{e})-\varSigma_{n}^{{\scriptscriptstyle t-}}(\mathfrak{e})
\end{eqnarray}
we obtain the following equations on the real axis
\begin{eqnarray}
B^{s}(\mathfrak{e},\omega) & = & \hspace{-0.18cm}\int\!\!\mbox{d}\mathfrak{e}^{\prime}{\varDelta_{s}^{{\scriptscriptstyle \text{s}}}}^{\prime}\mathcal{B}_{-}(\mathfrak{e},\mathfrak{e}^{\prime},\omega)\label{eq:ManyBodyDOSWithKSSelfEnergyParingTermsB}\\
B^{s\star}(\mathfrak{e},\omega) & = & \hspace{-0.18cm}\int\!\!\mbox{d}\mathfrak{e}^{\prime}{\varDelta_{s}^{{\scriptscriptstyle \text{s}}\prime}}^{\ast}\mathcal{B}_{-}(\mathfrak{e},\mathfrak{e}^{\prime},\omega)\\
B^{t}(\mathfrak{e},\omega) & = & \hspace{-0.18cm}\int\!\!\mbox{d}\mathfrak{e}^{\prime}{\varDelta_{s}^{{\scriptscriptstyle \text{s}}}}^{\prime}\mathcal{B}_{+}(\mathfrak{e},\mathfrak{e}^{\prime},\omega)\\
B^{t\star}(\mathfrak{e},\omega) & = & \hspace{-0.18cm}\int\!\!\mbox{d}\mathfrak{e}^{\prime}{\varDelta_{s}^{{\scriptscriptstyle \text{s}}\prime}}^{\ast}\mathcal{B}_{+}(\mathfrak{e},\mathfrak{e}^{\prime},\omega)
\end{eqnarray}
and thus, Eq.~(\ref{eq:MBPhiTerm}) becomes on the real axis (omitting
the arguments $\mathfrak{e},\omega$) 
\begin{eqnarray}
\mathfrak{\mathfrak{F}}_{\sigma} & = & \Bigl(\bigl(\varepsilon+\varSigma^{\varepsilon}+\mbox{sign}(\sigma)A^{\omega z}\bigr)^{2}\nonumber \\
 &  & \hspace{-0.5cm}+\bigl(B^{s}+\mbox{sign}(\sigma)B^{t}\bigr)\bigl(B^{s\star}+\mbox{sign}(\sigma)B^{t\star}\bigr)\Bigr)^{\frac{1}{2}}\,.
\end{eqnarray}
Now we can finally obtain the retarded GF with the equations from
Eqs.~(\ref{eq:ManyBodyWithKSSelfEnergyG11}) to (\ref{eq:ManyBodyWithKSSelfEnergyG21})
together with Eq.~(\ref{eq:MBPhiTerm}) for $\mathfrak{F}_{n\sigma}(\mathfrak{e})$
in terms of $B$ and the corresponding SE parts constructed from real
and imaginary part close to the real axis. Then we can evaluate the
DOS according to
\begin{equation}
\rho_{\sigma\alpha}(\omega)=-2\int\hspace{-0.18cm}\mbox{d}\mathfrak{e}\Im\bigl(\lim_{\eta\rightarrow0}\lim_{i\omega_{n}\rightarrow\omega+\mbox{i}\eta}\bar{G}_{n\sigma}^{{\scriptscriptstyle \alpha,\alpha}}(\mathfrak{e})\bigr)\varrho(\mathfrak{e})\label{eq:ManyBodyDOSKSSelfEnergy}
\end{equation}
We obtain the local DOS $\rho_{\sigma\alpha}(\mathbf{r},\omega)$
simply by replacing $\varrho(\mathfrak{e})$ with the local double
DOS $\varrho_{\sigma}(\mathfrak{e},\boldsymbol{r})=\hat{I}_{k\sigma}(\mathfrak{e})\vert\varphi_{k}(\boldsymbol{r}\sigma)\vert^{2}$.

\bibliographystyle{apsrev4-1}
\bibliography{BibTeX,references}

\end{document}